\documentclass[%
 reprint,
 amsmath,amssymb,
 aps,
]{revtex4-2}

\usepackage[utf8]{inputenc}
\usepackage{CJKutf8}

\usepackage{amsthm}
\usepackage{amsmath}
\usepackage{amssymb}
\usepackage{mathrsfs}
\usepackage{bbm}
\usepackage{bm}
\usepackage{quantikz}
\usepackage{graphicx}% Include figure files
\usepackage{dcolumn}% Align table columns on decimal point
\usepackage{hyperref}
\usepackage{xcolor,soul}

\usepackage[english]{babel}
\usepackage{algorithm}
\usepackage{algpseudocode}

\floatname{algorithm}{Protocol}

\renewcommand{\proj}[1]{| #1 \rangle\!\langle #1 |}

\DeclareMathOperator{\Tr}{Tr}

\DeclareMathOperator{\erf}{erf}

\DeclareMathOperator*{\argmax}{arg\,max}

\DeclareMathOperator{\polylog}{polylog}

\def \A {\normalfont{\textsf{A}}}
\def \B {\normalfont{\textsf{B}}}

\def \P {\normalfont{\textsf{P}}}
\def \Q {\normalfont{\textsf{Q}}}
\def \R {\normalfont{\textsf{R}}}
\def \CNOT {\normalfont{\textsf{CNOT}}}

\def \Cl {\normalfont{\texttt{Cl}}}
\def \Stab {\normalfont{\texttt{Stab}}}
\def \CBS {\normalfont{\texttt{CBS}}}
\def \MUBS {\normalfont{\texttt{MUBS}}}
\def \Pauli {\normalfont{\texttt{Pauli}}}
\def \UDS {\normalfont{\texttt{UDS}}}

\def \ic {\mathcal{I}}
\def \id {\mathbbm{I}}
\def \iu {\mathrm{i}}
\def \d {\mathrm{d}}

\def \ket#1{|#1\rangle}
\def \ketbra#1#2{|#1\rangle\! \langle #2|}
\def \sandwich#1#2#3{\langle#1|#2|#3\rangle}

\def \mod {\rm \ mod \ }
\def \sym {\textnormal{sym}}

\def \cor {\textnormal{cor}}
\def \res {\textnormal{res}}
\def \suc {\textnormal{suc}}
\def \fak {\textnormal{fak}}
\def \Sch {\textnormal{Sch}}

\def \F {\mathcal{F}}
\def \N {\mathcal{N}}

\usepackage[normalem]{ulem} 
\usepackage{color}

\newtheorem{theorem}{Theorem}

\newtheorem{corollary}[theorem]{Corollary}
\newtheorem{lemma}[theorem]{Lemma}

\newenvironment{remark} {\begin{proof}[Remark]} {\phantom\qedhere\end{proof}}
\begin{document}
\begin{CJK*}{UTF8}{gbsn}

\preprint{APS/123-QED}

\title{Fundamental limits on quantum cloning from the no-signalling principle}% Force line breaks with \\

\author{Yanglin Hu (胡杨林)}
\email{yanglin.hu@u.nus.edu}
\affiliation{%
 Centre for Quantum Technologies, 
 National University of Singapore, Singapore 117543, Singapore
}%

\author{Marco Tomamichel}%
\email{marco.tomamichel@nus.edu.sg}
\affiliation{%
 Centre for Quantum Technologies, 
 National University of Singapore, Singapore 117543, Singapore
}%
\affiliation{%
 Department of Electrical and Computer Engineering, 
 National University of Singapore, Singapore 117583, Singapore
}%
%\altaffiliation[Also at ]{Physics Department, XYZ University.}

\date{\today}

\begin{abstract}
    The no-cloning theorem is a cornerstone of quantum cryptography. 
    %\st{Here we generalize and rederive under weaker assumptions various upper bounds on the maximum achievable fidelity of probabilistic and deterministic cloning machines. Building on ideas by Gisin [Phys.~Lett.~A, 1998], our results hold even for cloning machines that do not obey the laws of quantum mechanics, as long as remote state preparation is possible and the no-signalling principle holds. } 
    Here we generalize and rederive in a unified framework various upper bounds on the maximum achievable fidelity of probabilistic and deterministic cloning machines. Building on ideas by Gisin [Phys.~Lett.~A, 1998], our result starts from the fact that remote state preparation is possible and the no-signalling principle holds. We apply our general theorem to several subsets of states that are of interest in quantum cryptography. 
\end{abstract}

\maketitle
\end{CJK*}

\section{Introduction} 

The no-cloning theorem states that within the framework of quantum mechanics there does not exist any universal procedure that is able to replicate an unknown quantum state reliably. 
The no-cloning theorem has rich implications in quantum theory, and its quantitative expressions are particularly useful in quantum cryptography. It is the cornerstone for the security of various primitives such as quantum money~\cite{Wiesner_1983,Aaronson_2012,Monila_2013}, quantum key distribution~\cite{Bennett_2014}, quantum secret sharing~\cite{Hillery_1999,Cleve_1999,Gottesman_2000} and uncloneable primitives~\cite{Ananth_2023,Broadbent_2020a,Broadbent_2020b,Georgiou_2020}. The security of these primitives relies heavily on the fact that quantum states cannot be cloned, or the closely related concept of entanglement monogamy (i.e.\ the fact that it is not possible to clone half of an entangled state).

The no-cloning theorem for $1$-to-$2$ perfect cloning for general states was proposed in~\cite{Wootters_1982,Dieks_1982}. Later it was generalized to various situations, such as imperfect cloning~\cite{Gisin_1997a}, $n$-to-$m$ cloning~\cite{Gisin_1997b,Werner_1998}, cloning for subsets of states~\cite{Fan_2001,Buscemi_2005,Demkowicz-Dobrzanski_2004}, and cloning procedures that only succeeds with non-vanishing probability~\cite{Duan_1998,Fiurasek_2004}. 
The latter is generally easier than deterministic cloning. For certain subsets of states, it is possible to produce more copies from some copies, or ``super-replicate'', with a non-vanishing probability~\cite{Chiribella_2013,Chiribella_2015,Yang_2018}. However, probabilistic quantum cloning machines perform no better than deterministic quantum cloning machines as long as the states to be cloned have a strong enough symmetry. There is a counterpart of the no-cloning theorem for quantum channels. It states that use of an unknown quantum channel $n$ times does not allow to simulate $n^2$ accesses to the same channel. No-cloning theorems for channels have been proved for clock unitaries~\cite{Dur_2015} and general unitaries~\cite{Chiribella_2015}.

In spite of its wide use, quantum mechanics fails to resolve certain fundamental problems~\mbox{\cite{Peres_2004,Bassi_2013}} without modifications~\mbox{\cite{Kaplan_2022,Polkovnikov_2023,Ghirardi_1986,Pearle_1989,Ghirardi_1990,Diosi_1989,Bassi_2005}}. It is essential to question whether the no-cloning theorem and the security of quantum cryptographic primitives relying on it still hold even if quantum mechanics will eventually need to be amended. One physically arguably more fundamental law is the no-signalling principle, which states that information cannot be transmitted between two separate parties without the transfer of a physical particle, and thus in particular information cannot be transmitted instantaneously. The no-signalling principle is respected by both quantum mechanics and relativity~\mbox{\cite{Gisin_1998}}, and thus widely believed to remain true even if quantum mechanics needs to be modified. As demonstrated in~\mbox{\cite{Simon_2001}}, the dynamical properties of quantum mechanics, and thus the no-cloning theorem, can be derived from the no-signaling principle, combining with two assumptions on static properties properties of the theory: (i) all states are described by vectors $\ket{\psi}$ in a Hilbert space, and (ii) all measurements follow the Born rule, i.e. $\Pr[x|\psi] = \langle \psi | M_x \ket{\psi}$ for some positive semi-definite $M_x$.

%\st{It has not gone unnoticed that some no-cloning theorems remain valid under these assumptions. However, prior works restrict themselves to specific situations, e.g. on $1$-to-$2$ cloning~\mbox{\cite{Gisin_1998,Masanes_2006}}, probabilistic cloning which produces perfect copies on success~\mbox{\cite{Hardy_1999,Pati_2000}} or clock state cloning~\mbox{\cite{Sekatski_2015}}. }

It has not gone unnoticed that some no-cloning theorems can be reproduced directly under the no-signalling principle and assumptions (i) and (ii) without the need for an intermediate step through quantum mechanics~\mbox{\cite{Gisin_1998,Hardy_1999,Pati_2000,Sekatski_2015}}. However, prior works restrict themselves to specific scenarios, e.g., on $1$-to-$2$ universal cloning~\mbox{\cite{Gisin_1998}}, probabilistic cloning which produces perfect copies on success~\mbox{\cite{Hardy_1999,Pati_2000}}, or clock state cloning~\mbox{\cite{Sekatski_2015}}, without offering a unified framework.

%\st{In contrast, our work provides a very general framework that allows us to rederive existing results in a unified matter and in some cases strengthen them. More precisely, we construct a general remote state preparation protocol in which $n$ identical quantum states are remotely prepared. Based on this remote state preparation protocol, we propose a general scheme to derive $n$-to-$m$ no-cloning bounds for states from the no-signalling principle. Our scheme is versatile in that it also applies to various subsets of states, e.g., multi-phase states, spin-coherent states, stabilizer states and subsets of Choi states. Notably, the no-cloning bound for Choi states can also be interpreted as a no-cloning bound for unitaries. While our bounds hold more generally, they in some cases exactly reproduce the best known no-cloning bounds for quantum mechanical cloning channels. }

In this letter, we introduce a general framework that not only re-derives existing results directly from the no-signalling principle and assumptions (i) and (ii) in a unified manner but also strengthens them in some cases. More precisely, we propose a remote state preparation protocol in which one party can remotely prepare $n$ identical quantum states from a set of quantum states for another party if the set satisfies a symmetric property. By integrating the remote state preparation protocol and a hypothetical cloning oracle that creates $m$ copies from $n$ copies of a quantum state, we show that the no-signaling principle is respected only if the no-cloning bound holds. We emphasize that our assumptions are equivalent to the full quantum mechanics framework, although we do not use the dynamical principles of quantum mechanics explicitly.

Our method is versatile, applicable to the no-cloning bound of a wide range of subsets of states with high symmetry. These subsets include general states, multi-phase states, spin-coherent states, Choi states, and stabilizer states, as well as the clock states discussed previously~\mbox{\cite{Sekatski_2015}}. Notably, the no-cloning bound for Choi states can also be interpreted as a no-cloning bound for channels. Our no-signalling bounds in certain cases exactly reproduce the best-known quantum bounds for the cloning oracles. 

The no-cloning bounds for various sets of quantum states can find their applications in quantum information tasks such as quantum purification~\mbox{\cite{Cirac_1999,Fiurasek_2004,Masullo_2005}}, quantum metrology~\mbox{\cite{Chiribella_2011,Chiribella_2013,Chiribella_2015,Sekatski_2015}} and quantum cryptography~\mbox{\cite{Wiesner_1983,Aaronson_2012,Monila_2013,Bennett_2014}}. A unified framework to derive these no-cloning bounds will enhance these applications. As an example, we apply our no-cloning bound for Choi states to Clifford unitaries to analyze circuit privacy~\cite{Hu_2023} of the quantum homomorphic encryption protocol in~\cite{Ouyang_2018}. Our no-cloning bound proves a lower bound for circuit privacy; i.e., we show that the corresponding quantum homomorphic encryption protocol cannot be circuit private.

The remainder of the paper is organized as follows. In Section~\ref{sec:general_scheme}, we present our general scheme to derive no-cloning theorems from the no-signalling principle. We introduce the remote state preparation protocol in Section~\ref{sec:remote_state_prep}, apply the no-signalling principle to the remote state preparation protocol in Section~\ref{sec:no_signalling} and derive the no-cloning bound for states and unitary channels in Section~\ref{sec:no_cloning_state} and \ref{sec:no_cloning_unitary} respectively. In Section~\ref{sec:examples}, we apply our general schemes to derive no-cloning theorems for various examples. This includes no-cloning bounds for general states (Section~\ref{sec:example_general}), multi-phase states (Section~\ref{sec:example_multi_phase}), Choi states for general unitaries (Section~\ref{sec:example_choi_general}), spin coherent states (Section~\ref{sec:example_spin_coherent}), stabilizer states (Section~\ref{sec:example_stabilizer}) and Choi states for Cliffords (Section~\ref{sec:example_choi_clifford}). In Section~\ref{sec:example_qhe}, we apply our no-cloning bound to lower bound the circuit privacy of the quantum homomorphic encryption protocol in~\cite{Ouyang_2018}. 

\section{General scheme for states}\label{sec:general_scheme}

Let $\mathbbm{C}^d$ be the $d$-dimensional Hilbert space and $\mathcal{S}(d)\subset \mathbbm{C}^d$ be the set of pure quantum states on $\mathbbm{C}^d$. A probabilistic $n$-to-$m$ cloning oracle $\mathcal{C}_{n,m,p}^{\mathcal{S}}$ for a subset of pure states $\mathcal{S}\subset \mathcal{S}(d)$ maps $n$ identical pure states from $\mathcal{S}$ into $m$ identical states approximately with a non-zero success probability $p\in(0,1]$. More precisely, it satisfies
\begin{align}\label{eqn:cloning_oracle}
    \mathcal{C}_{n,m,p}^{\mathcal{S}}(\proj{\psi}^{\otimes n}) & = p \mathcal{R}_{n,m}^{\mathcal{S}}(\proj{\psi}^{\otimes n}) \otimes \proj{0} \nonumber\\ 
    & \ \ \ + (1-p) \rho_{\perp} \otimes \proj{1}, 
\end{align}
where $\psi\in \mathcal{S}$ is the input state to clone, $\mathcal{R}_{n,m}^{\mathcal{S}}(\proj{\psi}^{\otimes n})$ is the output state which is supposed to approximate $m$ identical quantum states $\proj{\psi}^{\otimes m}$ well, $\proj{0}$ and $\proj{1}$ denote the success and failure of the probabilistic cloning oracle respectively, and $p$ is the success probability. 

%\st{The cloning oracle $\mathcal{C}_{n,m,p}^{\mathcal{S}}$ we discuss here may not follow quantum mechanics. It can be non-linear, non-positive or non-trace preserving on general quantum states, as long as $\mathcal{R}_{n,m}^{\mathcal{S}}$ follows certain requirements. In the exact version of our main theorem (Theorem~\mbox{\ref{thm:bound_states_exact}}), we assume that $\mathcal{R}_{n,m}^{\mathcal{S}}$ maps $n$ identical pure quantum states to quantum states. In the approximate version of our main theorem (Theorem~\mbox{\ref{thm:bound_states_approximate}}), we make a little stronger assumption that $\mathcal{R}_{n,m}^{\mathcal{S}}$ maps quantum states on the $n$-fold symmetric subspace to quantum states. }

Here we make the assumptions (i') and (ii') as follows: (i') $\mathcal{C}_{n,m,p}^{\mathcal{S}}$ maps quantum states on $\mathbbm{C}^{d^{n}}$ to quantum states on $\mathbbm{C}^{d^{m}}$, and (ii') all measurements on the input and output sides must obey the Born rule. These conditions are equivalent to assumptions (i) and (ii). When combined with the no-signalling principle, we effectively require that $\mathcal{C}_{n,m,p}^{\mathcal{S}}$ behaves quantum mechanically, i.e., it is linear, completely positive, and trace-preserving~\mbox{\cite{Simon_2001}}. However, we will not explicitly rely on the quantum mechanical behaviour of $\mathcal{C}_{n,m,p}^{\mathcal{S}}$ in this section. Instead, we will discuss the impossibility of weakening the assumption (i') or (ii') to allow non-quantum mechanical behaviours og $\mathcal{C}_{n,m,p}^{\mathcal{S}}$ in Section~\mbox{\ref{sec:discussion}}.

The figure of merit of the cloning oracle on a set $\mathcal{S}$ is the worst case global cloning fidelity 
\begin{align}\label{eqn:worst_case_fidelity}
    F_{\rm wc}(\mathcal{R}_{n,m}^{\mathcal{S}}) = \min_{\psi\in \mathcal{S}} F\left(\mathcal{R}_{n,m}^{\mathcal{S}}(\proj{\psi}^{\otimes n}), \proj{\psi}^{\otimes m}\right), 
\end{align}
where $F(\rho,\sigma)=(\Tr|\sqrt{\rho}\sqrt{\sigma}|)^2$ denotes the Uhlmann fidelity. For any two quantum states $\rho,\sigma$, we have
\begin{align}\label{eqn:bound_fidelity}
    0\leq F(\rho,\sigma)\leq 1. 
\end{align}

The no-signalling principle will pose fundamental limits on the above cloning oracle. Let $\mathcal{V}$ be a group generating $\mathcal{S}$, i.e., $\mathcal{S} = \{V\ket{\phi_0}, V\in\mathcal{V}\}$. Let $\mathcal{W}$ be a discrete subset of $\mathcal{S}$ and $p_\phi$ a probability distribution over $\mathcal{W}$. We denote 
\begin{align}
    & \sigma_\mathcal{V} = \int_{\mathcal{V}} \d\mu(V)(V^\dagger)^{\otimes m} \sigma_{\mathcal{R}} V^{\otimes m}, \label{eqn:def_sigma_V} \\
    & \rho_{\mathcal{W}}^{l} = \sum_{\phi\in\mathcal{W}} p_\phi \proj{\phi}^{\otimes l}, \label{eqn:def_rho_W}
\end{align}
where the integration is over the Haar measure on $\mathcal{V}$ and $\sigma_{\mathcal{R}}$ is a constant state that arises from the no-signalling principle and only depends on the cloning oracle (we will make it clear in Eq.~\eqref{eqn:no_signaling_states_exact} and \eqref{eqn:bound_states_approximate} later). Our main theorems are: 

\setcounter{theorem}{2}
\begin{theorem}
    If there exists a constant state $\rho_0$ such that 
    \begin{align}\label{eqn:pre_rq_exact}
        \rho_0 = V^{\otimes n} \rho_{\mathcal{W}}^{n} (V^{\dagger})^{\otimes n}, \forall V\in\mathcal{V}, 
    \end{align}
    {where $\rho_{\mathcal{W}}^m$ is defined in Eq.~\mbox{\eqref{eqn:def_rho_W}},} then $F_{\rm wc}(\mathcal{R}_{n,m}^{\mathcal{S}})$ is upper bounded by 
    \begin{align}\label{eqn:pre_bound_states_exact}
        F_{\rm wc}(\mathcal{R}_{n,m}^{\mathcal{S}}) \leq F(\sigma_{\mathcal{V}},\rho_{\mathcal{W}}^{m}), 
    \end{align}
    {where $\sigma_{\mathcal{V}}$ is in the form of Eq.~\mbox{\eqref{eqn:def_sigma_V}} and obeys Eq.~\mbox{\eqref{eqn:no_signaling_states_exact}}.}
\end{theorem}

\begin{theorem}
    If there exists a constant state $\rho_0$ such that 
    \begin{align}\label{eqn:pre_rq_approximate}
        (1-\epsilon)\rho_0 \leq V^{\otimes n} \rho_{\mathcal{W}}^{n} (V^{\dagger})^{\otimes n}\leq (1+\epsilon)\rho_0, \forall V\in\mathcal{V}, 
    \end{align}
    {where $\rho_{\mathcal{W}}^m$ is defined in Eq.~\mbox{\eqref{eqn:def_rho_W}},} then $F_{\rm wc}(\mathcal{R}_{n,m}^{\mathcal{S}})$ is upper bounded by 
    \begin{align}\label{eqn:pre_bound_states_approximate}
        F_{\rm wc}(\mathcal{R}_{n,m}^{\mathcal{S}}) \leq (1+\epsilon)^2 F(\sigma_{\mathcal{V}},\rho_{\mathcal{W}}^{m}), 
    \end{align}
    {where $\sigma_{\mathcal{V}}$ is in the form of Eq.~\mbox{\eqref{eqn:def_sigma_V}} and obeys Eq.~\mbox{\eqref{eqn:no_signaling_states_approx}}.}
\end{theorem}

$\sigma_{\mathcal{V}}$ has $m$-fold symmetry due to Eq.~\eqref{eqn:def_sigma_V}, but $\rho_{\mathcal{W}}^{m}$ only has $n$-fold symmetry due to Eq.~\eqref{eqn:pre_rq_exact} and Eq.~\eqref{eqn:pre_rq_approximate}. Therefore, Eq.~\eqref{eqn:pre_bound_states_exact} and Eq.~\eqref{eqn:pre_bound_states_approximate} provide meaningful upper bounds on $F_{\rm wc}(\mathcal{R}_{n,m}^{\mathcal{S}})$ in general.

\setcounter{theorem}{0}

The proof for the main theorems is divided into three steps: 

1. We propose a remote state preparation protocol that allows Bob to prepare identical quantum states for Alice without any communication. This protocol utilizes shared entangled states and local measurements, guaranteeing that Alice receives identical copies of the quantum states. 

2. We demonstrate that the no-signalling principle prohibits any signalling from Bob to Alice, even if they use a cloning oracle for post-processing. Therefore, the no-signalling condition for the remote state preparation protocol, combined with the use of a cloning oracle, is derived. 

3. We establish an explicit no-cloning bound by exploiting the symmetry of the subset of states to be cloned and the no-signalling condition, which has significant implications for the field of quantum information processing.

%\st{We make weaker assumptions on $\mathcal{R}_{n,m}^{\mathcal{S}}$ for the exact version of our main theorem than for the approximate version, despite their similar derivations. The exact version Theorem~\mbox{\ref{thm:bound_states_exact}} assumes that the inputs to the cloning oracle are $n$ identical pure quantum states, and thus $\mathcal{R}_{n,m}^{\mathcal{S}}$ only needs to map $n$ identical pure quantum states to quantum states. The approximate version Theorem~\mbox{\ref{thm:bound_states_approximate}} assumes that the inputs to the cloning oracle are (possibly mixed) quantum states on the symmetric subspace (because of the residue part $\rho_{\res,V}$ as is required in Eq.~\mbox{\eqref{eqn:no_signaling_states_approx}}) and thus $\mathcal{R}_{n,m}^{\mathcal{S}}$ needs to map a (possibly mixed) quantum state on the symmetric subspace to quantum states. $\mathcal{R}_{n,m}^{\mathcal{S}}$ is free to be non-linear, non-positive and non-trace preserving for other inputs.}

Note that in our proof, we only explicitly use the no-signalling principle and assumptions (i') and (ii') instead of the properties of quantum channels.

\subsection{Remote identical state preparation}\label{sec:remote_state_prep}

\begin{lemma}[Remote state preparation without errors]\label{lemma:remote_exact}
    If there is a constant density matrix $\rho_0$ such that
    \begin{align}\label{eqn:rq_exact}
        \rho_0 = V^{\otimes n} \rho_{\mathcal{W}}^n (V^\dagger)^{\otimes n},\forall V\in\mathcal{V}, 
    \end{align}
    {where $\rho_{\mathcal{W}}^n$ is defined in Eq.~\mbox{\eqref{eqn:def_rho_W}},} there exists a remote identical state preparation protocol in which Bob remotely prepares $(V\proj{\phi}V^\dagger)^{\otimes n}$ for Alice, where $V$ depends on Bob's choice of measurement and $\phi$ depends on Bob's measurement outcome. 
\end{lemma}

\begin{figure*}
    \centering
    \includegraphics[scale=0.4]{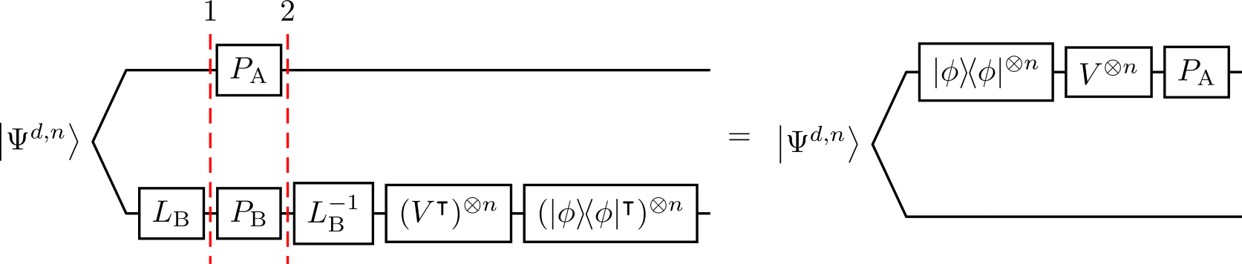}
    \caption{The state when Protocol~\ref{prot:remote_preparation} succeeds. We use squares to denote matrices rather than gates applied to states. On the left hand side, before the first slice, entangled state preparation; between the first and second slices, Alice's and Bob's measurements determine whether to proceed; after the second slice, Bob's measurement. On the right hand side, using $M \otimes \mathbbm{I} \ket{\Psi^{d,n}} = \mathbbm{I}\otimes M^{\intercal} \ket{\Psi^{d,n}}$, $P_{\A} = (L_{\B} P_{\B} L_{\B}^{-1})^{\intercal}$ and $P_{\A}^2 = P_{\A}$, we obtain Alice's post-measurement state. }
    \label{fig:remote_preparation}
\end{figure*}

\begin{algorithm}[H]
    \caption{Remote identical state preparation}
    \label{prot:remote_preparation}
    \begin{algorithmic}[1]
        \Require Alice and Bob share $\ket{\Phi^{n}}_{\A\B}$ 
        \Ensure Alice gets $(V\proj{\phi}V^\dagger)^{\otimes n}$ if Bob gets $\phi$, or both abort 
        \State Alice measures $\{P_{\A}, \id - P_{\A}\}$ on $\A$ 
        \State Bob measures $\{P_{\B}, \id - P_{\B}\}$ on $\B$
        \If{Alice and Bob obtain $P_{\A}$ and $P_{\B}$ respectively}
            \State Both know the success of the protocol
            \State Bob chooses $V\in\mathcal{V}$
            \State Bob measures $\phi$ using $ p_{\phi} (L_{\B}^{-1})^\dagger ((V\proj{\phi}V^\dagger)^{\otimes n} )^{\intercal}L_{\B}^{-1}$ 
            \Else
            \State Both know the failure of the protocol
        \EndIf
    \end{algorithmic}
\end{algorithm}    

\begin{proof}
    The remote state preparation protocol is summarized as Protocol~\ref{prot:remote_preparation} and its intuition is illustrated in Figure~\ref{fig:remote_preparation}.

    Let $\A\simeq \B \simeq \mathbbm{C}^{d^n}$ denote Alice's and Bob's local system respectively. $\ket{\Psi^{d,n}}_{\A\B} = \ket{\Psi^{d}}^{\otimes n}$ is the maximally entangled state on $\A\B$. We also define the transpose with respect to the computational basis $\ket{i}$. It holds that $M\otimes\mathbbm{I}\ket{\Psi^{d,n}}_{\A\B} = \mathbbm{I}\otimes M^{\intercal} \ket{\Psi^{d,n}}_{\A\B}$. We diagonalize $\rho_0$ as
    \begin{align}\label{eqn:rho_eigen}
        \rho_0 = \sum_x \lambda_x \proj{\lambda_x}, 
    \end{align}
    where $\lambda_x$ are eigenvalues and $\ket{\lambda_x}$ the corresponding eigenvectors. We let 
    \begin{align}
        & P_{\A} = \sum_{x} \proj{\lambda_x}, \quad P_{\B} = \sum_{x} (\proj{\lambda_x})^\intercal, \label{eqn:P_eigen}\\
        & L_{\B} =  \sum_{x} \sqrt{\lambda_x} (\proj{\lambda_x})^{\intercal} + c_{\perp} P_{\perp}^{\intercal}, \label{eqn:L_eigen}
    \end{align}
    where $P_{\perp}= \mathbbm{I} - P_{\A}$ and $c_{\perp}$ is a non-zero constant to choose freely (the choice of $c_{\perp}$ does not change our derivations and results, but makes $L_{\B}$ simpler).  
       
    Initially, Alice and Bob share an entangled state 
    \begin{align}
        \ket{\Phi^n}_{\A\B} = C (\id \otimes L_{\B}) \ket{\Psi^{d,n}}_{\A\B},
    \end{align}
    where $C = \sqrt{\frac{d^n}{\Tr(L_{\B}^\dagger L_{\B})}} $ is the normalization factor. Alice and Bob then perform local projective measurements $\{P_{\A}, \id - P_{\A} \}$ and $\{P_{\B}, \id - P_{\B}\}$ on $\A$ and $\B$ respectively. Alice's measurement outcome is perfectly correlated with Bob's measurement outcome because 
    \begin{align}
        & P_{\A}\otimes \mathbbm{I} \ket{\Phi^n}_{\A\B} = C P_{\A}\otimes L_{\B} \ket{\Psi^{d,n}}_{\A\B} \nonumber \\ 
        & = C \mathbbm{I}\otimes L_{\B} P_{\A}^{\intercal} \ket{\Psi^{d,n}}_{\A\B} = C \mathbbm{I}\otimes (L_{\B} P_{\A}^{\intercal}  L_{\B}^{-1})  L_{\B} \ket{\Psi^{d,n}}_{\A\B} \nonumber\\
        & = \mathbbm{I}\otimes (L_{\B} P_{\A}^{\intercal}  L_{\B}^{-1}) \ket{\Phi^n}_{\A\B} =  \mathbbm{I}\otimes P_{\B} \ket{\Phi^n}_{\A\B}. \nonumber  
    \end{align}
    When Alice and Bob obtain $P_{\A}$ and $P_{\B}$ respectively, the protocol succeeds and both parties proceed. Bob further chooses a unitary $V\in \mathcal{V}$ according to {some measure on $\mathcal{V}$}. Based on $V$, Bob performs the positive operator-valued measure $\phi\mapsto p_{\phi}(L_{\B}^{-1})^\dagger \left((V\proj{\phi} V^\dagger)^{\otimes n}\right)^{\intercal} L_{\B}^{-1}, \forall \phi \in \mathcal{W}$. This is indeed a positive operator-valued measure on the subspace $P_{\B}$ projects onto for all $V\in\mathcal{V}$ because 
    \begin{align}
        & P_{\B} = (L_{\B}^{-1})^\dagger \rho_0^{\intercal} L_{\B}^{-1} = (L_{\B}^{-1})^\dagger \left(V^{\otimes n}\rho_{\mathcal{W}}^{n} (V^\dagger)^{\otimes n}\right)^{\intercal} L_{\B}^{-1} \nonumber \\
        & = \sum_{\phi \in\mathcal{W}} p_{\phi} (L_{\B}^{-1})^\dagger  \left((V\proj{\phi} V^\dagger)^{\otimes n}\right)^{\intercal} L_{\B}^{-1}, \forall V\in\mathcal{V} \nonumber. 
    \end{align}
    Because $P_{\A}$ is a projector onto the support of $\rho_0$ and $(V\ket{\phi})^{\otimes n}$ is also in the support of $\rho_0$, i.e.,
    \begin{align}
        P_{\A} (V\ket{\phi})^{\otimes n} = (V\ket{\phi})^{\otimes n}, \forall V\in\mathcal{V}, \forall \phi\in\mathcal{W}. 
    \end{align}
    Let us compute Alice's marginal state conditioned on Bob obtaining $\phi$ as follows:  
    \begin{widetext}
    \begin{align}
        \rho_{\A}(\phi,V) &  = \frac{P_{\A} \Tr_{\B}\left(\Phi^n \mathbbm{I} \otimes \left(p_\phi (L_{\B}^{-1})^\dagger \left((V\proj{\phi}V^\dagger)^{\otimes n}\right)^{\intercal}  L_{\B}^{-1}\right)\right)P_{\A}}{\Tr\left(\Phi^n P_{\A}  \otimes \left(p_\phi (L_{\B}^{-1})^\dagger \left((V\proj{\phi}V^\dagger)^{\otimes n}\right)^{\intercal}  L_{\B}^{-1}\right)\right)} = \frac{P_{\A}\Tr_{\B}\left(\Psi^{d,n} \mathbbm{I} \otimes  \left((V\proj{\phi}V^\dagger)^{\otimes n}\right)^{\intercal} \right) P_{\A} }{\Tr\left(\Psi^{d,n}  P_{\A} \otimes  \left((V\proj{\phi}V^\dagger)^{\otimes n}\right)^{\intercal} \right)}\nonumber  \\
        & = \frac{ P_{\A}\Tr_{\B}(\Psi^{d,n} (V\proj{\phi}V^\dagger)^{\otimes n} \otimes \mathbbm{I}) P_{\A}}{\Tr(\Psi^{d,n} (V\proj{\phi}V^\dagger)^{\otimes n}  P_{\A} \otimes \mathbbm{I} )}  = \frac{ P_{\A} (V\proj{\phi}V^\dagger)^{\otimes n} P_{\A} }{\Tr ( (V\proj{\phi}V^\dagger)^{\otimes n}P_{\A}  )} = (V\proj{\phi} V^\dagger)^{\otimes n}. \nonumber  
    \end{align}
    \end{widetext}
    Here $\Phi^n = \proj{\Phi^n}_{\A\B}$ and $\Psi^{d,n} = \proj{\Psi^{d,n}}_{\A\B}$. Therefore, it is guaranteed that Bob remotely prepares identical states for Alice, i.e., Alice obtains $(V\proj{\phi}V^\dagger)^{\otimes n}$ when Bob obtains $\phi$. 
    
    The probability that the remote state preparation protocol succeeds and the probability that Alice obtains $(V\proj{\phi}V^\dagger)^{\otimes n}$ conditioned on success are, respectively, 
    \begin{align}\label{eqn:success_prob}
        p_{\suc}=\frac{1}{ \Tr(L_{\B}^\dagger L_{\B})} ,\quad p_{\phi|\suc}= p_\phi. 
    \end{align}
    which are independent of $V$. 
\end{proof}
\begin{remark}
    {For the second measurement, Bob can choose $V\in\mathcal{V}$ according to any measure on $\mathcal{V}$. We will fix the measure to be the Haar measure later in this work. }
\end{remark}

\begin{remark}
    From Eq.~\mbox{\eqref{eqn:L_eigen}} and~\mbox{\eqref{eqn:success_prob}}, $p_{\suc}$ is solely determined by $c_{\perp}$ via
    \begin{align}
        p_{\suc}= \frac{1}{1+ c_\perp^2 |P_{\perp}|^2}.
    \end{align}
    As $c_{\perp}$ can be freely chosen in $(0,+\infty)$, $p_{\suc}$ can also be freely chose in $(0,1)$. The choice of $c_\perp$ and thus $p_{\suc}$ will not change {the post-measurement state of the first measurement conditioned on success and thus} our later derivations, as we mainly focus on states and probabilities conditioned on success. 
\end{remark}

\begin{remark}
    {The first measurement can be understood as probabilistic initial entangled state preparation with a success probability of $p_{\suc}$. We can omit the first measurement by changing the initial entangled state shared by Alice and Bob. That is, instead of sharing the old initial state and performing the first measurement, Alice and Bob can share the new initial state which is the same as the post-measurement state of the first measurement conditioned on success. $c_\perp$ and $p_{\suc}$ are irrelevant to the new initial state and thus later derivations. }
\end{remark}

The condition Eq.~\eqref{eqn:rq_exact}  in Lemma~\ref{lemma:remote_exact} is sometimes too difficult to satisfy. Notably, the remote identical state preparation protocol is robust even if the condition is satisfied approximately. This fact greatly improves our remote identical state preparation protocol.

\begin{lemma}[Remote state preparation with errors]\label{lemma:remote_approximate}
    If there is a constant density matrix $\rho_0$ such that
    \begin{align}\label{eqn:rq_approximate}
        (1-\epsilon)\rho_0 \leq V^{\otimes n} \rho_{\mathcal{W}}^n (V^\dagger)^{\otimes n}\leq (1+\epsilon)\rho_0,\forall V\in\mathcal{V}, 
    \end{align}
   {where $\rho_{\mathcal{W}}^n$ is defined in Eq.~\mbox{\eqref{eqn:def_rho_W}}}, there exists a remote identical state preparation protocol in which Bob remotely prepares $(V\proj{\phi}V^\dagger)^{\otimes n}$ for Alice with a probability of at least $\frac{1}{1+\epsilon}$, where $V$ depends on Bob's choice of measurement and $\phi$ depends on Bob's measurement outcome. 
\end{lemma}

\begin{proof}
    The protocol remains the same except for the positive-operator valued measure Bob performs after the protocol succeeds and both party proceed. We denote 
    \begin{align}
        & p_{\res}  = \frac{\epsilon}{1+\epsilon}, \\
        & \tau_{\res,V}  = \frac{1+\epsilon}{\epsilon} P_{\B} - \frac{1}{\epsilon} (L_{\B}^{-1})^\dagger ( V^{\otimes n} \rho_{\mathcal{W}}^n (V^\dagger)^{\otimes n})^{\intercal} L_{\B}^{-1}. 
    \end{align}
    To clarify, $p_\res$ only represents a parameter for the convenience of writing rather than the probability to obtain $\res$. Taking the transpose of Eq.~\eqref{eqn:rq_approximate}, applying on each side $(L_{\B}^{-1})^\dagger$ from the left and $L_{\B}^{-1}$ from the right and substituting Eq.~\eqref{eqn:rho_eigen}, \eqref{eqn:P_eigen} and \eqref{eqn:L_eigen}, we obtain 
    \begin{align}\label{eqn:bound_res}
        0\leq p_{\res} \tau_{\res,V} \leq \frac{2\epsilon}{1+\epsilon} P_{\B}.  
    \end{align}
    The positive operator-valued measure is replaced by $\phi\mapsto (1-p_{\res}) p_\phi (L_{\B}^{-1})^\dagger \left((V\proj{\phi} V^\dagger)^{\otimes n}\right)^{\intercal} L_{\B}^{-1}$ and $ \res \mapsto p_{\res}\tau_{\res,V}$. Here $\res$ and $p_{\res}\tau_{\res,V}$ correspond to a ``fake success'', as Alice does not obtain identical states but she is not aware of it if Bob does not communicate with her. It is indeed a positive operator-valued measure on the subspace $P_{\B}$ projects onto for all $V\in\mathcal{V}$ because
    \begin{align}
        P_{\B}  =  \sum_{\phi\in\mathcal{W}} (1-p_{\res})  p_\phi & (L_{\B}^{-1})^\dagger  \left((V\proj{\phi} V^\dagger)^{\otimes n}\right)^{\intercal} L_{\B}^{-1}  \nonumber\\ 
        & + p_{\res} \tau_{\res,V}, \forall V\in\mathcal{V}. 
    \end{align}
    After replacing the positive operator-valued measure, Alice's state remains $(V\proj{\phi}V^\dagger)^{\otimes n}$ when Bob obtains $\phi$ as before, but her state becomes $\rho_{\res,V}$ when Bob obtains $\res$. We only need to check that conditioned on success, the probability $p_{\fak|\suc}$ of a fake success is negligible. Indeed, this is the case. The probability that the remote state preparation protocol succeeds and the probability that Alice obtains $(V\proj{\phi}V^\dagger)^{\otimes n}$ conditioned on success are, respectively
    \begin{align}
        p_{\suc}=\frac{1}{\Tr(L_{\B}^\dagger L_{\B})}, \quad p_{\phi|\suc}= (1-p_\res) p_{\phi}. 
    \end{align}
    Note that 
    \begin{align}
        \sum_{\phi\in\mathcal{W}} p_{\phi|\suc} + p_{\fak|\suc} = 1.  
    \end{align}
    The probability of a fake success conditioned on success is thus
    \begin{align}
        p_{\fak|\suc} = p_{\res}. 
    \end{align}
    Notably, $p_{\suc}$, $p_{\phi|\suc}$ and $p_{\fak|\suc}$ are independent of $V$. 
\end{proof}

\subsection{No-signalling condition}\label{sec:no_signalling}

Let us first focus on a remote identical state preparation protocol without errors. Conditioned on success, Alice obtains $n$ identical states $(V\proj{\phi}V^\dagger)^{\otimes n}$ with a probability of $p_\phi$. As $V\ket{\phi}\in \mathcal{S}, \forall V\in\mathcal{V}, \phi\in\mathcal{W}$, Alice inputs $(V\proj{\phi}V^\dagger)^{\otimes n}$ into the cloning oracle. The output of the cloning oracle approximates $(V\proj{\phi}V^\dagger)^{\otimes m}$. On the one hand, Bob remotely prepares different sets of states $\{(V\proj{\phi}V^\dagger)^{\otimes n},\phi\in\mathcal{W}\}$ for Alice by choosing different unitaries $V$. On the other hand, the no-signalling principle requires that Alice's state is independent of Bob's choice of unitary. Thus, the no-signalling condition for the remote state preparation protocol without errors is 
\begin{align}\label{eqn:no_signaling_states_exact}
    \sum_{\phi\in\mathcal{W}} p_{\phi} \mathcal{R}_{n,m}^{\mathcal{S}}\left( (V\proj{\phi}V^\dagger)^{\otimes n}\right) = \sigma_{\mathcal{R}}, 
\end{align}
where $\sigma_{\mathcal{R}}$ is a constant state independent of $V$. 

The no-signalling condition for the remote identical states with errors is derived via the same argument as
\begin{align}\label{eqn:no_signaling_states_approx}
   \sum_{\phi\in\mathcal{W}} (1-p_{\res}) p_{\phi} \mathcal{R}_{n,m}^{\mathcal{S}}\left( (V\proj{\phi}V^\dagger)^{\otimes n}\right) & \nonumber \\
    + p_{\res} \mathcal{R}_{n,m}^{\mathcal{S}}(\rho_{\res,V}) & = \sigma_{\mathcal{R}}, 
\end{align}
where $\sigma_{\mathcal{R}}$ is a constant state independent of $V$. 
%\end{lemma}

\subsection{No-cloning theorem for states}\label{sec:no_cloning_state}
With the remote state preparation protocol in Lemma~\ref{lemma:remote_exact} and \ref{lemma:remote_approximate}, and the no-signalling condition Eq.~\eqref{eqn:no_signaling_states_exact} and \eqref{eqn:no_signaling_states_approx}, we proceed to prove our main theorems. 

\begin{theorem}[restated]\label{thm:bound_states_exact}
    If there exists a constant density matrix $\rho_0$ such that 
    \begin{align}
        \rho_0 = V^{\otimes n} \rho_{\mathcal{W}}^{n} (V^{\dagger})^{\otimes n}, \forall V\in\mathcal{V}, 
    \end{align}
    {where $\rho_{\mathcal{W}}^n$ is defined in Eq.~\mbox{\eqref{eqn:def_rho_W}},} then $F_{\rm wc}(\mathcal{R}_{n,m}^{\mathcal{S}})$ is upper bounded by 
    \begin{align}\label{eqn:bound_states_exact}
        F_{\rm wc}(\mathcal{R}_{n,m}^{\mathcal{S}}) \leq F(\sigma_{\mathcal{V}},\rho_{\mathcal{W}}^{n}), 
    \end{align} 
    {where $\sigma_{\mathcal{V}}$ is in the form of Eq.~\mbox{\eqref{eqn:def_sigma_V}} and obeys Eq.~\mbox{\eqref{eqn:no_signaling_states_exact}}.}
\end{theorem}
\begin{proof}
    We first use the joint concavity of the square root of the fidelity, or more directly the joint quasi-concavity of the fidelity, to upper bound the minimal cloning fidelity with the fidelity of the average state, i.e., 
    \begin{align}\label{eqn:derivation_exact}
        & F_{\rm wc}(\mathcal{R}_{n,m}^{\mathcal{S}})  \nonumber \\
        & \leq F\left(\sum_{\phi\in\mathcal{W}} p_\phi \mathcal{R}_{n,m}^{\mathcal{S}}\left((V\proj{\phi}V^\dagger)^{\otimes n}\right),V^{\otimes m} \rho_{\mathcal{W}}^m (V^\dagger)^{\otimes m} \right) . 
    \end{align}
    With the no-signalling condition Eq.~\eqref{eqn:no_signaling_states_exact}, we obtain
    \begin{align}
        F_{\rm wc}(\mathcal{R}_{n,m}^{\mathcal{S}}) \leq  F\left(\sigma_{\mathcal{R}}, V^{\otimes m} \rho_{\mathcal{W}}^{m} (V^{\dagger})^{\otimes m} \right).
    \end{align}
    Due to the unitary equivalence of the fidelity, 
    \begin{align}
        F\left(\sigma_\mathcal{R}, V^{\otimes m} \rho_{\mathcal{W}}^{m} (V^{\dagger})^{\otimes m} \right) = F\left((V^{\dagger})^{\otimes m} \sigma_{\mathcal{R}} V^{\otimes m} , \rho_{\mathcal{W}}^{m} \right). 
    \end{align}
    We now average the square root of the fidelity {according to the Haar measure} on $\mathcal{V}$ and use the joint concavity again to obtain 
    \begin{align}
        F_{\rm wc}(\mathcal{R}_{n,m}^{\mathcal{S}}) \leq F(\sigma_{\mathcal{V}},\rho_{\mathcal{W}}^{m}), 
    \end{align}
    which concludes the proof. 
\end{proof}

\begin{theorem}[restated]\label{thm:bound_states_approximate}
    If there exists a constant density matrix $\rho_0$ such that 
    \begin{align}
        (1-\epsilon)\rho_0 \leq V^{\otimes n} \rho_{\mathcal{W}}^{n} (V^{\dagger})^{\otimes n} \leq (1+\epsilon)\rho_0, \forall V\in\mathcal{V}, 
    \end{align}
    {where $\rho_{\mathcal{W}}^n$ is defined in Eq.~\mbox{\eqref{eqn:def_rho_W}},} then $F_{\rm wc}(\mathcal{R}_{n,m}^{\mathcal{S}})$ is upper bounded by 
    \begin{align}\label{eqn:bound_states_approximate}
        F_{\rm wc}(\mathcal{R}_{n,m}^{\mathcal{S}}) \leq (1+\epsilon)^2 F(\sigma_{\mathcal{V}},\rho_{\mathcal{W}}), 
    \end{align} 
    {where $\sigma_{\mathcal{V}}$ is in the form of Eq.~\mbox{\eqref{eqn:def_sigma_V}} and obeys Eq.~\mbox{\eqref{eqn:no_signaling_states_approx}}.}
\end{theorem}
\begin{proof}
    The first step is again averaging over $\mathcal{W}\cup\{\res\}$ and using the joint concavity, 
    \begin{widetext}
    \begin{align}\label{eqn:derivation_approximate}
        & (1-p_\res)\sqrt{F_{\rm wc}(\mathcal{R}_{n,m}^{\mathcal{S}})} + p_{\res} \sqrt{F(\mathcal{R}_{n,m}^{\mathcal{S}}(\rho_{\res,V}),V^{\otimes m}\rho_{\mathcal{W}} (V^\dagger)^{\otimes m})} \nonumber \\
        & \leq \sqrt{F\left( \sum_{\phi\in\mathcal{W}} (1-p_\res)p_\phi \mathcal{R}_{n,m}^{\mathcal{S}}\left((V\proj{\phi}V^\dagger)^{\otimes n} \right)+ p_{\res}\mathcal{R}_{n,m}^{\mathcal{S}}(\rho_{\res,V}) ,V^{\otimes m}\rho_{\mathcal{W}}^{m} (V^\dagger)^{\otimes m}\right)}. 
    \end{align}
    \end{widetext}
    We now use the no-signalling condition Eq.~\eqref{eqn:no_signaling_states_approx}, as well as Eq.\eqref{eqn:bound_fidelity}, Eq.~\eqref{eqn:bound_res} and the unitary equivalence to obtain
    \begin{align}
        F_{\rm wc}(\mathcal{R}_{n,m}^{\mathcal{S}})\leq (1+\epsilon)^2 F\left((V^{\dagger})^{\otimes m} \sigma_{\mathcal{R}} V^{\otimes m}, \rho_{\mathcal{W}}^{m}\right).
    \end{align}
    We now average {according to the Haar measure} over $\mathcal{V}$ and use the joint concavity again to obtain 
    \begin{align}
        F_{\rm wc}(\mathcal{R}_{n,m}^{\mathcal{S}}) \leq (1+\epsilon)^2 F(\sigma_{\mathcal{V}},\rho_{\mathcal{W}}^{m}), 
    \end{align}
    which completes the proof. 
\end{proof}
\begin{remark} 
    {The above proof works the same for other geometric measures such as trace distance. Due to historical reasons, we focus on the fidelity. }
\end{remark}
\begin{remark}
    {To obtain the no-cloning bound, one needs to optimize over $\sigma_{\mathcal{V}}$ and $\rho_{\mathcal{W}}^n$. Because the free parameter $c_{\perp}$ and the success probability $p_\suc$ in the remote identical state preparation protocol does not appear in $\sigma_{\mathcal{V}}$ and $\sigma_{\mathcal{W}}$, one does not optimize over $c_{\perp}$ or $p_{\suc}$. Therefore, $c_{\perp}$ or $p_{\suc}$ are irrelevant.}
\end{remark}

\subsection{No-cloning theorem for unitaries}\label{sec:no_cloning_unitary}

The Choi-Jamiolkowski isomorphism is a mapping between a quantum channel and its corresponding Choi state. The Choi-Jamiolkowski isomorphism allows us to relate the average fidelity between quantum channels and the entanglement fidelity between them, i.e., the fidelity of their corresponding Choi states~\cite{Nielsen_2002}. 

In the context of cloning, the worst case global cloning average fidelity for quantum channels is equivalent to the worst case global cloning fidelity for their corresponding Choi states. This is because one can prepare $n$ identical Choi states of a channel given $n$ accesses to the channel, and conversely, $n$ identical Choi states allow one to probabilistically implement the corresponding channel $n$ times via gate teleportation. Therefore, a cloning oracle for a set of unitary channels is equivalent to a cloning oracle for their corresponding Choi states.

By applying the general scheme to derive the no-cloning bound from the no-signalling principle for the set of Choi states for a set of channels, one obtains the no-cloning bound for the channels themselves. 

\section{Example applications}\label{sec:examples}
Our general scheme for states is versatile to various subsets of quantum states. These subsets include general states, multi-phase states, spin-coherent states, stabilizer states, Choi states for general unitary channels, Choi states for multi-phase unitary channels and Choi states for Cliffords. 

\subsection{General states}\label{sec:example_general}
In this subsection, we consider the no-cloning bound for the set of $d$-dimensional pure states $\mathcal{S}(d)$. The $d$-dimensional unitary group $\mathcal{U}(d)$ generates $\mathcal{S}(d)$ from $\ket{0}$.   

It is well-known that averaging all $l$-fold tensor product states over the Haar measure results in the projector onto the symmetric subspace~\cite{Harrow_2013}, 
\begin{align}\label{eqn:state_haar_measure}
    \int \d \phi \proj{\phi}^{\otimes l} = \rho_0^l, 
\end{align}
where $\rho_0^l =  \frac{1}{d[l]}P_{\sym}^{l}$, $P_{\sym}^{l}$ denotes projector onto the symmetric subspace $\mathbbm{P}_{\sym}^{l}$ on $\mathbbm{C}^{d^l}$ and $d[l]=\binom{d+l-1}{l}$ denotes the dimension of the symmetric subspace. In this section, we will use the case where $l=n$ and $l=m$. 

At the same time, a weighted exact quantum $n$-design $(\{p_\phi\},\mathcal{W})$ is defined {as a probability distribution $\{p_\phi\}$ on a discrete subset $\mathcal{W}$ of $\mathcal{S}(d)$ such that when replacing the integration over the Haar measure on $\mathcal{S}(d)$ with the average over the probability distribution on $\mathcal{W}$, Eq.~\mbox{\eqref{eqn:state_haar_measure}} still holds exactly for the corresponding $\rho_0^n$~\mbox{\cite{Hayashi_2005,Ambainis_2007}}, i.e.,}
\begin{align}
    \rho_{\mathcal{W}}^n = \rho_0^n, 
\end{align}
{Here $\rho_{\mathcal{W}}^n$ is defined in Eq.~\mbox{\eqref{eqn:def_rho_W}}, i.e., }
\begin{align}
    \rho_{\mathcal{W}}^n = \sum_{\phi\in\mathcal{W}} p_\phi \proj{\phi}^{\otimes n}. 
\end{align}
{$\rho_{\mathcal{W}}^n$} remains the same under $\mathcal{U}(d)$ as $P_{\sym}^n$ is invariant under $\mathcal{U}(d)$, 
\begin{align}
    V^{\otimes n} \rho_{\mathcal{W}}^n (V^{\dagger})^{\otimes n} = \rho_0^n, \forall V\in\mathcal{U}(d). 
\end{align}

Similarly, a weighted $\epsilon$-approximate quantum $n$-design $(\{p_\phi\},\mathcal{W})$ reproduces Eq.~\eqref{eqn:state_haar_measure} approximately~\cite{Ambainis_2007}, i.e.,
\begin{align}
    (1-\epsilon)\rho_0^n \leq \rho_{\mathcal{W}}^n \leq  (1+\epsilon)\rho_0^n. 
\end{align}
The unitary invariance remains true, i.e.,
\begin{align}
    (1-\epsilon) \rho_0^n \leq  V^{\otimes n} \rho_{\mathcal{W}}^n (V^{\dagger})^{\otimes n}\leq  (1+\epsilon)\rho_0^n, \forall V\in\mathcal{U}(d). 
\end{align}

We apply the weighted quantum $n$-design $(\{p_\phi\}, \mathcal{W})$ and construct the remote state preparation protocol {following Eq.~\mbox{\eqref{eqn:P_eigen}} and~\mbox{\eqref{eqn:L_eigen}}} with
\begin{align}
    P_{\A} = P_{\B} = P_{\sym}^{n}, \quad L_{\B} = \frac{1}{\sqrt{d[n]}} \id. 
\end{align}
{Note that $c_\perp$ does not matter for the bound and can be chosen arbitrarily. We have set $c_\perp = \frac{1}{\sqrt{d[n]}}$ for convenience.} A no-cloning bound then follows from Theorem~\ref{thm:bound_states_exact} or Theorem~\ref{thm:bound_states_approximate}. 

\begin{corollary}\label{cor:general_state}
    Consider a weighted $\epsilon$-approximate quantum $n$-design $(\{p_\phi\},\mathcal{W})$ with $|\mathcal{W}|$ elements. Then, 
    \begin{align}
        F_{\rm wc}\left(\mathcal{R}_{n,m}^{\mathcal{S}(d)}\right) \leq (1+\epsilon)^2 \frac{|\mathcal{W}|}{d[m]}. 
    \end{align}
\end{corollary}
\begin{proof}
    We will compute the upper bound from Theorem~\ref{thm:bound_states_approximate} with
    \begin{align}
         F_{\rm wc}\left(\mathcal{R}_{n,m}^{\mathcal{S}(d)}\right) \leq (1+\epsilon)^2 F(\sigma_{\mathcal{U}(d)},\rho_{\mathcal{W}}^{m}), 
    \end{align}
    {where $\sigma_{\mathcal{U}(d)}$ is defined in Eq.~\mbox{\eqref{eqn:def_sigma_V}}, i.e.,}
    \begin{align}
        \sigma_{\mathcal{U}(d)} = \int \d \mu(V) (V^\dagger)^{\otimes m} \sigma_{\mathcal{R}} V^{\otimes m}, 
    \end{align}
    {and $\rho_{\mathcal{W}}^m$ is defined in Eq.~\mbox{\eqref{eqn:def_rho_W}}, i.e.,}
    \begin{align}
        \rho_{\mathcal{W}}^m = \sum_{\ket{\phi}\in\mathcal{W}} p_\phi \proj{\phi}^{\otimes m}. 
    \end{align}
    Because $\ket{\phi}^{\otimes n}$ is a quantum state on the symmetric subspace $\mathbbm{P}_{\sym}^m$, we have
    \begin{align}
        P_{\sym}^{m} \rho_{\mathcal{W}}^m P_{\sym}^m = \rho_{\mathcal{W}}^m. 
    \end{align}
    Therefore,
    \begin{align}
        F(\sigma_{\mathcal{U}(d)},\rho_{\mathcal{W}}^{m}) = F(P_{\sym }^{m} \sigma_{\mathcal{U}(d)} P_{\sym}^{m},\rho_{\mathcal{W}}^{m}). 
    \end{align}
    Recall that $\sigma_{\mathcal{U}(d)}$ is defined by an integration over the Haar measure on $\mathcal{U}(d)$, and thus $[V^{\otimes m},\sigma_{\mathcal{U}(d)}]=0$. Moreover, the symmetric subspace $\mathbbm{P}_{\sym}^m$ is an invariant subspace of $V^{\otimes m}$, thus $[V^{\otimes m},P_{\sym}^m]=0$. Combining both, 
    \begin{align}
        [V^{\otimes m}, P_{\sym}^{m} \sigma_{\mathcal{U}(d)} P_{\sym}^{m} ]=0, \forall V\in\mathcal{U}(d). 
    \end{align}
    Because the symmetric subspace is an irreducible representation (irrep) of $\mathcal{U}(d)$, we obtain by Schur's lemma that
    \begin{align}
        P_{\sym}^{m} \sigma_{\mathcal{U}(d)} P_{\sym}^{m} = c P_{\sym}^m  \leq \rho_0^m, 
    \end{align} 
    where we have used $c \leq \frac{1}{d[m]}$ and $\rho_0^m = \frac{1}{d[m]} P_{\sym}^m$. As is shown by~\cite[Theorem 3.25]{Watrous_2018}, $F(\rho,\sigma)$ monotonically increases with $\rho$ or $\sigma$. Therefore, 
    \begin{align}
        F(\sigma_{\mathcal{U}(d)},\rho_{\mathcal{W}}^{m})\leq F\left(\rho_0^m,\rho_{\mathcal{W}}^{m}\right). 
    \end{align}
    We apply Eq.~\eqref{eqn:rank_inequality} in Appendix~\ref{apd:rank} to derive
    \begin{align}
        F\left(\rho_0^m,\rho_{\mathcal{W}}^{m}\right) \leq |\mathcal{W}| \max_{\phi\in\mathcal{W}} F(\rho_0^m,\proj{\phi}^{\otimes m}) = \frac{|\mathcal{W}|}{d[m]},
    \end{align}
    where we have used $F(\rho_0^m,\proj{\phi}^{\otimes m}) = \frac{1}{d[m]}$. Then by Theorem~\ref{thm:bound_states_approximate}, we obtain the desired bound. 
\end{proof}

In~\cite{Ambainis_2007}, the authors construct $\epsilon$-approximate quantum $n$-designs with $|\mathcal{W}|= O(d^n \polylog d )$ in the asymptotic limit for large $d$ and fixed $n$. We thus get an upper bound in the asymptotic limit for large $d$ and fixed $n$ and $m$  
\begin{align}
    F_{\rm wc}\left(\mathcal{R}_{n,m}^{\mathcal{S}(d)}\right) \leq O(d^{n-m} \polylog d).
\end{align}
Up to a polylogarithm function, it reproduces the best known bound that assumes quantum mechanical cloning oracle $F_{\rm wc}\left(\mathcal{R}_{n,m}^{\mathcal{S}(d)}\right)\leq \frac{d[n]}{d[m]}\to O(d^{n-m})$ in the asymptotic limit for large $d$ and fixed $n$ and $m$~\cite{Werner_1998}.  

\subsection{Multi-phase states}\label{sec:example_multi_phase}

The set of $d$-dimensional multi-phase states and the group of multi-phase unitaries are defined as 
\begin{align}
    \mathcal{M}(d) & = \left\{\ket{\phi_\theta} = \frac{1}{\sqrt{d}}\sum_i e^{\iu \theta_i } \ket{i}, \forall \theta\in [0,2\pi)^{\times d} \right\},\\
    \mathcal{T}(d) & =\left\{V_\theta=\sum_i e^{\iu \theta_i}\proj{i}, \forall \theta \in[0,2\pi)^{\times d}\right\}. 
\end{align}
$\mathcal{T}(d)$ generates $\mathcal{M}(d)$ from $\ket{\phi_0} = \frac{1}{\sqrt{d}}\sum_i\ket{i}$. The set of Choi states for $\mathcal{T}(d)$ is 
\begin{align}
    \mathcal{J}_{\mathcal{T}(d)} = \left\{\ket{J_\theta} = \frac{1}{\sqrt{d}}\sum_i e^{\iu \theta_i } \ket{i}\otimes \ket{i}, \forall \theta\in [0,2\pi)^{\times d} \right\}.
\end{align}
$\mathcal{J}_{\mathcal{T}(d)}$ is equivalent to $\mathcal{M}(d)$ via 
\begin{align}
    \ket{J_\theta} = \CNOT (\ket{\phi_\theta}\ket{0} ), 
\end{align}
where $\CNOT$ is the $d$-dimensional gate defined by
\begin{align}
    \CNOT = \sum_{ij} \proj{i} \otimes \ketbra{(i+j)\,{\rm mod}\,d}{j}. 
\end{align}
And thus, we derive a no-cloning bound for $\mathcal{M}(d)$ and $\mathcal{J}_{\mathcal{T}(d)}$ simultaneously. 

We consider the weighted subset $(\{p_k\},\mathcal{W}) $ of $\mathcal{M}(d)$, where 
\begin{align}
    \mathcal{W}=\left\{\ket{\phi_k} = \frac{1}{\sqrt{d}} \sum_i e^{\iu \theta_0 k_i}\ket{i}, \forall k\in\mathbbm{Z}_{n+1}^{\times d} \right\},
\end{align}
where $\theta_0=\frac{2\pi}{n+1}$ and $\{p_k\}$ is uniform distribution on $\mathcal{W}$. Let $\alpha\in \mathbbm{Z}_{n+1}^{\times d}$ with $\sum_i \alpha_i = n$. We also denote $\alpha!= \alpha_1!...\alpha_d!$. We define $\ket{\alpha} = D \sum_{\pi\in \mathcal{P}(n)} \pi (\ket{i_1}...\ket{i_n}) $ on $\mathbbm{C}^{d^n}$ where $\mathcal{P}(n)$ is the permutation group, $\ket{i_1}...\ket{i_n}$ is a state with $\alpha_i$ subsystems in state $\ket{i}$ and $D$ is a normalization factor. Let $\beta\in \mathbbm{Z}_{m+1}^{\times d}$ with $\sum_i \beta_i = m$. We similarly denote $\beta!$ and $\ket{\beta}$ on $\mathbbm{C}^{{d^m}}$. $\{\ket{\alpha}\}$ and $\{\ket{\beta}\}$ form a complete orthonormal basis of $\mathbbm{P}_{\sym}^n$ and $\mathbbm{P}_{\sym}^m$ respectively. A simple calculation shows that 
\begin{align}\label{eqn:phase_constant_equality}
    \sum_{k_i=0}^{n} e^{\iu (\alpha_i-\alpha_i') k_i \theta_0 }  = (n+1)\delta_{\alpha_i=\alpha_i'\mod (n+1)}. 
\end{align}
{Let $\rho_{\mathcal{W}}^n$ be a state defined in Eq.~\mbox{\eqref{eqn:def_rho_W}}, i.e.,}
\begin{align}
    \rho_{\mathcal{W}}^n = \sum_{\phi_k\in\mathcal{W}} p_k \proj{\phi_k}^{\otimes n}.
\end{align}
{Using Eq.~\mbox{\eqref{eqn:phase_constant_equality}},} it holds that 
\begin{align}
    \rho_{\mathcal{W}}^{n} & = \sum_{k} \sum_{\alpha,\alpha'} \frac{n! e^{\iu (\alpha-\alpha')\cdot k \theta_0 }}{(n+1)^d d^n \sqrt{\alpha!\alpha'!}} \ketbra{\alpha}{\alpha'} \nonumber \\
    & = \sum_{\alpha} \frac{n!}{d^n \alpha!} \proj{\alpha} = \rho_0^n. 
\end{align}
Because $\rho_0^n$ is invariant under $\mathcal{T}(d)$, we obtain
\begin{align}
    V_{\theta}^{\otimes n} \rho_{\mathcal{W}}^{n} (V_{\theta}^\dagger)^{\otimes n} = \rho_0^n,\forall V_\theta\in\mathcal{T}(d). 
\end{align}
We apply the weighted subset $(\{p_k\},\mathcal{W})$ and the group of multi-phase unitaries $\mathcal{T}(d)$ and construct the remote state preparation protocol {following Eq.~\mbox{\eqref{eqn:P_eigen}} and~\mbox{\eqref{eqn:L_eigen}}:} 
\begin{align}
    & P_{\A} = P_{\B}=P_{\sym}^{n}, \\\
    & L_{\B}= \sum_\alpha \sqrt{\frac{n!}{d^n\alpha!}} \proj{\alpha} + \mathbbm{I}-P_{\sym}^{n}. 
\end{align}
{We have set the free parameter $c_{\perp}=1$.} The no-cloning bound follows Theorem~\ref{thm:bound_states_exact}. 

\begin{theorem}
    Let $\nu\in\mathbbm{Z}_{n+1}^{\times d}$. $F_{\rm wc}\left(\mathcal{R}_{n,m}^{\mathcal{M}(d)}\right)$ is upper bounded by  
    \begin{align}\label{eqn:bound_multi_phase}
        F_{\rm wc}\left(\mathcal{R}_{n,m}^{\mathcal{M}(d)}\right) \leq \sum_{\nu} \frac{1}{d^m} \frac{m!}{\beta^\nu!},  
    \end{align}
    where 
    \begin{align}
        \beta^\nu = \argmax_{\beta=\nu\mod(n+1)} \frac{m!}{\beta!}. 
    \end{align}
\end{theorem}
\begin{proof}
    We need to compute the upper bound from Theorem~\ref{thm:bound_states_exact} with
    \begin{align}
        F_{\rm wc}\left(\mathcal{R}_{n,m}^{\mathcal{M}(d)}\right)\leq F(\sigma_{\mathcal{T}(d)},\rho_{\mathcal{W}}^m), 
    \end{align}
    {where $\sigma_{\mathcal{T}(d)}$ is defined in Eq.~\mbox{\eqref{eqn:def_sigma_V}}, i.e., }
     \begin{align}
        \sigma_{\mathcal{T}(d)} = \int \frac{\d \theta}{(2\pi)^d} (V_\theta^\dagger)^{\otimes m} \sigma_{\mathcal{R}} V_\theta^{\otimes m}, 
    \end{align}
    {and $\rho_{\mathcal{W}}^m$ is defined in Eq.~\mbox{\eqref{eqn:def_rho_W}}, i.e.,}
    \begin{align}
        \rho_{\mathcal{W}}^m = \sum_{\phi_k\in\mathcal{W}} p_k \proj{\phi_k}^{\otimes m}. 
    \end{align}

    Because $\ket{\phi}^{\otimes m}$ is a quantum state on the symmetric subspace $\mathbbm{P}_{\sym}^m$, we have
    \begin{align}
        P_{\sym}^m \sigma_{\mathcal{T}(d)} P_{\sym}^m = \sigma_{\mathcal{T}(d)},
    \end{align}
    and as a result, 
     \begin{align}
        F(\sigma_{\mathcal{T}(d)}, \rho_{\mathcal{W}}^{m}) = F(P_{\sym}^m \sigma_{\mathcal{T}(d)} P_{\sym}^m, \rho_{\mathcal{W}}^{m}).  
    \end{align}
       
    Since $\sigma_{\mathcal{T}(d)}$ is defined by an integration over the Haar measure over $\mathcal{T}(d)$ and since $\mathbbm{P}_{\sym}^m$ is an invariant subspace of $\mathcal{T}(d)$, we have 
    \begin{align}
        P_{\sym}^{m} \sigma_{\mathcal{T}(d)} P_{\sym}^m = (V^\dagger)^{\otimes m} P_{\sym}^{m} \sigma_{\mathcal{T}(d)} P_{\sym}^m V^{\otimes m}, \nonumber  \\
        \forall V_\theta\in \mathcal{T}(d). 
    \end{align}
    Integrating both sides over $\mathcal{T}(d)$, it yields
    \begin{align}
        P_{\sym}^{m} \sigma_{\mathcal{T}(d)} P_{\sym}^m = \sum_\beta b_\beta \proj{\beta}. 
    \end{align}
    Let $\sigma_{\mathcal{T}(d)}^{\sym}$ be a density matrix in the form of 
    \begin{align}
        \sigma_{\mathcal{T}(d)}^{\sym} = \sum_{\beta} c_{\beta }\proj{\beta}, 
    \end{align}
    where $c_\beta = \frac{b_\beta}{\sum_{\beta'} b_{\beta'}}$. It's true that 
    \begin{align}
        P_{\sym}^{m} \sigma_{\mathcal{T}(d)} P_{\sym}^m \leq \sigma_{\mathcal{T}(d)}^{\sym}. 
    \end{align}
    Since $F(\rho,\sigma)$ monotonically increases with $\rho$ or $\sigma$, we obtain 
    \begin{align}
        F(\sigma_{\mathcal{T}(d)}, \rho_{\mathcal{W}}^{m}) \leq F(\sigma_{\mathcal{T}(d)}^{\sym}, \rho_{\mathcal{W}}^{m}).  
    \end{align}
    Further, we have
    \begin{align}
        \rho_{\mathcal{W}}^{m} = \sum_{k} \sum_{\beta,\beta'} \frac{m! e^{\iu (\beta-\beta')\cdot k \theta_0 }}{(n+1)^d d^m \sqrt{\beta!\beta'!}} \ketbra{\beta}{\beta'} = \sum_{\nu} d_\nu \proj{\varphi_\nu}, 
    \end{align}
    where 
    \begin{align}
        d_\nu  = \sum_{\beta=\nu \mod (n+1) }\frac{m!}{d^m \beta!},
    \end{align}
    and
    \begin{align}
        \ket{\varphi_\nu}  = \frac{1}{\sqrt{d^m d_\nu}} \sum_{\nu} \sqrt{\frac{m!}{\beta!}} \ket{\beta}. 
    \end{align}
    Each $\ket{\beta}$ has non-zero overlap with only one $\ket{\varphi_\nu}$, thus 
    \begin{align}
        F(\sigma_{\mathcal{T}(d)}^{\sym}, \rho_{\mathcal{W}}^{m}) = \left(\sum_{\nu} \sqrt{d_\nu \sum_{\beta} c_\beta |\!\braket{\varphi_\nu}{\beta}\!|^2}\right)^2.
    \end{align}
    Therefore, $F(\sigma_{\mathcal{T}(d)}^{\sym}, \rho_{\mathcal{W}}^{m})$ is upper bounded by the case where $c_{\beta^{\nu}}\neq 0$ only if $\beta^\nu=\argmax |\!\braket{\varphi_\nu}{\beta}\!|^2$, i.e.,
    \begin{align}
        F(\sigma_{\mathcal{T}(d)}^{\sym}, \rho_{\mathcal{W}}^{m}) \leq  \left(\sum_{\nu} \sqrt{ c_{\beta^\nu} \frac{m!}{d^m\beta^\nu!}}\right)^2.
    \end{align}
    Lastly, we use the Cauchy-Schwarz inequality to obtain 
    \begin{align}
        F(\sigma_{\mathcal{T}(d)}^{\sym}, \rho_{\mathcal{W}}^{m})\leq \sum_{\nu}\frac{m!}{d^m\beta^\nu!}, 
    \end{align}
    and we complete our proof. 
\end{proof}

In the asymptotic limit for large $n$ and $m$ and fixed $d$, when $m\gg dn^2$, the right-hand side of Eq.~\eqref{eqn:bound_multi_phase} is approximately upper-bounded by the error function (see Appendix~\ref{apd:multinomial}). Therefore, we obtain
\begin{align}
    F_{\rm wc}\left(\mathcal{R}_{n,m}^{\mathcal{M}(d)}\right)\leq \sqrt{d(1-\frac{1}{d})^{d-1}} \erf\left(\frac{dn}{2\sqrt{2(d-1)m}}\right)^{d-1} . 
\end{align}
Our bound shows that multi-phase states can be probabilistically cloned only if $m=O(n^2)$ for fixed $d$. This implies $n$-to-$m$ cloning for multi-phase states is  possible only if $m=O(n^2)$ for fixed $d$, as has been found in~\cite{Chiribella_2013}. 

The construction for multi-phase states in our paper generalizes the construction for clock states in~\cite{Sekatski_2015}. 

\subsection{Choi states for general unitaries}\label{sec:example_choi_general}

The set of Choi states $\mathcal{J}_{\mathcal{U}(d)}$ for the unitary group $\mathcal{U}(d)$ is defined as
\begin{align}
    \mathcal{J}_{\mathcal{U}(d)} =\Big\{\ket{J_U} = \frac{1}{\sqrt{d}}\sum_i U\ket{i}\otimes \ket{i} , \forall U\in\mathcal{U}(d) \Big\}. 
\end{align}
%We write $J_U = \proj{J_U}$. 
$\mathcal{U}(d)$ generates $\mathcal{J}_{\mathcal{U}(d)}$ from the maximally entangled state $\ket{\Psi^d}$. We now consider the no-cloning bound for $\mathcal{J}_{\mathcal{U}(d)}$. 

We use Schur-Weyl duality~\cite{Goodman_2000} 
\begin{align}\label{eqn:schur_weyl}
    (\mathbbm{C}^{d})^{\otimes l} \simeq \bigoplus_{\lambda} \mathbbm{P}_{\lambda}^{l} \otimes \mathbbm{Q}_{\lambda}^{l},
\end{align}
where $\mathbbm{P}_\lambda^l$ are irreps of the unitary group $\mathcal{U}(d)$, $\mathbbm{Q}_\lambda^{l}$ are irreps of the permutation group $\mathcal{P}(l)$ and $\lambda$ are the indices running over all irreps. We choose the Schur basis in~\cite{Bacon_2006}, which have real coefficients in the computational basis. We denote the Schur basis by $\{\ket{p_\lambda^l,q_\lambda^l,\lambda},\forall \lambda, p_\lambda^l, q_\lambda^l\}$. Due to the isomorphism Eq.~\eqref{eqn:schur_weyl}, we find a basis {$\{\ket{p_\lambda^l,\lambda}\}_{p_\lambda^l}$} of $\mathbbm{P}_\lambda^{l}$ and a basis {$\{\ket{q_\lambda^l,\lambda}\}_{q_\lambda^l}$} of $\mathbbm{Q}_\lambda^{l}$ such that $\ket{p_\lambda^l,q_\lambda^l, \lambda} \simeq \ket{p_\lambda^l,\lambda}\ket{q_\lambda^l,\lambda},\forall \lambda, p_\lambda^l, q_\lambda^l$. We will mainly work in the isomorphic space $\bigoplus_\lambda \mathbbm{P}_\lambda^l \otimes \mathbbm{Q}_\lambda^l$ and only return to $\mathbbm{C}^{d^l}$ in the final step. We denote the projector on $\mathbbm{P}_\lambda^l$ and $\mathbbm{Q}_\lambda^l$ by $P_\lambda^l$ and $Q_\lambda^l$ as well as the dimension of $\mathbbm{P}_\lambda^l$ and $\mathbbm{Q}_\lambda^l$ by $|\mathbbm{P}_\lambda^l|$ and $|\mathbbm{Q}_\lambda^l|$. The Choi states are quantum states on $\mathbbm{C}^{d^l}\otimes \mathbbm{C}^{d^l}$. We denote the first and second systems with subscripts $\R_1$ and $\R_2$ respectively. We apply Schur-Weyl duality to both systems. We denote the irreps of the unitary and permutation groups in the first system with subscripts $\P_1$, $\Q_1$ and $\P_2$, $\Q_2$ respectively. We show in Appendix~\ref{apd:haar_measure} that integrating $\proj{J_U}^{\otimes l}$ over the Haar measure on $\mathcal{U}(d)$ results in   
\begin{align}\label{eqn:choi_haar_measure}
    \int \d U \proj{J_U}^{\otimes l} = \rho_0^l, 
\end{align}
where
\begin{align}
    \rho_0^l & = \sum_{\lambda} \frac{|\mathbbm{Q}_\lambda^l|}{d^l |\mathbbm{P}_\lambda^l|} \Pi_\lambda^l,\label{eqn:rho_0_Choi} \\
    \Pi_\lambda^l & = P_{\lambda,\P_1}^l \otimes P_{\lambda,\P_2}^l \otimes \proj{\Psi_\lambda^l}_{\Q_1\Q_2}, \\
    \ket{\Psi_\lambda^l} & = \frac{1}{\sqrt{|\mathbbm{Q}_\lambda^l|}} \sum_{q_\lambda^l} \ket{q_\lambda^l,\lambda}_{\Q_1}\otimes \ket{q_\lambda^l,\lambda}_{\Q_2}. 
\end{align}
We omit the subscripts for systems because $\Pi_\lambda^l$ and $\ket{\Psi_\lambda^l}$ are symmetric for both systems. In this section, we will delve into the case where $l=n$ and $l=m$. 

Let quantum channels  $\mathscr{G}_{\mathcal{U}(d),n}$ and $\mathscr{G}_{\mathcal{U}_n}$ be 
\begin{align}
    \mathscr{G}_{\mathcal{U}(d),n}(\rho) = \int \d \mu(U) U^{\otimes n} \rho (U^{\dagger})^{\otimes n}, 
\end{align}
where the integration is over the Haar measure and 
\begin{align}
    \mathscr{G}_{\mathcal{U}_n}(\rho) = \sum_{U\in\mathcal{U}_n}p_U U^{\otimes n} \rho (U^{\dagger})^{\otimes n}. 
\end{align}
{where $\{p_U\}$ is a probability distribution on a discrete subset $\mathcal{U}_n$ of $\mathcal{U}(d)$}. Following the definition in~\cite{Dankert_2009,Low_2010,Nakata_2021}, a weighted exact unitary $n$-design $(\{p_U\},\mathcal{U}_n)$ satisfies
\begin{align}\label{eqn:exact_unitary_n_design}
    \mathscr{G}_{\mathcal{U}_n} = \mathscr{G}_{\mathcal{U}(d),n}.
\end{align}
Let $\mathcal{W} = \{\ket{J_U},U\in\mathcal{U}_n\}$. We apply {both $\mathscr{G}_{\mathcal{U}_n}\otimes \mathcal{I}$ and $\mathscr{G}_{\mathcal{U}(d),n}\otimes \mathcal{I}$ where $\mathcal{I}$ is the identity channel} in Eq.~\eqref{eqn:exact_unitary_n_design} to the maximally entangled {states} 
$\ket{\Psi^{d,n}}$ to get
\begin{align}
    \rho_{\mathcal{W}}^n = \rho_0^n, 
\end{align}
where $\rho_{\mathcal{W}}^n$ is defined according to Eq.~\eqref{eqn:def_rho_W}, i.e.,
\begin{align}
    \rho_{\mathcal{W}}^n = \sum_{\ket{J_U}\in\mathcal{U}_n} p_U \proj{J_U}.
\end{align}
Because $\rho_0^n$ is invariant under $\mathcal{U}(d)$, the above {state} is invariant under $\mathcal{U}(d)$ b i.e., 
\begin{align}
    (V^{\otimes n}\otimes \mathbbm{I}) \rho_{\mathcal{W}}^n ((V^\dagger)^{\otimes n}\otimes \mathbbm{I}) = \rho_0^n, \forall V\in\mathcal{U}(d). 
\end{align}
Similarly, a weighted relative $\epsilon$-approximate unitary $n$-design $(\{p_U\},\mathcal{U}_n)$ is defined as~\cite{Haferkamp_2022}
\begin{align}\label{eqn:approx_unitary_n_design}
    (1-\epsilon)\mathscr{G}_{\mathcal{U}(d),n}  \preccurlyeq  \mathscr{G}_{\mathcal{U}_n} \preccurlyeq (1+\epsilon)\mathscr{G}_{\mathcal{U}(d),n},
\end{align}
where $\mathcal{A}\preccurlyeq \mathcal{B}$ signifies for that $\mathcal{B}- \mathcal{A}$ is a completely positive map. Applying Eq.~\eqref{eqn:approx_unitary_n_design} to the maximally entangled state, we have 
\begin{align}
    (1-\epsilon) \rho_0^n \leq \rho_{\mathcal{W}}^n \leq (1+\epsilon) \rho_0^n, 
\end{align}
and 
\begin{align}
    (1-\epsilon) \rho_0^n  \leq (V^{\otimes n}\otimes \mathbbm{I}) \rho_{\mathcal{W}}^n ((V^{\dagger})^{\otimes n}\otimes \mathbbm{I}) \nonumber \\
    \leq (1+\epsilon) \rho_0^n, \forall V\in\mathcal{U}(d). 
\end{align}

We thus choose the set of Choi states for the weighted unitary $n$-design $(\{p_{U}\},\mathcal{W})$. {Following Eq.~\mbox{\eqref{eqn:P_eigen}} and~\mbox{\eqref{eqn:L_eigen}},} the remote state preparation protocol is constructed by
\begin{align}
    & P_{\A} = P_{\B} = P_{\cor}^n = \sum_\lambda \Pi_\lambda^n, \\
    & L_{\B} = \sum_\lambda \sqrt{\frac{|\mathbbm{Q}_\lambda^n|}{d^n|\mathbbm{P}_\lambda^n|}} \Pi_\lambda^n + (\mathbbm{I} - P_{\cor}^n). 
\end{align}
{We have set the free parameter $c_{\perp}=1$.} The no-cloning bound then follows from Theorem~\ref{thm:bound_states_approximate}. 

\begin{corollary}\label{cor:general_choi}
    Let $(\{p_U\},\mathcal{U}_n)$ be a weighted relative $\epsilon$-approximate unitary $n$-design with size $|\mathcal{U}_n|$. Then, 
    \begin{align}\label{eqn:bound_unitary}
        F_{\rm wc}\left(\mathcal{R}_{n,m}^{\mathcal{J}_{\mathcal{U}(d)}}\right) \leq (1+\epsilon)^2 \frac{|\mathcal{U}_n|}{d^{2m}}\sum_\lambda |\mathbbm{Q}_\lambda^m|^2. 
    \end{align}
\end{corollary}
\begin{proof}
    Following from Theorem~\ref{thm:bound_states_approximate}, we will work in the isomorphic space $\bigoplus_\lambda \mathbbm{P}_\lambda^m \otimes \mathbbm{Q}_\lambda^m$ to explicitly compute 
    \begin{align}
        F_{\rm wc}\left(\mathcal{R}_{n,m}^{\mathcal{J}_{\mathcal{U}(d)}}\right) \leq F(\sigma_{\mathcal{U}(d)},\rho_{\mathcal{W}}^m). 
    \end{align}
    {where $\sigma_{\mathcal{U}(d)}$ and $\rho_{\mathcal{W}}^m$ are defined in Eq.~\mbox{\eqref{eqn:def_sigma_V}} and Eq.~\mbox{\eqref{eqn:def_rho_W}} respectively, i.e.,}
    \begin{align}
        \sigma_{\mathcal{U}(d)}=\int \d \mu(V) ((V^{\dagger})^{\otimes m}\otimes \mathbbm{I})\sigma_\mathcal{R}(V^{\otimes m}\otimes \mathbbm{I}), 
    \end{align}
    and 
    \begin{align}
        \rho_{\mathcal{W}}^m = \sum_{U\in\mathcal{U}_n} p_U\proj{J_U}^{\otimes m}. 
    \end{align}

    By Schur-Weyl duality~\cite{Goodman_2000}, we write the $m$-fold unitary as
    \begin{align}
        U^{\otimes m} \simeq \bigoplus_\lambda U_\lambda^m \otimes Q_\lambda^m, 
    \end{align}
    where $U_\lambda^m$ and $Q_\lambda^m$ are on $\mathbbm{P}_\lambda^m$ and $\mathbbm{Q}_\lambda^m$ respectively. Because the Schur transform which maps computational basis to the Schur basis is real~\cite{Bacon_2006}, the $m$-fold maximally entangled state is 
    \begin{align}
        \ket{\Psi^{d,m}} = \sum_\lambda \sqrt{\frac{|\mathbbm{P}_\lambda^m||\mathbbm{Q}_\lambda^m|}{d^m}} \ket{\Phi_\lambda^m}\otimes \ket{\Psi_\lambda^m},
    \end{align}
    where  
    \begin{align}
        \ket{\Phi_\lambda^m} & = \frac{1}{\sqrt{|\mathbbm{P}_{\lambda}^m|}} \sum_{p_\lambda^m} \ket{p_\lambda^m,\lambda} \otimes \ket{p_\lambda^m,\lambda}, \\
        \ket{\Psi_\lambda^m} & = \frac{1}{\sqrt{|\mathbbm{Q}_{\lambda}^m|}} \sum_{q_\lambda^m} \ket{q_\lambda^m,\lambda} \otimes \ket{q_\lambda^m,\lambda}. 
    \end{align}
    As a result, we rewrite $\ket{J_U}^{\otimes m}$ as 
    \begin{align}
        & \ket{J_U}^{\otimes m} = U^{\otimes m}\ket{\Psi^{d,m}} \nonumber \\ 
        & = \bigoplus_\lambda \sqrt{\frac{|\mathbbm{P}_\lambda^m||\mathbbm{Q}_\lambda^m|} {d^m}} (U_{\lambda,\P_1}^m\otimes P_{\lambda,\P_2}^m)\ket{\Phi_\lambda^m} \otimes \ket{\Psi_\lambda^m}. 
    \end{align}
    $\ket{J_U}^{\otimes m}$ is indeed a quantum state on the subspace that $P_{\cor}^m$ projects onto, so it follows that
    \begin{align}
        P_{\cor}^m \rho_{\mathcal{W}}^m P_{\cor}^m = \rho_{\mathcal{W}}^m. 
    \end{align}
    Thus 
    \begin{align}\label{eqn:intermediate_1}
        F(\sigma_{\mathcal{U}(d)},\rho_{\mathcal{W}}^m) = F(P_{\cor}^m \sigma_{\mathcal{U}(d)} P_{\cor}^m,\rho_{\mathcal{W}}^m).
    \end{align}

    Because $\sigma_{\mathcal{U}(d)}$ is defined as an integration over the Haar measure on $\mathcal{U}(d)$, we have the unitary equivalence, i.e., 
    \begin{align}
        \sigma_{\mathcal{U}(d)}  = (V^{\otimes m}\otimes \mathbbm{I}) \sigma_{\mathcal{U}(d)} ((V^\dagger)^{\otimes m}\otimes \mathbbm{I}), \forall V\in\mathcal{U}(d). 
    \end{align}
    We integrate both sides over $\mathcal{U}(d)$ to obtain
    \begin{align}\label{eqn:sigma_U(d)}
        \sigma_{\mathcal{U}(d)} = \bigoplus_\lambda \frac{1}{|\mathbbm{P}_\lambda^m|} c_\lambda P_{\lambda,\P_1}^m \otimes \sigma_{\lambda, \Q_1 \R_2},
    \end{align}
    where $\sigma_{\lambda, \Q_1 \R_2}$ denotes the state on $\mathbbm{Q}_{\lambda,\Q_1}^m\otimes \mathbbm{C}^{d^m}_{\R_2}$. 
    
    Now recall that $\sigma_{\mathcal{U}(d)}$ is a Choi state. The partial trace on $\R_1$ of a Choi state is an identity on $\R_2$, i.e.,
    \begin{align}
        \Tr_{\R_1}(\sigma_{\mathcal{U}(d)}) = \sum_\lambda c_\lambda \Tr_{\Q_1}(\sigma_{\lambda,\Q_1\R_2}) =\frac{1}{d^m} \mathbbm{I}. 
    \end{align}
    As a result, the following inequality holds:  
    \begin{align}
         c_\lambda \Tr_{\Q_1}(\sigma_{\lambda,\Q_1\R_2}) \leq \frac{1}{d^m} \mathbbm{I}. 
    \end{align}
    Due to a well known result (originally proved in~\cite{Tomamichel_2012}, see Appendix~\ref{apd:pinching} for an alternative proof), we have
    \begin{align}
        \rho_{\A\B} \leq \dim(\A) \mathbbm{I}_{\A} \otimes \rho_{\B}. 
    \end{align}
    Therefore, it holds that
    \begin{align}
        c_\lambda \sigma_{\lambda,\Q_1\R_2} \leq \frac{|\mathbbm{Q}_\lambda^m|}{d^m}  Q_{\lambda,\Q_1}^m  \otimes \mathbbm{I}. 
    \end{align}
    Substituting into Eq.~\eqref{eqn:sigma_U(d)}, we get
    \begin{align}
        \sigma_{\mathcal{U}(d)} \leq \bigoplus_\lambda \frac{|\mathbbm{Q}_\lambda^m|}{d^m|\mathbbm{P}_\lambda^m|} P_{\lambda,\P_1}^m \otimes  Q_{\lambda,\Q_1}^m \otimes \mathbbm{I}. 
    \end{align}
    As a result, 
    \begin{align}
        P_{\cor}^m \sigma_{\mathcal{U}(d)} P_{\cor}^m \leq \sum_{\lambda} \frac{|\mathbbm{Q}_\lambda^m|}{d^m|\mathbbm{P}_\lambda^m|} \Pi_\lambda^m = \rho_0^m. 
    \end{align}
    Because $F(\rho,\sigma)$ increases as $\rho$ or $\sigma$ increases, 
    \begin{align}\label{eqn:intermediate_2}
        F(P_{\cor}^m \sigma_{\mathcal{U}(d)} P_{\cor}^m,\rho_{\mathcal{W}}^m) \leq F(\rho_0^m,\rho_{\mathcal{W}}^m). 
    \end{align}
    We now apply Eq.~\eqref{eqn:rank_inequality} in Appendix~\ref{apd:rank} to obtain
    \begin{align}
        F(\rho_0^m,\rho_{\mathcal{W}}^m) \leq |\mathcal{U}_n| F(\rho_0^m,\proj{J_U}^{\otimes m} ) = \frac{|\mathcal{U}_n|}{d^{2m}} \sum_\lambda |\mathbbm{Q}_\lambda^m|^2,
    \end{align}
    where we have used
    \begin{align}
        F(\rho_0^m,\proj{J_U}^{\otimes m}) = F(\rho_0^m,\proj{\Psi^{d,m}}) = \sum_\lambda \frac{|\mathbbm{Q}_\lambda^m|^2}{d^{2m}}, 
    \end{align}
    as $\rho_0^m$ is invariant under $V^{\otimes m} \otimes \mathbbm{I}$. 
\end{proof}

The authors are not aware of good estimations for $|\mathcal{W}|$ and $\sum_\lambda|\mathbbm{Q}_\lambda^m|^2$ so they could not analyze asymptotic behaviors for Eq.~\eqref{eqn:bound_unitary}. However, we use the following example to show that our method reproduces meaningful upper bounds. 

\begin{corollary}\label{cor:bound_Choi}
    $F_{\rm wc}\left(\mathcal{R}_{1,2}^{\mathcal{J}_{\mathcal{U}(2^r)}}\right)$ is upper bounded by 
    \begin{align}
        F_{\rm wc}\left(\mathcal{R}_{1,2}^{\mathcal{J}_{\mathcal{U}(2^r)}}\right)\leq  \frac{2^r+\sqrt{2^{2r}-1}}{2^{3r}}.
    \end{align}
\end{corollary}
\begin{proof}
    Let $d=2^r$, $n=1$ and $m=2$. Let the multi-qubit Pauli group be $\mathcal{U}_1$ and such that the Choi state of the multi-qubit Pauli group is $\mathcal{W}$. We decompose $\mathbbm{C}^{ 2^{2r}}\simeq (\mathbbm{P}_0^2\otimes \mathbbm{Q}_0^2 )\oplus (\mathbbm{P}_1^2\otimes \mathbbm{Q}_1^2)$, where $\lambda=0$ and $\lambda=1$ stand for the trivial and sign representation of the permutation group, respectively. Note:  
    \begin{align}
        & |\mathbbm{P}_0^2| = 2^{r-1}(2^r+1), \\
        & |\mathbbm{P}_1^2|=2^{r-1}(2^r-1), \\
        & |\mathbbm{Q}_0^2| = |\mathbbm{Q}_1^2|=1.
    \end{align}
    Using Eq.~\eqref{eqn:intermediate_1} and~\eqref{eqn:intermediate_2} and Theorem~\ref{thm:bound_states_approximate}, we can compute the upper bound with
    \begin{align}
        F_{\rm wc}\left(\mathcal{R}_{1,2}^{\mathcal{J}_{\mathcal{U}(2^r)}}\right)\leq F(\rho_0^2,\rho_{\mathcal{W}}^2). 
    \end{align}
    According to Eq.~\eqref{eqn:rho_0_Choi}, $\rho_0^2$ is explicitly written as 
    \begin{align}
        \rho_0^2 = \frac{P_{0}^2 \otimes P_{0}^2}{2^{3r-1}(2^r+1)}  \oplus \frac{P_{1}^2 \otimes P_{1}^2}{2^{3r-1}(2^r-1)} . 
    \end{align}
    At the same time, we find a complete orthonormal basis for $\mathbbm{C}^{2^{2r}}$ other than the Schur basis in~\cite{Bacon_2006}. We denote 
    \begin{align}
        \sigma_{k,l} & = \bigotimes_{i=1}^{r} X^{k_i} Z^{l_i},\\ 
        \ket{\Psi_{k,l}} & = \frac{1}{\sqrt{2^r}} (\sigma_{k,l}\otimes \mathbbm{I}) (\ket{00}+\ket{11})^{\otimes r}, 
    \end{align}
    where $k,l\in\mathbbm{Z}_{2}^{\times r}$. $\{\ket{\Psi_{k,l}}\}$ forms a complete orthonormal basis of $\mathbbm{C}^{2^{2r}}$, i.e.,
    \begin{align}
        \braket{\Psi_{k,l}}{\Psi_{k',l'}} = \delta_{kk'} \delta_{ll'}. 
    \end{align}
    Furthermore, $\{\ket{\Psi_{k,l}}, k\cdot l =0\}$ and $\{\ket{\Psi_{k,l}}, k\cdot l = 1\}$ form orthonormal basis of $\mathbbm{P}_{0}^2$ and  $\mathbbm{P}_{1}^2$ respectively. $\ket{\Psi_{k,l}}$ are also eigenstates of the $2$-fold Pauli group, i.e.,
    \begin{align}
        \sigma_{k',l'} \otimes \sigma_{k',l'} \ket{\Psi_{k,l}} = (-1)^{k\cdot l' + k'\cdot l} \ket{\Psi_{k,l}}. 
    \end{align}
    Within this complete orthonormal basis, $\rho_{\mathcal{W}}^2$ is given by
    %\begin{widetext}
    \begin{align}
        \rho_{\mathcal{W}}^2 = \frac{1}{4^{r}} \sum_{k,l} \proj{\Psi_{k,l}} \otimes \proj{\Psi_{k,l}}. 
    \end{align}
    %\end{widetext}
    Using $\sandwich{\Psi_{k,l}}{P_{\lambda}^2}{\Psi_{k,l}} = \delta_{\lambda,k\cdot l}$, we obtain 
    \begin{align}
        F(\rho_0^2,\rho_{\mathcal{W}}^2) = \Big(\frac{\sqrt{2^r+1} + \sqrt{2^r-1}}{2^{\frac{3r+1}{2}}}\Big)^2 = \frac{2^r + \sqrt{2^{2r}-1}}{2^{3r}}, 
    \end{align}
    which completes our proof.    
\end{proof}

Our upper bound $F_{\rm wc}\left(\mathcal{R}_{1,2}^{\mathcal{J}_{\mathcal{U}(2^r)}}\right)\leq  \frac{2^r+\sqrt{2^{2r}-1}}{2^{3r}}$ exactly reproduces the best known upper bound assuming quantum mechanics $F_{\rm wc}\left(\mathcal{R}_{1,2}^{\mathcal{J}_{\mathcal{U}(d)}}\right)\leq \frac{d+\sqrt{d^2-1}}{d^3}$ for $1$-to-$2$ cloning~\cite{Chiribella_2008} when $d=2^r$. 

For $d\neq 2^r$, however, it is not easy to  compute the upper bound analytically as we did in Corollary~\ref{cor:bound_Choi}. We numerically compute our upper bounds with sub-optimal unitary $1$-designs for small $d$. Numerical results indicate that our upper bounds with sub-optimal unitary $1$-designs from the no-signalling principle are close to but slightly larger than the best known upper bound from quantum mechanics in~\cite{Chiribella_2008}. Our upper bounds with the optimal unitary $1$-design from the no-signalling principle could potentially achieve the upper bound from quantum mechanics. 

\subsection{Spin coherent states}\label{sec:example_spin_coherent}

Let $d=2s+1$. The $z$-, $x$- and $y$- spin operators are {denoted} by $S_z$, $S_x$ and $S_y$. We denote the state with total spin $s$ and $z$-spin $p$ by $\ket{s,p}$. The set of spin coherent states with total spin $s$ (thus dimension $d=2s+1$) is defined as~\cite{Arecchi_1972}
\begin{align}
    \mathcal{L}(s) & = \{ \ket{s,\theta,\phi,z,\gamma}  = U_{\theta,\phi,z,\gamma}\ket{s,-s}, \forall U_{\theta,\phi,z,\gamma}\in \mathcal{A}(s) \}, \\
    \mathcal{A}(s) & = \{U_{\theta,\phi,z,\gamma} =e^{i\gamma} e^{\iu \theta (S_y\cos\phi-S_x\sin\phi)} e^{\iu z S_z}, \nonumber \\
    &\quad \forall (\theta,\phi,z,\gamma) \in[0,\pi)\times[0,2\pi)\times [0,2\pi)\times [0,2\pi) \}. 
\end{align}
Notice that $\ket{s,\theta,\phi,z,\gamma}$ corresponds to the same quantum states denoted by $\ket{s,\theta,\phi} = \ket{s,\theta,\phi,0,0}$ up to an irrelevant phase. $\mathcal{A}(s)$ generates $\mathcal{L}(s)$ from $\ket{s,-s}$. Moreover, $\mathcal{A}(s)$ is isomorphic to $\mathcal{U}(2)$, which is shown in the Euler angle representation: 
\begin{align}
    e^{\iu \gamma }e^{\iu \theta (S_y\cos\phi-S_x\sin\phi)} e^{\iu z S_z} = e^{\iu \gamma } e^{-\iu \phi S_z} e^{\iu \theta S_y} e^{\iu (z+\phi) S_z}. 
\end{align} 
The completeness of spin-$s$ coherent states is 
\begin{align}
    \frac{2s+1}{4\pi}\int\d\Omega \proj{s,\theta,\phi} = P_{s}, 
\end{align}
where $\d\Omega =\d\phi \d\theta \sin\theta$ and $P_{s} = \sum_{p=-s}^{s} \proj{s,p}$. 
We will work on the no-cloning bound for $\mathcal{L}(s)$.  

Notably, $l$-fold spin-$s$ coherent states are equivalent to spin-$ls$ coherent states. Indeed, denoting $S_\mu'= \sum_{i}S_{\mu,i}$ where $S_{\mu,i}$ for $S_\mu$ acting on the $i$th system, we obtain
\begin{align}
    & \ket{s,\theta,\phi}^{\otimes l} = \big(e^{\iu\theta(S_y\cos\phi-S_x\sin\phi  )}\ket{s,-s}\big)^{\otimes l} \nonumber \\ 
    & = e^{\iu\theta (S_y'\cos\phi - S_x'\sin\phi )} \ket{ls,-ls} = 
    \ket{ls,\theta,\phi}. 
\end{align}
The completeness of $l$-fold spin-$s$ coherent states is equivalent to the completeness of spin-$ls$ coherent states, i.e., 
\begin{align}
    \frac{2ls+1}{4\pi} \int \d\Omega \proj{s,\theta,\phi}^{\otimes l} = P_{ls}. 
\end{align}
Borrowing the idea in~\cite{Hayashi_2005}, we show in Appendix~\ref{apd:spin_design} that there exists a discrete proper subset $\mathcal{W}\subsetneq \mathcal{L}(s)$ with at most $\frac{(2ls+1)^2}{2}$ elements such that 
\begin{align}\label{eqn:spin_coherent_rq_2}
    \sum_{\ket{s,\theta,\phi}\in \mathcal{W}} p_{\theta,\phi}\proj{s,\theta,\phi}^{\otimes l} = \rho_0^l, 
\end{align}
where $\rho_0^l=\frac{1}{2ls+1} P_{ls}$. We call such subsets spin coherent designs. Better spin coherent designs are possible, but we only focus on its existence here. We will deal with the cases where $l=n$ and $l=m$. 

We choose a spin coherent $n$-design $\mathcal{W}$ and $\mathcal{A}(s)$ to construct the remote identical state preparation protocol {following Eq.~\mbox{\eqref{eqn:P_eigen}} and~\mbox{\eqref{eqn:L_eigen}}:}
\begin{align}
    P_{\A}=P_{\B}=P_{ns}, \quad L_{\B}=\frac{1}{\sqrt{2ns+1}}\mathbbm{I}, 
\end{align}
where we have set $c_{\perp}=\frac{1}{\sqrt{2ns+1}}$. 
\begin{corollary}\label{cor:spin_coherent}
    $F_{\rm wc}\left(\mathcal{R}_{n,m}^{\mathcal{L}(s)}\right)$ is upper bounded by 
    \begin{align}
        F_{\rm wc}\left(\mathcal{R}_{n,m}^{\mathcal{L}(s)}\right)\leq (1+\epsilon)^2\frac{|\mathcal{W}|}{2ms+1}. 
    \end{align}
\end{corollary}
\begin{proof}
    We again apply Theorem~\ref{thm:bound_states_approximate} to obtain 
    \begin{align}
        F_{\rm wc}\left(\mathcal{R}_{n,m}^{\mathcal{L}(s)}\right)\leq F(\sigma_{\mathcal{A}(s)},\rho_{\mathcal{W}}^m). 
    \end{align}
    {where $\sigma_{\mathcal{A}(s)}$ and $\rho_{\mathcal{W}}^m$ are defined in Eq.~\mbox{\eqref{eqn:def_sigma_V}} and Eq.~\mbox{\eqref{eqn:def_rho_W}},i.e.,}
    \begin{align}
        \sigma_{\mathcal{A}(s)} = \int \frac{\d \Omega}{4\pi} (U_{\theta,\phi,z,\gamma}^\dagger)^{\otimes m} \sigma_{\mathcal{R}} U_{\theta,\phi,z,\gamma}^{\otimes m}, 
    \end{align}
    and  
    \begin{align}
        \rho_{\mathcal{W}}^m = \sum_{\ket{s,\theta,\phi}\in\mathcal{W}} p_{\theta,\phi} \proj{s,\theta,\phi}^{\otimes m}. 
    \end{align}
    
    Because $\ket{s,\theta,\phi}^{\otimes m}$ is a quantum state on the subspace $P_{ms}$ projects onto, we have
    \begin{align}
        P_{ms}\rho_{\mathcal{W}}^m P_{ms} = \rho_{\mathcal{W}}^m.
    \end{align}
    Therefore, 
    \begin{align}
        F(\sigma_{\mathcal{A}(s)},\rho_{\mathcal{W}}^m) = F(P_{ms}\sigma_{\mathcal{A}(s)}P_{ms},\rho_{\mathcal{W}}^m).
    \end{align}
    
    Since $\sigma_{\mathcal{A}(s)}$ is an integration over the Haar measure on $\mathcal{A}(s)$, $[U_{\theta,\phi,z,\gamma}^{\otimes m},\sigma_{\mathcal{A}(s)}]=0$. Besides, the subspace that $P_{ms}$ projects onto is an invariant subspace of $U_{\theta,\phi,z,\gamma}^{\otimes m}$, thus $[U_{\theta,\phi,z,\gamma}^{\otimes m},P_{ms}]=0$. Combining both, we get
    \begin{align}
        [U_{\theta,\phi,z,\gamma}^{\otimes m}, P_{ms}\sigma_{\mathcal{A}(s)}P_{ms} ]=0, \forall U_{\theta,\phi,z,\gamma}\in\mathcal{A}(s). 
    \end{align}
    
    Because the subspace $P_{ms}$ projects onto is an irrep of $\mathcal{U}(2)$ and $\mathcal{A}(s)$ is isomorphic to ${\rm U(2)}$, we obtain by Schur's lemma that 
    \begin{align}
         P_{ms}\sigma_{\mathcal{A}(s)}P_{ms} = c P_{ms} \leq \rho_0^m. 
    \end{align}
    By the monotonicity of $F(\rho,\sigma)$ with $\rho$ or $\sigma$, we obtain 
    \begin{align}
        F(P_{ms}\sigma_{\mathcal{A}(s)}P_{ms},\rho_{\mathcal{W}}^m) \leq  F(\rho_0^m,\rho_{\mathcal{W}}^m), 
    \end{align}
    By applying \eqref{eqn:rank_inequality} in Appendix~\ref{apd:rank}, we conclude that
    \begin{align}
        F(\rho_0^m,\rho_{\mathcal{W}}^m) \leq |\mathcal{W}| F(\rho_0^m,\proj{s,\theta,\phi}^{\otimes m}) \leq \frac{|\mathcal{W}|}{2ms+1}, 
    \end{align}
    which is exactly the desired bound. 
\end{proof}
The spin coherent design in Appendix~\ref{apd:spin_design} corresponds to an upper bound 
\begin{align}
    F_{\rm wc}\left(\mathcal{R}_{n,m}^{\mathcal{L}(s)}\right)\leq \frac{(2ns+1)^2}{2(2ms+1)}. 
\end{align}
From our derivation, the same upper bound also holds for the average case global cloning fidelity $F_{\rm ac}\left(\mathcal{R}_{n,m}^{\mathcal{L}(s)}\right)$. 

The $n$-to-$m$ cloning of spin coherent states has been studied in~\cite{Demkowicz-Dobrzanski_2004, Yang_2014}. In~\cite{Yang_2014}, the authors construct a measure-and-prepare based probabilistic cloning oracle for spin coherent states that achieves the average case global cloning fidelity $F_{\rm ac}\left(\mathcal{R}_{n,m}^{\mathcal{L}(s)}\right) = \frac{2ns+1}{2(n+m)s+1}$. 
The construction establishes a lower bound on $F_{\rm ac}$. 

It remains an open question to find the tight upper and lower bound for cloning spin coherent states. Better upper bounds can be obtained with our scheme by constructing a better subset $\mathcal{W}$ with a smaller size $|\mathcal{W}|$. A possible method is to replace the exact version with the approximate version, which usually reduces the size of the spin coherent design significantly. However, we will leave this for future studies. 

\subsection{Stabilizer states}\label{sec:example_stabilizer}

Our scheme may also apply to meaningful discrete subsets, e.g. the set of multi-qubit stabilizer states. 

Let $r$ be the number of qubits. Multi-qubit stabilizer states are the $+1$ eigenstates of stabilizers~\cite{Gottesman_1998}. The set $\Stab(2^r)$ of $r$-qubit stabilizer states is generated from $\ket{0}^{\otimes r}$ by the $r$-qubit Clifford group $\Cl(2^r)$. We consider the no-cloning bound for $\Stab(2^r)$ for $n=1$ and $n=2$. 

Multi-qubit stabilizer states form a quantum $3$-design~\cite{Kueng_2015}. For $n=1$ and $n=2$, it's possible to form a quantum $n$-design $\mathcal{W}\subsetneq \Stab(2^r)$ and meanwhile ensure that $\mathcal{W}$ varies under $\Cl(2^r)$. In that case, we set 
\begin{align}
    P_{\A} = P_{\B}=P_{\sym}^{n}, \quad  L_{\B}= \frac{1}{\sqrt{d[n]}}\mathbbm{I}, 
\end{align}
which is the same as what we did for general states. 

For $n=1$, $\mathcal{W}$ is the computational basis $\CBS(2^r)$. Because multi-qubit Cliffords form a unitary $3$-design~\cite{Zhu_2017}, $\sigma_{\Cl(2^r)} = \sigma_{\mathcal{U}(2^r)} = P_{\sym}^{2^r,m}$ holds for $m\leq 3$ but not for $m\geq 4$. That means the no-cloning bound obtained via our method satisfies 
\begin{align}
    & F_{\rm wc}\left(\mathcal{R}_{1,m}^{\Stab(2^r)}\right) = F_{\rm wc}\left(\mathcal{R}_{1,m}^{\mathcal{S}(2^r)}\right), \quad m\leq 3, \\
    & F_{\rm wc}\left(\mathcal{R}_{1,m}^{\Stab(2^r)}\right) \geq F_{\rm wc}\left(\mathcal{R}_{1,m}^{\mathcal{S}(2^r)}\right), \quad m\geq 4. 
\end{align}

For $n=2$, $\mathcal{W}$ is the maximal set of multi-qubit mutually unbiased basis $\MUBS(2^r)$, because $\MUBS(2^r)\subset \Stab(2^r)$~\cite{Lawrence_2002}, $|\MUBS(2^r)| =2^r(2^r+1)< 2^r\prod_{i=1}^{r}(2^i+1)=|\Stab(2^r)|$~\cite{Wootters_1989,Gross_2006} for $r\geq 2$, and $\MUBS(2^r)$ is a quantum $2$-design~\cite{Klappenecker_2005}. Again, because Cliffords form a unitary $3$-design, the no-cloning bound obtained via our method satisfies
\begin{align}
    & F_{\rm wc}\left(\mathcal{R}_{2,m}^{\Stab(2^r)}\right) = F_{\rm wc}\left(\mathcal{R}_{2,m}^{\mathcal{S}(2^r)}\right), \quad m\leq 3, \\
    & F_{\rm wc}\left(\mathcal{R}_{2,m}^{\Stab(2^r)}\right) \geq F_{\rm wc}\left(\mathcal{R}_{2,m}^{\mathcal{S}(2^r)}\right), \quad m\geq 4. 
\end{align}

\subsection{Choi states for Cliffords}\label{sec:example_choi_clifford}
Our scheme is also applied to discrete subsets of Choi states. An example is the set of Choi states for multi-qubit Cliffords. 

Again let $r$ be the number of qubits that Cliffords act on. The set $\mathcal{J}_{\Cl(2^r)}$ of Choi states of $r$-qubit Cliffords is generated from $\frac{1}{\sqrt{2^r}}(\ket{00}+\ket{11})^{\otimes r}$ by the $r$-qubit Clifford group $\Cl(2^r)$. We consider the no-cloning bound for $\mathcal{J}_{\Cl(2^r)}$ for $n=1$ and $n=2$. 

Multi-qubit Cliffords form a unitary $3$-design~\cite{Zhu_2017} but fail to form a unitary $4$-design~\cite{Zhu_2016}. For $n=1$ and $n=2$, it's possible to find a unitary $n$-design $\mathcal{U}_n\subsetneq \Cl(2^r)$ and a corresponding set $\mathcal{W}=\mathcal{J}_{\mathcal{U}_n}$ of Choi states and meanwhile ensure that $\mathcal{J}_{\mathcal{U}_n}$ varies under $\Cl(2^r)$. In that case, we make the same choice as in Section~\ref{sec:example_choi_general}: 
\begin{align}
    & P_{\A} = P_{\B} = P_{\cor}^{n} = \sum_{\lambda} \Pi_\lambda^n, \\
    & L_{\B} = \sum_\lambda \sqrt{\frac{|\mathbbm{Q}_\lambda^n|}{2^{rn}|\mathbbm{P}_\lambda^n|}} \Pi_\lambda^n +(\mathbbm{I} - \sum_\lambda \Pi_\lambda^n). 
\end{align}

For $n=1$, $\mathcal{W}$ is the Choi states for the Pauli group $\Pauli(2^r)$. Multi-qubit Cliffords form a unitary $3$-design, therefore,
\begin{align}
    & F_{\rm wc}\left(\mathcal{R}_{1,m}^{\mathcal{J}_{\Cl(2^r)}}\right) = F_{\rm wc}\left(\mathcal{R}_{1,m}^{\mathcal{J}_{\mathcal{U}(2^r)}}\right), \quad m\leq 3, \\
    & F_{\rm wc}\left(\mathcal{R}_{1,m}^{\mathcal{J}_{\Cl(2^r)}}\right) \geq F_{\rm wc}\left(\mathcal{R}_{1,m}^{\mathcal{J}_{\mathcal{U}(2^r)}}\right), \quad m\geq 4. 
\end{align}

For $n=2$, $\mathcal{W}$ are the Choi states for any subset $\UDS(2^r)$ of Cliffords that still forms a unitary $2$-design. The existence of $\UDS(2^r)$ is proved by Chau in~\cite{Chau_2005}. Example constructions are given in~\cite{Cleve_2016,Bravyi_2022}. Because Cliffords form a unitary $3$-design, we have
\begin{align}
    & F_{\rm wc}\left(\mathcal{R}_{2,m}^{\mathcal{J}_{\Cl(2^r)}}\right) = F_{\rm wc}\left(\mathcal{R}_{2,m}^{\mathcal{J}_{\mathcal{U}(2^r)}}\right), \quad m\leq 3, \\
    & F_{\rm wc}\left(\mathcal{R}_{2,m}^{\mathcal{J}_{\Cl(2^r)}} \right) \geq F_{\rm wc}\left(\mathcal{R}_{2,m}^{\mathcal{J}_{\mathcal{U}(2^r)}}\right), \quad m\geq 4. 
\end{align}

\subsection{Application in quantum homomorphic encryption}\label{sec:example_qhe}
Quantum homomorphic encryption is a fundamental primitive out of which more complex quantum cryptography protocols can be built. It allows computation by a server for a user directly on encrypted data. We apply our no-cloning bound for Choi states for Cliffords to analyze the circuit privacy~\cite{Hu_2023} for quantum homomorphic encryption whose logical operators for Cliffords use transversal Cliffords. Especially, the quantum homomorphic encryption protocol which allows Cliffords in~\cite{Ouyang_2018} is such an example. 

In~\cite{Hu_2023}, the circuit privacy for quantum homomorphic encryption is defined within the real-world ideal-world simulation scheme. That is, we consider an actual protocol which the user and the server performs potentially maliciously and an ideal protocol which a third party performs for the user and the server honestly. If we can simulate the output of the actual protocol with the output of the ideal protocol, then the actual protocol is secure. The circuit privacy $\epsilon_c$ is explicitly defined as 
\begin{align}\label{circuit_privacy}
    \epsilon_c = \max_{\psi} \min_{\N,\phi} \max_{\F} \|\hat{\F}\otimes \ic(\psi) - \N \circ (\F \otimes \ic)(\phi) \|_{\Tr}. 
\end{align}
where $\psi$ is the purification of the actual state sent to the server, $\hat{\F}$ is the actual computation the server performs, $\phi$ is the purification of the ideal state sent to the server, $\F$ is the ideal computation the server performs, $\N$ is the simulator which simulates the actual protocol with the ideal protocol, and $\|\,\cdot\,\|_{\Tr}$ denotes the trace distance. 

\begin{theorem}
     Consider a quantum homomorphic encryption protocol which applies transversal logical operators $\hat{\F} = \F^{\otimes m} \otimes U_{\F}$ for all Cliffords $\F$, where $U_{\F}$ are transversal unitary but are not identical applications of $\F$. The circuit privacy $\epsilon_c$ for the quantum homomorphic encryption protocol is lower-bounded by the worst case cloning fidelity $F_{\rm wc}\left(\mathcal{R}_{1,m}^{\mathcal{J}_{\Cl(2^r)}}\right)$ for Choi states for Cliffords via 
    \begin{align}
        \epsilon_c  \geq  1 - \sqrt{F_{\rm wc}\left(\mathcal{R}_{1,m}^{\mathcal{J}_{\Cl(2^r)}}\right)}.
    \end{align}
\end{theorem}

\begin{proof} 
    Suppose that the quantum homomorphic encryption protocol encodes $r$ logical qubits. We denote the system that $U_{\F}$ acts on by $\rm aux$. We fix $\ket{\psi} = \ket{\Psi}^{\otimes m} \otimes \ket{\rm aux}$ where $\ket{\Psi}$ is the maximally entangled state and $\ket{\rm aux}$ is a pure state on $\rm aux$. Tracing out $\rm aux$, we obtain by the data processing inequality
    \begin{align}
        \epsilon_c \geq & \min_{\N,\phi} \max_{\F} \|\proj{J_\F}^{\otimes m} - \Tr_{\rm aux} \circ \N \circ (\F\otimes\ic)(\phi)\|_{\Tr}.
    \end{align}
    Noting that one access to $\F$ can be achieved probabilistically by one Choi state for $\F$ via gate teleportation, we can construct from $\mathcal{N}$ a probabilistic $1$-to-$m$ cloning oracle for Choi states for Cliffords, i.e.,
    \begin{align}
        \mathcal{C}_{1,m,p}^{\mathcal{J}_{\Cl(2^r)}}(\proj{J_\F}) & = p \mathcal{R}_{1,m}^{\mathcal{J}_{\Cl(2^r)}}(\proj{J_{\F}}) \otimes \proj{0} \nonumber \\
        &\ \ \ \  + (1-p) \rho_{\perp} \otimes \proj{1},  
    \end{align}
    which $\epsilon_c$-closely clones $\proj{J_\F}$ to $\proj{J_\F}^{\otimes m}$ in trace distance for all Cliffords $\F$ on success. The cloning oracle takes $\proj{J_\F}$ and appends $\phi$, then performs gate teleportation, after that declares $\proj{0}$ if gate teleportation is successful and otherwise $\proj{1}$, and finally applies $\Tr_{\rm aux} \circ \N$. Thus, by minimizing $\mathcal{R}_{1,m}^{\mathcal{J}_{\Cl(2^r)}}$ in the full set, $\epsilon_c$ is lower-bounded by 
    \begin{align}
        \epsilon_c \geq \min_{\mathcal{R}}\max_{\F}\left\|\proj{J_{\F}}^{\otimes m} -\mathcal{R}_{1,m}^{\mathcal{J}_{\Cl(2^r)}}(\proj{J_{\F}}) \right\|_{\Tr}. 
    \end{align}
    Notice that
    \begin{align}
        & F_{\rm wc}\left(\mathcal{R}_{1,m}^{\mathcal{J}_{\Cl(2^r)}}\right) \nonumber\\
        & = \max_{\mathcal{R}} \min_{\F} F\left(\proj{J_\F}^{\otimes m}, \mathcal{R}_{1,m}^{\mathcal{J}_{\Cl(2^r)}}(\proj{J_{\F}}) \right). 
    \end{align}
    As $\proj{J_{\F}}^{\otimes m}$ is pure, we further use the inequality when at least one of $\rho$ and $\sigma$ is pure (see Appendix~\ref{apd:trace_fidelity} for proof), i.e.,
    \begin{align}
        \|\rho-\sigma\|_{\Tr} \geq 1 - F(\rho,\sigma), 
    \end{align}
    to obtain
    \begin{align}
        \epsilon_c & \geq \min_{\mathcal{R}} \max_{\F} \left\| \proj{J_{\F}}^{\otimes m} -\mathcal{R}_{1,m}^{\mathcal{J}_{\Cl(2^r)}}(\proj{J_{\F}}) \right\|_{\Tr } \nonumber\\ 
        & \geq 1 - \max_{\mathcal{R}} \min_{\F} F\left(\proj{J_\F}^{\otimes m},\mathcal{R}_{1,m}^{\mathcal{J}_{\Cl(2^r)}}(\proj{J_{\F}})\right)\nonumber \\
        & \geq 1 - F_{\rm wc}\left(\mathcal{R}_{1,m}^{\mathcal{J}_{\Cl(2^r)}}\right). 
    \end{align}
\end{proof}

Due to the no-cloning bound for Choi states for Cliffords, $F_{\rm wc}\left(\mathcal{R}_{1,m}^{\mathcal{J}_{\Cl(2^r)}}\right)$ cannot be close to $1$ and $\epsilon_c$ cannot be close to $0$. More specifically, noting $F_{\rm wc} \left( \mathcal{R}_{1,m}^{\mathcal{J}_{\Cl(2^r)}} \right)$ decreases monotonically with respect to $m$ because of the data processing inequality, we have 
\begin{align}
    \epsilon_c & \geq 1 - F_{\rm wc} \left( \mathcal{R}_{1,m}^{\mathcal{J}_{\Cl(2^r)}} \right) \geq 1 - F_{\rm wc} \left( \mathcal{R}_{1,2}^{\mathcal{J}_{\Cl(2^r)}} \right) \nonumber\\
    & = 1 - F_{\rm wc} \left( \mathcal{R}_{1,2}^{\mathcal{J}_{\mathcal{U}(2^r)}} \right) \geq 1 - \frac{2^r+\sqrt{2^{2r}-1}}{2^{3r}}, 
\end{align}
for $m \geq 2$. Therefore, quantum homomorphic encryption which applies logical operators $\hat{\F} = \F^{\otimes m} \otimes U_{\F}$ for all Cliffords $\F$ cannot be circuit private. 

As an example, the quantum homomorphic encryption protocol which allows Cliffords in~\cite{Ouyang_2018} satisfies the above property. It encodes each logical qubit into $m = 2t$ physical qubits, where $t=4k+1$. It applies logical operators $\hat{\F}=\F^{\otimes m}$ for Cliffords $\F$. It can achieve perfect correctness $\epsilon=0$. When $t\to\infty$, it can achieve perfect data privacy $\epsilon_d \to 0$. The no-cloning bound reveals that it can only achieve circuit privacy $\epsilon_c \geq 0.533$. The aforementioned result aligns well with the trade-off bound established in~\cite{Hu_2023}, which states that $\epsilon_d +\epsilon_c +4\sqrt{\epsilon} \geq \frac{1}{2}$.

\section{Discussion and conclusion}\label{sec:discussion}

In this paper, we propose a general scheme discuss the performance of the cloning oracle
\begin{align}
    \mathcal{C}_{n,m,p}^{\mathcal{S}} & = p \mathcal{R}_{n,m}^{\mathcal{S}}(\proj{\psi}^{\otimes n}) \otimes \proj{0} \nonumber\\ 
    & \ \ \ + (1-p) \rho_{\perp} \otimes \proj{1}.
\end{align}

%\st{which need not be a quantum channel, but must satisfy the following properties. Conditioned on the success of the cloning oracle, $\mathcal{R}_{n,m}^{\mathcal{S}}$ has to map $n$ identical pure quantum states to quantum states for Theorem~\mbox{\ref{thm:bound_states_exact}} (or quantum states on the $n$-fold symmetric subspace to quantum states for Theorem~\mbox{\ref{thm:bound_states_approximate}}). The cloning oracle is allowed to be non-linear, non-positive, or non-trace preserving when it fails.}

By assuming the no-signalling condition as well as assumptions (i') and (ii'), we effectively assume $\mathcal{C}_{n,m,p}^{\mathcal{S}}$ to be quantum mechanical (linear, completely positive and trace preserving). 

A question naturally arises: is it possible to relax either (i') or (ii') such that the cloning oracle can fit into our framework while does not necessarily obey quantum mechanics? The answer is no, because an physical implementation of our framework requires that the quantum mechanical formalism holds on all possible input and output states, which exactly reproduces assumption (i') and (ii').

As we pointed out, our scheme is based on the full quantum mechanical framework. Consequently, our no-signalling bounds are either greater than or equal to optimal quantum mechanical bounds. As ~\mbox{\eqref{eqn:bound_states_exact}} and~\mbox{\eqref{eqn:bound_states_approximate}} imply, the extent to which our no-signalling bounds reproduce optimal quantum mechanical bounds depends on the choice of the subset $\mathcal{W}$ to (approximately) emulate the full set $\mathcal{S}$ up to the $n$-th order. Corollary~\mbox{\ref{cor:general_state}},~\mbox{\ref{cor:general_choi}} and~\mbox{\ref{cor:spin_coherent}} reveal the dependence of our no-signalling bounds on the particular subset $\mathcal{W}$. In general, subsets $\mathcal{W}$ with smaller sizes $|\mathcal{W}|$ tend to yield correspondingly smaller no-signalling bounds, thereby approaching optimal quantum mechanical bounds more closely. 

There are examples for this relation in various scenarios. For instance,  when discussing the no-cloning bound for general states, we choose an efficient algorithm~\mbox{\cite{Ambainis_2007}} to generate approximate quantum $n$-designs with small sizes with respect to the dimension $d$. In this way, we ultimately asymptotically approach the optimal quantum mechanical bound with the no-signalling bound. Similarly, regarding general Choi states, we choose the multi-qubit Pauli group as our subset $\mathcal{W}$ because their sizes achieve the lower bound on those of unitary $1$-designs~\mbox{\cite{Roy_2009}}. This allows our no-signalling bound to achieve the optimal quantum mechanical bound on the $1$-to-$2$ cloning for general Choi states. In conclusion, while our scheme is based on the full quantum mechanics, the convergence of our no-signalling bound and the optimal quantum mechanical bound relies on choosing the subset $\mathcal{W}$ wisely.

Our general scheme to derive the no-cloning bound from the no-signalling principle is versatile in that it applies to various subsets of states with symmetries. We have applied our scheme to continuous sets of states generated by a continuous group, including the set of general states, multi-phase states, Choi states for general unitaries, and spin coherent states. Our scheme also applies to discrete sets of states generated by a discrete group for certain cases, including the set of stabilizer states and the Choi states for Cliffords.

\begin{acknowledgments}
We thank David Gross for valuable suggestions on quantum and unitary designs. We also thank Reviewers' insightful comments, especially on the relation between quantum mechanics and the no-signalling condition. MT and YH are supported by the National Research Foundation, Singapore and A*STAR under its CQT Bridging Grant. They are also funded by the Quantum Engineering programme grant NRF2021-QEP2-01-P06.
\end{acknowledgments}

\bibliography{apssamp}% Produces the bibliography via BibTeX.

\begin{thebibliography}{83}%
\makeatletter
\providecommand \@ifxundefined [1]{%
 \@ifx{#1\undefined}
}%
\providecommand \@ifnum [1]{%
 \ifnum #1\expandafter \@firstoftwo
 \else \expandafter \@secondoftwo
 \fi
}%
\providecommand \@ifx [1]{%
 \ifx #1\expandafter \@firstoftwo
 \else \expandafter \@secondoftwo
 \fi
}%
\providecommand \natexlab [1]{#1}%
\providecommand \enquote  [1]{``#1''}%
\providecommand \bibnamefont  [1]{#1}%
\providecommand \bibfnamefont [1]{#1}%
\providecommand \citenamefont [1]{#1}%
\providecommand \href@noop [0]{\@secondoftwo}%
\providecommand \href [0]{\begingroup \@sanitize@url \@href}%
\providecommand \@href[1]{\@@startlink{#1}\@@href}%
\providecommand \@@href[1]{\endgroup#1\@@endlink}%
\providecommand \@sanitize@url [0]{\catcode `\\12\catcode `\$12\catcode
  `\&12\catcode `\#12\catcode `\^12\catcode `\_12\catcode `\%12\relax}%
\providecommand \@@startlink[1]{}%
\providecommand \@@endlink[0]{}%
\providecommand \url  [0]{\begingroup\@sanitize@url \@url }%
\providecommand \@url [1]{\endgroup\@href {#1}{\urlprefix }}%
\providecommand \urlprefix  [0]{URL }%
\providecommand \Eprint [0]{\href }%
\providecommand \doibase [0]{https://doi.org/}%
\providecommand \selectlanguage [0]{\@gobble}%
\providecommand \bibinfo  [0]{\@secondoftwo}%
\providecommand \bibfield  [0]{\@secondoftwo}%
\providecommand \translation [1]{[#1]}%
\providecommand \BibitemOpen [0]{}%
\providecommand \bibitemStop [0]{}%
\providecommand \bibitemNoStop [0]{.\EOS\space}%
\providecommand \EOS [0]{\spacefactor3000\relax}%
\providecommand \BibitemShut  [1]{\csname bibitem#1\endcsname}%
\let\auto@bib@innerbib\@empty
%</preamble>
\bibitem [{\citenamefont {Wiesner}(1983)}]{Wiesner_1983}%
  \BibitemOpen
  \bibfield  {author} {\bibinfo {author} {\bibfnamefont {S.}~\bibnamefont
  {Wiesner}},\ }\bibfield  {title} {\bibinfo {title} {Conjugate coding},\
  }\href {https://doi.org/10.1145/1008908.1008920} {\bibfield  {journal}
  {\bibinfo  {journal} {SIGACT News}\ }\textbf {\bibinfo {volume} {15}},\
  \bibinfo {pages} {78–88} (\bibinfo {year} {1983})}\BibitemShut {NoStop}%
\bibitem [{\citenamefont {Aaronson}\ \emph {et~al.}(2012)\citenamefont
  {Aaronson}, \citenamefont {Farhi}, \citenamefont {Gosset}, \citenamefont
  {Hassidim}, \citenamefont {Kelner},\ and\ \citenamefont
  {Lutomirski}}]{Aaronson_2012}%
  \BibitemOpen
  \bibfield  {author} {\bibinfo {author} {\bibfnamefont {S.}~\bibnamefont
  {Aaronson}}, \bibinfo {author} {\bibfnamefont {E.}~\bibnamefont {Farhi}},
  \bibinfo {author} {\bibfnamefont {D.}~\bibnamefont {Gosset}}, \bibinfo
  {author} {\bibfnamefont {A.}~\bibnamefont {Hassidim}}, \bibinfo {author}
  {\bibfnamefont {J.}~\bibnamefont {Kelner}},\ and\ \bibinfo {author}
  {\bibfnamefont {A.}~\bibnamefont {Lutomirski}},\ }\bibfield  {title}
  {\bibinfo {title} {Quantum money},\ }\href
  {https://doi.org/10.1145/2240236.2240258} {\bibfield  {journal} {\bibinfo
  {journal} {Commun. ACM}\ }\textbf {\bibinfo {volume} {55}},\ \bibinfo {pages}
  {84–92} (\bibinfo {year} {2012})}\BibitemShut {NoStop}%
\bibitem [{\citenamefont {Molina}\ \emph {et~al.}(2013)\citenamefont {Molina},
  \citenamefont {Vidick},\ and\ \citenamefont {Watrous}}]{Monila_2013}%
  \BibitemOpen
  \bibfield  {author} {\bibinfo {author} {\bibfnamefont {A.}~\bibnamefont
  {Molina}}, \bibinfo {author} {\bibfnamefont {T.}~\bibnamefont {Vidick}},\
  and\ \bibinfo {author} {\bibfnamefont {J.}~\bibnamefont {Watrous}},\
  }\bibfield  {title} {\bibinfo {title} {Optimal counterfeiting attacks and
  generalizations for wiesner's quantum money},\ }in\ \href
  {https://doi.org/10.1007/978-3-642-35656-8_4} {\emph {\bibinfo {booktitle}
  {Theory of Quantum Computation, Communication, and Cryptography}}},\ \bibinfo
  {editor} {edited by\ \bibinfo {editor} {\bibfnamefont {K.}~\bibnamefont
  {Iwama}}, \bibinfo {editor} {\bibfnamefont {Y.}~\bibnamefont {Kawano}},\ and\
  \bibinfo {editor} {\bibfnamefont {M.}~\bibnamefont {Murao}}}\ (\bibinfo
  {publisher} {Springer Berlin Heidelberg},\ \bibinfo {address} {Berlin,
  Heidelberg},\ \bibinfo {year} {2013})\ pp.\ \bibinfo {pages}
  {45--64}\BibitemShut {NoStop}%
\bibitem [{\citenamefont {Bennett}\ and\ \citenamefont
  {Brassard}(2014)}]{Bennett_2014}%
  \BibitemOpen
  \bibfield  {author} {\bibinfo {author} {\bibfnamefont {C.~H.}\ \bibnamefont
  {Bennett}}\ and\ \bibinfo {author} {\bibfnamefont {G.}~\bibnamefont
  {Brassard}},\ }\bibfield  {title} {\bibinfo {title} {Quantum cryptography:
  Public key distribution and coin tossing},\ }\href
  {https://doi.org/10.1016/j.tcs.2014.05.025} {\bibfield  {journal} {\bibinfo
  {journal} {Theoretical Computer Science}\ }\textbf {\bibinfo {volume}
  {560}},\ \bibinfo {pages} {7} (\bibinfo {year} {2014})},\ \bibinfo {note}
  {theoretical Aspects of Quantum Cryptography – celebrating 30 years of
  BB84}\BibitemShut {NoStop}%
\bibitem [{\citenamefont {Hillery}\ \emph {et~al.}(1999)\citenamefont
  {Hillery}, \citenamefont {Bu\ifmmode~\check{z}\else \v{z}\fi{}ek},\ and\
  \citenamefont {Berthiaume}}]{Hillery_1999}%
  \BibitemOpen
  \bibfield  {author} {\bibinfo {author} {\bibfnamefont {M.}~\bibnamefont
  {Hillery}}, \bibinfo {author} {\bibfnamefont {V.}~\bibnamefont
  {Bu\ifmmode~\check{z}\else \v{z}\fi{}ek}},\ and\ \bibinfo {author}
  {\bibfnamefont {A.}~\bibnamefont {Berthiaume}},\ }\bibfield  {title}
  {\bibinfo {title} {Quantum secret sharing},\ }\href
  {https://doi.org/10.1103/PhysRevA.59.1829} {\bibfield  {journal} {\bibinfo
  {journal} {Phys. Rev. A}\ }\textbf {\bibinfo {volume} {59}},\ \bibinfo
  {pages} {1829} (\bibinfo {year} {1999})}\BibitemShut {NoStop}%
\bibitem [{\citenamefont {Cleve}\ \emph {et~al.}(1999)\citenamefont {Cleve},
  \citenamefont {Gottesman},\ and\ \citenamefont {Lo}}]{Cleve_1999}%
  \BibitemOpen
  \bibfield  {author} {\bibinfo {author} {\bibfnamefont {R.}~\bibnamefont
  {Cleve}}, \bibinfo {author} {\bibfnamefont {D.}~\bibnamefont {Gottesman}},\
  and\ \bibinfo {author} {\bibfnamefont {H.-K.}\ \bibnamefont {Lo}},\
  }\bibfield  {title} {\bibinfo {title} {How to share a quantum secret},\
  }\href {https://doi.org/10.1103/PhysRevLett.83.648} {\bibfield  {journal}
  {\bibinfo  {journal} {Phys. Rev. Lett.}\ }\textbf {\bibinfo {volume} {83}},\
  \bibinfo {pages} {648} (\bibinfo {year} {1999})}\BibitemShut {NoStop}%
\bibitem [{\citenamefont {Gottesman}(2000)}]{Gottesman_2000}%
  \BibitemOpen
  \bibfield  {author} {\bibinfo {author} {\bibfnamefont {D.}~\bibnamefont
  {Gottesman}},\ }\bibfield  {title} {\bibinfo {title} {Theory of quantum
  secret sharing},\ }\href {https://doi.org/10.1103/PhysRevA.61.042311}
  {\bibfield  {journal} {\bibinfo  {journal} {Phys. Rev. A}\ }\textbf {\bibinfo
  {volume} {61}},\ \bibinfo {pages} {042311} (\bibinfo {year}
  {2000})}\BibitemShut {NoStop}%
\bibitem [{\citenamefont {Ananth}\ \emph {et~al.}(2023)\citenamefont {Ananth},
  \citenamefont {Kaleoglu},\ and\ \citenamefont {Liu}}]{Ananth_2023}%
  \BibitemOpen
  \bibfield  {author} {\bibinfo {author} {\bibfnamefont {P.}~\bibnamefont
  {Ananth}}, \bibinfo {author} {\bibfnamefont {F.}~\bibnamefont {Kaleoglu}},\
  and\ \bibinfo {author} {\bibfnamefont {Q.}~\bibnamefont {Liu}},\ }\href@noop
  {} {\bibinfo {title} {Cloning games: A general framework for unclonable
  primitives}} (\bibinfo {year} {2023}),\ \Eprint
  {https://arxiv.org/abs/2302.01874} {arXiv:2302.01874 [quant-ph]} \BibitemShut
  {NoStop}%
\bibitem [{\citenamefont {Broadbent}\ and\ \citenamefont
  {Islam}(2020)}]{Broadbent_2020a}%
  \BibitemOpen
  \bibfield  {author} {\bibinfo {author} {\bibfnamefont {A.}~\bibnamefont
  {Broadbent}}\ and\ \bibinfo {author} {\bibfnamefont {R.}~\bibnamefont
  {Islam}},\ }\bibfield  {title} {\bibinfo {title} {Quantum encryption with
  certified deletion},\ }in\ \href
  {https://doi.org/10.1007/978-3-030-64381-2_4} {\emph {\bibinfo {booktitle}
  {Theory of Cryptography}}},\ \bibinfo {editor} {edited by\ \bibinfo {editor}
  {\bibfnamefont {R.}~\bibnamefont {Pass}}\ and\ \bibinfo {editor}
  {\bibfnamefont {K.}~\bibnamefont {Pietrzak}}}\ (\bibinfo  {publisher}
  {Springer International Publishing},\ \bibinfo {address} {Cham},\ \bibinfo
  {year} {2020})\ pp.\ \bibinfo {pages} {92--122}\BibitemShut {NoStop}%
\bibitem [{\citenamefont {Broadbent}\ and\ \citenamefont
  {Lord}(2020)}]{Broadbent_2020b}%
  \BibitemOpen
  \bibfield  {author} {\bibinfo {author} {\bibfnamefont {A.}~\bibnamefont
  {Broadbent}}\ and\ \bibinfo {author} {\bibfnamefont {S.}~\bibnamefont
  {Lord}},\ }\bibfield  {title} {\bibinfo {title} {{Uncloneable Quantum
  Encryption via Oracles}},\ }in\ \href
  {https://doi.org/10.4230/LIPIcs.TQC.2020.4} {\emph {\bibinfo {booktitle}
  {15th Conference on the Theory of Quantum Computation, Communication and
  Cryptography (TQC 2020)}}},\ \bibinfo {series} {Leibniz International
  Proceedings in Informatics (LIPIcs)}, Vol.\ \bibinfo {volume} {158},\
  \bibinfo {editor} {edited by\ \bibinfo {editor} {\bibfnamefont {S.~T.}\
  \bibnamefont {Flammia}}}\ (\bibinfo  {publisher} {Schloss
  Dagstuhl--Leibniz-Zentrum f{\"u}r Informatik},\ \bibinfo {address} {Dagstuhl,
  Germany},\ \bibinfo {year} {2020})\ pp.\ \bibinfo {pages}
  {4:1--4:22}\BibitemShut {NoStop}%
\bibitem [{\citenamefont {Georgiou}\ and\ \citenamefont
  {Zhandry}(2020)}]{Georgiou_2020}%
  \BibitemOpen
  \bibfield  {author} {\bibinfo {author} {\bibfnamefont {M.}~\bibnamefont
  {Georgiou}}\ and\ \bibinfo {author} {\bibfnamefont {M.}~\bibnamefont
  {Zhandry}},\ }\href {https://eprint.iacr.org/2020/877} {\bibinfo {title}
  {Unclonable decryption keys}},\ \bibinfo {howpublished} {Cryptology ePrint
  Archive, Paper 2020/877} (\bibinfo {year} {2020})\BibitemShut {NoStop}%
\bibitem [{\citenamefont {Wootters}\ and\ \citenamefont
  {Zurek}(1982)}]{Wootters_1982}%
  \BibitemOpen
  \bibfield  {author} {\bibinfo {author} {\bibfnamefont {W.~K.}\ \bibnamefont
  {Wootters}}\ and\ \bibinfo {author} {\bibfnamefont {W.~H.}\ \bibnamefont
  {Zurek}},\ }\bibfield  {title} {\bibinfo {title} {A single quantum cannot be
  cloned},\ }\href {https://doi.org/10.1038/299802a0} {\bibfield  {journal}
  {\bibinfo  {journal} {Nature}\ }\textbf {\bibinfo {volume} {299}},\ \bibinfo
  {pages} {802} (\bibinfo {year} {1982})}\BibitemShut {NoStop}%
\bibitem [{\citenamefont {Dieks}(1982)}]{Dieks_1982}%
  \BibitemOpen
  \bibfield  {author} {\bibinfo {author} {\bibfnamefont {D.}~\bibnamefont
  {Dieks}},\ }\bibfield  {title} {\bibinfo {title} {Communication by epr
  devices},\ }\href {https://doi.org/10.1016/0375-9601(82)90084-6} {\bibfield
  {journal} {\bibinfo  {journal} {Physics Letters A}\ }\textbf {\bibinfo
  {volume} {92}},\ \bibinfo {pages} {271} (\bibinfo {year} {1982})}\BibitemShut
  {NoStop}%
\bibitem [{\citenamefont {Gisin}\ and\ \citenamefont
  {Huttner}(1997)}]{Gisin_1997a}%
  \BibitemOpen
  \bibfield  {author} {\bibinfo {author} {\bibfnamefont {N.}~\bibnamefont
  {Gisin}}\ and\ \bibinfo {author} {\bibfnamefont {B.}~\bibnamefont
  {Huttner}},\ }\bibfield  {title} {\bibinfo {title} {Quantum cloning,
  eavesdropping and bell's inequality},\ }\href
  {https://doi.org/10.1016/S0375-9601(97)00083-2} {\bibfield  {journal}
  {\bibinfo  {journal} {Physics Letters A}\ }\textbf {\bibinfo {volume}
  {228}},\ \bibinfo {pages} {13} (\bibinfo {year} {1997})}\BibitemShut
  {NoStop}%
\bibitem [{\citenamefont {Gisin}\ and\ \citenamefont
  {Massar}(1997)}]{Gisin_1997b}%
  \BibitemOpen
  \bibfield  {author} {\bibinfo {author} {\bibfnamefont {N.}~\bibnamefont
  {Gisin}}\ and\ \bibinfo {author} {\bibfnamefont {S.}~\bibnamefont {Massar}},\
  }\bibfield  {title} {\bibinfo {title} {Optimal quantum cloning machines},\
  }\href {https://doi.org/10.1103/PhysRevLett.79.2153} {\bibfield  {journal}
  {\bibinfo  {journal} {Phys. Rev. Lett.}\ }\textbf {\bibinfo {volume} {79}},\
  \bibinfo {pages} {2153} (\bibinfo {year} {1997})}\BibitemShut {NoStop}%
\bibitem [{\citenamefont {Werner}(1998)}]{Werner_1998}%
  \BibitemOpen
  \bibfield  {author} {\bibinfo {author} {\bibfnamefont {R.~F.}\ \bibnamefont
  {Werner}},\ }\bibfield  {title} {\bibinfo {title} {Optimal cloning of pure
  states},\ }\href {https://doi.org/10.1103/PhysRevA.58.1827} {\bibfield
  {journal} {\bibinfo  {journal} {Phys. Rev. A}\ }\textbf {\bibinfo {volume}
  {58}},\ \bibinfo {pages} {1827} (\bibinfo {year} {1998})}\BibitemShut
  {NoStop}%
\bibitem [{\citenamefont {Fan}\ \emph {et~al.}(2001)\citenamefont {Fan},
  \citenamefont {Matsumoto}, \citenamefont {Wang},\ and\ \citenamefont
  {Wadati}}]{Fan_2001}%
  \BibitemOpen
  \bibfield  {author} {\bibinfo {author} {\bibfnamefont {H.}~\bibnamefont
  {Fan}}, \bibinfo {author} {\bibfnamefont {K.}~\bibnamefont {Matsumoto}},
  \bibinfo {author} {\bibfnamefont {X.-B.}\ \bibnamefont {Wang}},\ and\
  \bibinfo {author} {\bibfnamefont {M.}~\bibnamefont {Wadati}},\ }\bibfield
  {title} {\bibinfo {title} {Quantum cloning machines for equatorial qubits},\
  }\href {https://doi.org/10.1103/PhysRevA.65.012304} {\bibfield  {journal}
  {\bibinfo  {journal} {Phys. Rev. A}\ }\textbf {\bibinfo {volume} {65}},\
  \bibinfo {pages} {012304} (\bibinfo {year} {2001})}\BibitemShut {NoStop}%
\bibitem [{\citenamefont {Buscemi}\ \emph {et~al.}(2005)\citenamefont
  {Buscemi}, \citenamefont {D'Ariano},\ and\ \citenamefont
  {Macchiavello}}]{Buscemi_2005}%
  \BibitemOpen
  \bibfield  {author} {\bibinfo {author} {\bibfnamefont {F.}~\bibnamefont
  {Buscemi}}, \bibinfo {author} {\bibfnamefont {G.~M.}\ \bibnamefont
  {D'Ariano}},\ and\ \bibinfo {author} {\bibfnamefont {C.}~\bibnamefont
  {Macchiavello}},\ }\bibfield  {title} {\bibinfo {title} {Economical
  phase-covariant cloning of qudits},\ }\href
  {https://doi.org/10.1103/PhysRevA.71.042327} {\bibfield  {journal} {\bibinfo
  {journal} {Phys. Rev. A}\ }\textbf {\bibinfo {volume} {71}},\ \bibinfo
  {pages} {042327} (\bibinfo {year} {2005})}\BibitemShut {NoStop}%
\bibitem [{\citenamefont {Demkowicz-Dobrza\ifmmode~\acute{n}\else
  \'{n}\fi{}ski}\ \emph {et~al.}(2004)\citenamefont
  {Demkowicz-Dobrza\ifmmode~\acute{n}\else \'{n}\fi{}ski}, \citenamefont
  {Ku\ifmmode~\acute{s}\else \'{s}\fi{}},\ and\ \citenamefont
  {W\'odkiewicz}}]{Demkowicz-Dobrzanski_2004}%
  \BibitemOpen
  \bibfield  {author} {\bibinfo {author} {\bibfnamefont {R.}~\bibnamefont
  {Demkowicz-Dobrza\ifmmode~\acute{n}\else \'{n}\fi{}ski}}, \bibinfo {author}
  {\bibfnamefont {M.}~\bibnamefont {Ku\ifmmode~\acute{s}\else \'{s}\fi{}}},\
  and\ \bibinfo {author} {\bibfnamefont {K.}~\bibnamefont {W\'odkiewicz}},\
  }\bibfield  {title} {\bibinfo {title} {Cloning of spin-coherent states},\
  }\href {https://doi.org/10.1103/PhysRevA.69.012301} {\bibfield  {journal}
  {\bibinfo  {journal} {Phys. Rev. A}\ }\textbf {\bibinfo {volume} {69}},\
  \bibinfo {pages} {012301} (\bibinfo {year} {2004})}\BibitemShut {NoStop}%
\bibitem [{\citenamefont {Duan}\ and\ \citenamefont {Guo}(1998)}]{Duan_1998}%
  \BibitemOpen
  \bibfield  {author} {\bibinfo {author} {\bibfnamefont {L.-M.}\ \bibnamefont
  {Duan}}\ and\ \bibinfo {author} {\bibfnamefont {G.-C.}\ \bibnamefont {Guo}},\
  }\bibfield  {title} {\bibinfo {title} {Probabilistic cloning and
  identification of linearly independent quantum states},\ }\href
  {https://doi.org/10.1103/PhysRevLett.80.4999} {\bibfield  {journal} {\bibinfo
   {journal} {Phys. Rev. Lett.}\ }\textbf {\bibinfo {volume} {80}},\ \bibinfo
  {pages} {4999} (\bibinfo {year} {1998})}\BibitemShut {NoStop}%
\bibitem [{\citenamefont {Fiur\'a\ifmmode~\check{s}\else
  \v{s}\fi{}ek}(2004)}]{Fiurasek_2004}%
  \BibitemOpen
  \bibfield  {author} {\bibinfo {author} {\bibfnamefont {J.}~\bibnamefont
  {Fiur\'a\ifmmode~\check{s}\else \v{s}\fi{}ek}},\ }\bibfield  {title}
  {\bibinfo {title} {Optimal probabilistic cloning and purification of quantum
  states},\ }\href {https://doi.org/10.1103/PhysRevA.70.032308} {\bibfield
  {journal} {\bibinfo  {journal} {Phys. Rev. A}\ }\textbf {\bibinfo {volume}
  {70}},\ \bibinfo {pages} {032308} (\bibinfo {year} {2004})}\BibitemShut
  {NoStop}%
\bibitem [{\citenamefont {Chiribella}\ \emph {et~al.}(2013)\citenamefont
  {Chiribella}, \citenamefont {Yang},\ and\ \citenamefont
  {Yao}}]{Chiribella_2013}%
  \BibitemOpen
  \bibfield  {author} {\bibinfo {author} {\bibfnamefont {G.}~\bibnamefont
  {Chiribella}}, \bibinfo {author} {\bibfnamefont {Y.}~\bibnamefont {Yang}},\
  and\ \bibinfo {author} {\bibfnamefont {A.~C.-C.}\ \bibnamefont {Yao}},\
  }\bibfield  {title} {\bibinfo {title} {Quantum replication at the heisenberg
  limit},\ }\href {https://doi.org/10.1038/ncomms3915} {\bibfield  {journal}
  {\bibinfo  {journal} {Nature communications}\ }\textbf {\bibinfo {volume}
  {4}},\ \bibinfo {pages} {2915} (\bibinfo {year} {2013})}\BibitemShut
  {NoStop}%
\bibitem [{\citenamefont {Chiribella}\ \emph {et~al.}(2015)\citenamefont
  {Chiribella}, \citenamefont {Yang},\ and\ \citenamefont
  {Huang}}]{Chiribella_2015}%
  \BibitemOpen
  \bibfield  {author} {\bibinfo {author} {\bibfnamefont {G.}~\bibnamefont
  {Chiribella}}, \bibinfo {author} {\bibfnamefont {Y.}~\bibnamefont {Yang}},\
  and\ \bibinfo {author} {\bibfnamefont {C.}~\bibnamefont {Huang}},\ }\bibfield
   {title} {\bibinfo {title} {Universal superreplication of unitary gates},\
  }\href {https://doi.org/10.1103/PhysRevLett.114.120504} {\bibfield  {journal}
  {\bibinfo  {journal} {Phys. Rev. Lett.}\ }\textbf {\bibinfo {volume} {114}},\
  \bibinfo {pages} {120504} (\bibinfo {year} {2015})}\BibitemShut {NoStop}%
\bibitem [{\citenamefont {Yang}(2018)}]{Yang_2018}%
  \BibitemOpen
  \bibfield  {author} {\bibinfo {author} {\bibfnamefont {Y.}~\bibnamefont
  {Yang}},\ }\bibfield  {title} {\bibinfo {title} {Compression and replication
  of quantum information},\ }\href {http://hdl.handle.net/10722/265308}
  {\bibfield  {journal} {\bibinfo  {journal} {HKU Theses Online (HKUTO)}\ }
  (\bibinfo {year} {2018})}\BibitemShut {NoStop}%
\bibitem [{\citenamefont {D\"ur}\ \emph {et~al.}(2015)\citenamefont {D\"ur},
  \citenamefont {Sekatski},\ and\ \citenamefont {Skotiniotis}}]{Dur_2015}%
  \BibitemOpen
  \bibfield  {author} {\bibinfo {author} {\bibfnamefont {W.}~\bibnamefont
  {D\"ur}}, \bibinfo {author} {\bibfnamefont {P.}~\bibnamefont {Sekatski}},\
  and\ \bibinfo {author} {\bibfnamefont {M.}~\bibnamefont {Skotiniotis}},\
  }\bibfield  {title} {\bibinfo {title} {Deterministic superreplication of
  one-parameter unitary transformations},\ }\href
  {https://doi.org/10.1103/PhysRevLett.114.120503} {\bibfield  {journal}
  {\bibinfo  {journal} {Phys. Rev. Lett.}\ }\textbf {\bibinfo {volume} {114}},\
  \bibinfo {pages} {120503} (\bibinfo {year} {2015})}\BibitemShut {NoStop}%
\bibitem [{\citenamefont {Peres}\ and\ \citenamefont
  {Terno}(2004)}]{Peres_2004}%
  \BibitemOpen
  \bibfield  {author} {\bibinfo {author} {\bibfnamefont {A.}~\bibnamefont
  {Peres}}\ and\ \bibinfo {author} {\bibfnamefont {D.~R.}\ \bibnamefont
  {Terno}},\ }\bibfield  {title} {\bibinfo {title} {Quantum information and
  relativity theory},\ }\href {https://doi.org/10.1103/RevModPhys.76.93}
  {\bibfield  {journal} {\bibinfo  {journal} {Rev. Mod. Phys.}\ }\textbf
  {\bibinfo {volume} {76}},\ \bibinfo {pages} {93} (\bibinfo {year}
  {2004})}\BibitemShut {NoStop}%
\bibitem [{\citenamefont {Bassi}\ \emph {et~al.}(2013)\citenamefont {Bassi},
  \citenamefont {Lochan}, \citenamefont {Satin}, \citenamefont {Singh},\ and\
  \citenamefont {Ulbricht}}]{Bassi_2013}%
  \BibitemOpen
  \bibfield  {author} {\bibinfo {author} {\bibfnamefont {A.}~\bibnamefont
  {Bassi}}, \bibinfo {author} {\bibfnamefont {K.}~\bibnamefont {Lochan}},
  \bibinfo {author} {\bibfnamefont {S.}~\bibnamefont {Satin}}, \bibinfo
  {author} {\bibfnamefont {T.~P.}\ \bibnamefont {Singh}},\ and\ \bibinfo
  {author} {\bibfnamefont {H.}~\bibnamefont {Ulbricht}},\ }\bibfield  {title}
  {\bibinfo {title} {Models of wave-function collapse, underlying theories, and
  experimental tests},\ }\href {https://doi.org/10.1103/RevModPhys.85.471}
  {\bibfield  {journal} {\bibinfo  {journal} {Rev. Mod. Phys.}\ }\textbf
  {\bibinfo {volume} {85}},\ \bibinfo {pages} {471} (\bibinfo {year}
  {2013})}\BibitemShut {NoStop}%
\bibitem [{\citenamefont {Kaplan}\ and\ \citenamefont
  {Rajendran}(2022)}]{Kaplan_2022}%
  \BibitemOpen
  \bibfield  {author} {\bibinfo {author} {\bibfnamefont {D.~E.}\ \bibnamefont
  {Kaplan}}\ and\ \bibinfo {author} {\bibfnamefont {S.}~\bibnamefont
  {Rajendran}},\ }\bibfield  {title} {\bibinfo {title} {Causal framework for
  nonlinear quantum mechanics},\ }\href
  {https://doi.org/10.1103/PhysRevD.105.055002} {\bibfield  {journal} {\bibinfo
   {journal} {Phys. Rev. D}\ }\textbf {\bibinfo {volume} {105}},\ \bibinfo
  {pages} {055002} (\bibinfo {year} {2022})}\BibitemShut {NoStop}%
\bibitem [{\citenamefont {Polkovnikov}\ \emph {et~al.}(2023)\citenamefont
  {Polkovnikov}, \citenamefont {Gramolin}, \citenamefont {Kaplan},
  \citenamefont {Rajendran},\ and\ \citenamefont {Sushkov}}]{Polkovnikov_2023}%
  \BibitemOpen
  \bibfield  {author} {\bibinfo {author} {\bibfnamefont {M.}~\bibnamefont
  {Polkovnikov}}, \bibinfo {author} {\bibfnamefont {A.~V.}\ \bibnamefont
  {Gramolin}}, \bibinfo {author} {\bibfnamefont {D.~E.}\ \bibnamefont
  {Kaplan}}, \bibinfo {author} {\bibfnamefont {S.}~\bibnamefont {Rajendran}},\
  and\ \bibinfo {author} {\bibfnamefont {A.~O.}\ \bibnamefont {Sushkov}},\
  }\bibfield  {title} {\bibinfo {title} {Experimental limit on nonlinear
  state-dependent terms in quantum theory},\ }\href
  {https://doi.org/10.1103/PhysRevLett.130.040202} {\bibfield  {journal}
  {\bibinfo  {journal} {Phys. Rev. Lett.}\ }\textbf {\bibinfo {volume} {130}},\
  \bibinfo {pages} {040202} (\bibinfo {year} {2023})}\BibitemShut {NoStop}%
\bibitem [{\citenamefont {Ghirardi}\ \emph {et~al.}(1986)\citenamefont
  {Ghirardi}, \citenamefont {Rimini},\ and\ \citenamefont
  {Weber}}]{Ghirardi_1986}%
  \BibitemOpen
  \bibfield  {author} {\bibinfo {author} {\bibfnamefont {G.~C.}\ \bibnamefont
  {Ghirardi}}, \bibinfo {author} {\bibfnamefont {A.}~\bibnamefont {Rimini}},\
  and\ \bibinfo {author} {\bibfnamefont {T.}~\bibnamefont {Weber}},\ }\bibfield
   {title} {\bibinfo {title} {Unified dynamics for microscopic and macroscopic
  systems},\ }\href {https://doi.org/10.1103/PhysRevD.34.470} {\bibfield
  {journal} {\bibinfo  {journal} {Phys. Rev. D}\ }\textbf {\bibinfo {volume}
  {34}},\ \bibinfo {pages} {470} (\bibinfo {year} {1986})}\BibitemShut
  {NoStop}%
\bibitem [{\citenamefont {Pearle}(1989)}]{Pearle_1989}%
  \BibitemOpen
  \bibfield  {author} {\bibinfo {author} {\bibfnamefont {P.}~\bibnamefont
  {Pearle}},\ }\bibfield  {title} {\bibinfo {title} {Combining stochastic
  dynamical state-vector reduction with spontaneous localization},\ }\href
  {https://doi.org/10.1103/PhysRevA.39.2277} {\bibfield  {journal} {\bibinfo
  {journal} {Phys. Rev. A}\ }\textbf {\bibinfo {volume} {39}},\ \bibinfo
  {pages} {2277} (\bibinfo {year} {1989})}\BibitemShut {NoStop}%
\bibitem [{\citenamefont {Ghirardi}\ \emph {et~al.}(1990)\citenamefont
  {Ghirardi}, \citenamefont {Pearle},\ and\ \citenamefont
  {Rimini}}]{Ghirardi_1990}%
  \BibitemOpen
  \bibfield  {author} {\bibinfo {author} {\bibfnamefont {G.~C.}\ \bibnamefont
  {Ghirardi}}, \bibinfo {author} {\bibfnamefont {P.}~\bibnamefont {Pearle}},\
  and\ \bibinfo {author} {\bibfnamefont {A.}~\bibnamefont {Rimini}},\
  }\bibfield  {title} {\bibinfo {title} {Markov processes in hilbert space and
  continuous spontaneous localization of systems of identical particles},\
  }\href {https://doi.org/10.1103/PhysRevA.42.78} {\bibfield  {journal}
  {\bibinfo  {journal} {Phys. Rev. A}\ }\textbf {\bibinfo {volume} {42}},\
  \bibinfo {pages} {78} (\bibinfo {year} {1990})}\BibitemShut {NoStop}%
\bibitem [{\citenamefont {Di\'osi}(1989)}]{Diosi_1989}%
  \BibitemOpen
  \bibfield  {author} {\bibinfo {author} {\bibfnamefont {L.}~\bibnamefont
  {Di\'osi}},\ }\bibfield  {title} {\bibinfo {title} {Models for universal
  reduction of macroscopic quantum fluctuations},\ }\href
  {https://doi.org/10.1103/PhysRevA.40.1165} {\bibfield  {journal} {\bibinfo
  {journal} {Phys. Rev. A}\ }\textbf {\bibinfo {volume} {40}},\ \bibinfo
  {pages} {1165} (\bibinfo {year} {1989})}\BibitemShut {NoStop}%
\bibitem [{\citenamefont {Bassi}(2005)}]{Bassi_2005}%
  \BibitemOpen
  \bibfield  {author} {\bibinfo {author} {\bibfnamefont {A.}~\bibnamefont
  {Bassi}},\ }\bibfield  {title} {\bibinfo {title} {Collapse models: analysis
  of the free particle dynamics},\ }\href
  {https://doi.org/10.1088/0305-4470/38/14/008} {\bibfield  {journal} {\bibinfo
   {journal} {Journal of Physics A: Mathematical and General}\ }\textbf
  {\bibinfo {volume} {38}},\ \bibinfo {pages} {3173} (\bibinfo {year}
  {2005})}\BibitemShut {NoStop}%
\bibitem [{\citenamefont {Gisin}(1998)}]{Gisin_1998}%
  \BibitemOpen
  \bibfield  {author} {\bibinfo {author} {\bibfnamefont {N.}~\bibnamefont
  {Gisin}},\ }\bibfield  {title} {\bibinfo {title} {Quantum cloning without
  signaling},\ }\href {https://doi.org/10.1016/S0375-9601(98)00170-4}
  {\bibfield  {journal} {\bibinfo  {journal} {Physics Letters A}\ }\textbf
  {\bibinfo {volume} {242}},\ \bibinfo {pages} {1} (\bibinfo {year}
  {1998})}\BibitemShut {NoStop}%
\bibitem [{\citenamefont {Simon}\ \emph {et~al.}(2001)\citenamefont {Simon},
  \citenamefont {Bu\ifmmode~\check{z}\else \v{z}\fi{}ek},\ and\ \citenamefont
  {Gisin}}]{Simon_2001}%
  \BibitemOpen
  \bibfield  {author} {\bibinfo {author} {\bibfnamefont {C.}~\bibnamefont
  {Simon}}, \bibinfo {author} {\bibfnamefont {V.}~\bibnamefont
  {Bu\ifmmode~\check{z}\else \v{z}\fi{}ek}},\ and\ \bibinfo {author}
  {\bibfnamefont {N.}~\bibnamefont {Gisin}},\ }\bibfield  {title} {\bibinfo
  {title} {No-signaling condition and quantum dynamics},\ }\href
  {https://doi.org/10.1103/PhysRevLett.87.170405} {\bibfield  {journal}
  {\bibinfo  {journal} {Phys. Rev. Lett.}\ }\textbf {\bibinfo {volume} {87}},\
  \bibinfo {pages} {170405} (\bibinfo {year} {2001})}\BibitemShut {NoStop}%
\bibitem [{\citenamefont {Hardy}\ and\ \citenamefont
  {Song}(1999)}]{Hardy_1999}%
  \BibitemOpen
  \bibfield  {author} {\bibinfo {author} {\bibfnamefont {L.}~\bibnamefont
  {Hardy}}\ and\ \bibinfo {author} {\bibfnamefont {D.~D.}\ \bibnamefont
  {Song}},\ }\bibfield  {title} {\bibinfo {title} {No signalling and
  probabilistic quantum cloning},\ }\href
  {https://doi.org/10.1016/S0375-9601(99)00448-X} {\bibfield  {journal}
  {\bibinfo  {journal} {Physics Letters A}\ }\textbf {\bibinfo {volume}
  {259}},\ \bibinfo {pages} {331} (\bibinfo {year} {1999})}\BibitemShut
  {NoStop}%
\bibitem [{\citenamefont {Pati}(2000)}]{Pati_2000}%
  \BibitemOpen
  \bibfield  {author} {\bibinfo {author} {\bibfnamefont {A.~K.}\ \bibnamefont
  {Pati}},\ }\bibfield  {title} {\bibinfo {title} {Probabilistic exact cloning
  and probabilistic no-signalling},\ }\href
  {https://doi.org/https://doi.org/10.1016/S0375-9601(00)00281-4} {\bibfield
  {journal} {\bibinfo  {journal} {Physics Letters A}\ }\textbf {\bibinfo
  {volume} {270}},\ \bibinfo {pages} {103} (\bibinfo {year}
  {2000})}\BibitemShut {NoStop}%
\bibitem [{\citenamefont {Sekatski}\ \emph {et~al.}(2015)\citenamefont
  {Sekatski}, \citenamefont {Skotiniotis},\ and\ \citenamefont
  {D\"ur}}]{Sekatski_2015}%
  \BibitemOpen
  \bibfield  {author} {\bibinfo {author} {\bibfnamefont {P.}~\bibnamefont
  {Sekatski}}, \bibinfo {author} {\bibfnamefont {M.}~\bibnamefont
  {Skotiniotis}},\ and\ \bibinfo {author} {\bibfnamefont {W.}~\bibnamefont
  {D\"ur}},\ }\bibfield  {title} {\bibinfo {title} {No-signaling bounds for
  quantum cloning and metrology},\ }\href
  {https://doi.org/10.1103/PhysRevA.92.022355} {\bibfield  {journal} {\bibinfo
  {journal} {Phys. Rev. A}\ }\textbf {\bibinfo {volume} {92}},\ \bibinfo
  {pages} {022355} (\bibinfo {year} {2015})}\BibitemShut {NoStop}%
\bibitem [{\citenamefont {Cirac}\ \emph {et~al.}(1999)\citenamefont {Cirac},
  \citenamefont {Ekert},\ and\ \citenamefont {Macchiavello}}]{Cirac_1999}%
  \BibitemOpen
  \bibfield  {author} {\bibinfo {author} {\bibfnamefont {J.~I.}\ \bibnamefont
  {Cirac}}, \bibinfo {author} {\bibfnamefont {A.~K.}\ \bibnamefont {Ekert}},\
  and\ \bibinfo {author} {\bibfnamefont {C.}~\bibnamefont {Macchiavello}},\
  }\bibfield  {title} {\bibinfo {title} {Optimal purification of single
  qubits},\ }\href {https://doi.org/10.1103/PhysRevLett.82.4344} {\bibfield
  {journal} {\bibinfo  {journal} {Phys. Rev. Lett.}\ }\textbf {\bibinfo
  {volume} {82}},\ \bibinfo {pages} {4344} (\bibinfo {year}
  {1999})}\BibitemShut {NoStop}%
\bibitem [{\citenamefont {Masullo}\ \emph {et~al.}(2005)\citenamefont
  {Masullo}, \citenamefont {Ricci},\ and\ \citenamefont
  {De~Martini}}]{Masullo_2005}%
  \BibitemOpen
  \bibfield  {author} {\bibinfo {author} {\bibfnamefont {L.}~\bibnamefont
  {Masullo}}, \bibinfo {author} {\bibfnamefont {M.}~\bibnamefont {Ricci}},\
  and\ \bibinfo {author} {\bibfnamefont {F.}~\bibnamefont {De~Martini}},\
  }\bibfield  {title} {\bibinfo {title} {Generalized universal cloning and
  purification in quantum information by multistep state symmetrization},\
  }\href {https://doi.org/10.1103/PhysRevA.72.060304} {\bibfield  {journal}
  {\bibinfo  {journal} {Phys. Rev. A}\ }\textbf {\bibinfo {volume} {72}},\
  \bibinfo {pages} {060304(R)} (\bibinfo {year} {2005})}\BibitemShut {NoStop}%
\bibitem [{\citenamefont {Chiribella}(2011)}]{Chiribella_2011}%
  \BibitemOpen
  \bibfield  {author} {\bibinfo {author} {\bibfnamefont {G.}~\bibnamefont
  {Chiribella}},\ }\bibfield  {title} {\bibinfo {title} {On quantum estimation,
  quantum cloning and finite quantum de finetti theorems},\ }in\ \href@noop {}
  {\emph {\bibinfo {booktitle} {Theory of Quantum Computation, Communication,
  and Cryptography}}},\ \bibinfo {editor} {edited by\ \bibinfo {editor}
  {\bibfnamefont {W.}~\bibnamefont {van Dam}}, \bibinfo {editor} {\bibfnamefont
  {V.~M.}\ \bibnamefont {Kendon}},\ and\ \bibinfo {editor} {\bibfnamefont
  {S.}~\bibnamefont {Severini}}}\ (\bibinfo  {publisher} {Springer Berlin
  Heidelberg},\ \bibinfo {address} {Berlin, Heidelberg},\ \bibinfo {year}
  {2011})\ pp.\ \bibinfo {pages} {9--25}\BibitemShut {NoStop}%
\bibitem [{\citenamefont {Hu}\ \emph {et~al.}(2023)\citenamefont {Hu},
  \citenamefont {Ouyang},\ and\ \citenamefont {Tomamichel}}]{Hu_2023}%
  \BibitemOpen
  \bibfield  {author} {\bibinfo {author} {\bibfnamefont {Y.}~\bibnamefont
  {Hu}}, \bibinfo {author} {\bibfnamefont {Y.}~\bibnamefont {Ouyang}},\ and\
  \bibinfo {author} {\bibfnamefont {M.}~\bibnamefont {Tomamichel}},\ }\bibfield
   {title} {\bibinfo {title} {Privacy and correctness trade-offs for
  information-theoretically secure quantum homomorphic encryption},\ }\href
  {https://doi.org/10.22331/q-2023-04-13-976} {\bibfield  {journal} {\bibinfo
  {journal} {{Quantum}}\ }\textbf {\bibinfo {volume} {7}},\ \bibinfo {pages}
  {976} (\bibinfo {year} {2023})}\BibitemShut {NoStop}%
\bibitem [{\citenamefont {Ouyang}\ \emph {et~al.}(2018)\citenamefont {Ouyang},
  \citenamefont {Tan},\ and\ \citenamefont {Fitzsimons}}]{Ouyang_2018}%
  \BibitemOpen
  \bibfield  {author} {\bibinfo {author} {\bibfnamefont {Y.}~\bibnamefont
  {Ouyang}}, \bibinfo {author} {\bibfnamefont {S.-H.}\ \bibnamefont {Tan}},\
  and\ \bibinfo {author} {\bibfnamefont {J.~F.}\ \bibnamefont {Fitzsimons}},\
  }\bibfield  {title} {\bibinfo {title} {Quantum homomorphic encryption from
  quantum codes},\ }\href {https://doi.org/10.1103/PhysRevA.98.042334}
  {\bibfield  {journal} {\bibinfo  {journal} {Phys. Rev. A}\ }\textbf {\bibinfo
  {volume} {98}},\ \bibinfo {pages} {042334} (\bibinfo {year}
  {2018})}\BibitemShut {NoStop}%
\bibitem [{\citenamefont {Nielsen}(2002)}]{Nielsen_2002}%
  \BibitemOpen
  \bibfield  {author} {\bibinfo {author} {\bibfnamefont {M.~A.}\ \bibnamefont
  {Nielsen}},\ }\bibfield  {title} {\bibinfo {title} {A simple formula for the
  average gate fidelity of a quantum dynamical operation},\ }\href
  {https://doi.org/https://doi.org/10.1016/S0375-9601(02)01272-0} {\bibfield
  {journal} {\bibinfo  {journal} {Physics Letters A}\ }\textbf {\bibinfo
  {volume} {303}},\ \bibinfo {pages} {249} (\bibinfo {year}
  {2002})}\BibitemShut {NoStop}%
\bibitem [{\citenamefont {Harrow}(2013)}]{Harrow_2013}%
  \BibitemOpen
  \bibfield  {author} {\bibinfo {author} {\bibfnamefont {A.~W.}\ \bibnamefont
  {Harrow}},\ }\href {https://doi.org/10.48550/arXiv.1308.6595} {\bibinfo
  {title} {The church of the symmetric subspace}} (\bibinfo {year} {2013}),\
  \Eprint {https://arxiv.org/abs/1308.6595} {arXiv:1308.6595 [quant-ph]}
  \BibitemShut {NoStop}%
\bibitem [{\citenamefont {Hayashi}\ \emph {et~al.}(2005)\citenamefont
  {Hayashi}, \citenamefont {Hashimoto},\ and\ \citenamefont
  {Horibe}}]{Hayashi_2005}%
  \BibitemOpen
  \bibfield  {author} {\bibinfo {author} {\bibfnamefont {A.}~\bibnamefont
  {Hayashi}}, \bibinfo {author} {\bibfnamefont {T.}~\bibnamefont {Hashimoto}},\
  and\ \bibinfo {author} {\bibfnamefont {M.}~\bibnamefont {Horibe}},\
  }\bibfield  {title} {\bibinfo {title} {Reexamination of optimal quantum state
  estimation of pure states},\ }\href
  {https://doi.org/10.1103/PhysRevA.72.032325} {\bibfield  {journal} {\bibinfo
  {journal} {Phys. Rev. A}\ }\textbf {\bibinfo {volume} {72}},\ \bibinfo
  {pages} {032325} (\bibinfo {year} {2005})}\BibitemShut {NoStop}%
\bibitem [{\citenamefont {Ambainis}\ and\ \citenamefont
  {Emerson}(2007)}]{Ambainis_2007}%
  \BibitemOpen
  \bibfield  {author} {\bibinfo {author} {\bibfnamefont {A.}~\bibnamefont
  {Ambainis}}\ and\ \bibinfo {author} {\bibfnamefont {J.}~\bibnamefont
  {Emerson}},\ }\bibfield  {title} {\bibinfo {title} {Quantum t-designs: t-wise
  independence in the quantum world},\ }in\ \href
  {https://doi.org/10.1109/CCC.2007.26} {\emph {\bibinfo {booktitle}
  {Twenty-Second Annual IEEE Conference on Computational Complexity
  (CCC'07)}}}\ (\bibinfo {year} {2007})\ pp.\ \bibinfo {pages}
  {129--140}\BibitemShut {NoStop}%
\bibitem [{\citenamefont {Watrous}(2018)}]{Watrous_2018}%
  \BibitemOpen
  \bibfield  {author} {\bibinfo {author} {\bibfnamefont {J.}~\bibnamefont
  {Watrous}},\ }\href {https://doi.org/10.1017/9781316848142} {\emph {\bibinfo
  {title} {The Theory of Quantum Information}}}\ (\bibinfo  {publisher}
  {Cambridge University Press},\ \bibinfo {year} {2018})\BibitemShut {NoStop}%
\bibitem [{\citenamefont {Goodman}\ and\ \citenamefont
  {Wallach}(2000)}]{Goodman_2000}%
  \BibitemOpen
  \bibfield  {author} {\bibinfo {author} {\bibfnamefont {R.}~\bibnamefont
  {Goodman}}\ and\ \bibinfo {author} {\bibfnamefont {N.~R.}\ \bibnamefont
  {Wallach}},\ }\href
  {https://www.cambridge.org/sg/academic/subjects/mathematics/algebra/representations-and-invariants-classical-groups?format=PB&isbn=9780521663489}
  {\emph {\bibinfo {title} {Representations and invariants of the classical
  groups}}}\ (\bibinfo  {publisher} {Cambridge University Press},\ \bibinfo
  {year} {2000})\BibitemShut {NoStop}%
\bibitem [{\citenamefont {Bacon}\ \emph {et~al.}(2006)\citenamefont {Bacon},
  \citenamefont {Chuang},\ and\ \citenamefont {Harrow}}]{Bacon_2006}%
  \BibitemOpen
  \bibfield  {author} {\bibinfo {author} {\bibfnamefont {D.}~\bibnamefont
  {Bacon}}, \bibinfo {author} {\bibfnamefont {I.~L.}\ \bibnamefont {Chuang}},\
  and\ \bibinfo {author} {\bibfnamefont {A.~W.}\ \bibnamefont {Harrow}},\
  }\bibfield  {title} {\bibinfo {title} {Efficient quantum circuits for schur
  and clebsch-gordan transforms},\ }\href
  {https://doi.org/10.1103/PhysRevLett.97.170502} {\bibfield  {journal}
  {\bibinfo  {journal} {Phys. Rev. Lett.}\ }\textbf {\bibinfo {volume} {97}},\
  \bibinfo {pages} {170502} (\bibinfo {year} {2006})}\BibitemShut {NoStop}%
\bibitem [{\citenamefont {Dankert}\ \emph {et~al.}(2009)\citenamefont
  {Dankert}, \citenamefont {Cleve}, \citenamefont {Emerson},\ and\
  \citenamefont {Livine}}]{Dankert_2009}%
  \BibitemOpen
  \bibfield  {author} {\bibinfo {author} {\bibfnamefont {C.}~\bibnamefont
  {Dankert}}, \bibinfo {author} {\bibfnamefont {R.}~\bibnamefont {Cleve}},
  \bibinfo {author} {\bibfnamefont {J.}~\bibnamefont {Emerson}},\ and\ \bibinfo
  {author} {\bibfnamefont {E.}~\bibnamefont {Livine}},\ }\bibfield  {title}
  {\bibinfo {title} {Exact and approximate unitary 2-designs and their
  application to fidelity estimation},\ }\href
  {https://doi.org/10.1103/PhysRevA.80.012304} {\bibfield  {journal} {\bibinfo
  {journal} {Phys. Rev. A}\ }\textbf {\bibinfo {volume} {80}},\ \bibinfo
  {pages} {012304} (\bibinfo {year} {2009})}\BibitemShut {NoStop}%
\bibitem [{\citenamefont {Low}(2010)}]{Low_2010}%
  \BibitemOpen
  \bibfield  {author} {\bibinfo {author} {\bibfnamefont {R.~A.}\ \bibnamefont
  {Low}},\ }\href@noop {} {\bibinfo {title} {Pseudo-randomness and learning in
  quantum computation}} (\bibinfo {year} {2010}),\ \Eprint
  {https://arxiv.org/abs/1006.5227} {arXiv:1006.5227 [quant-ph]} \BibitemShut
  {NoStop}%
\bibitem [{\citenamefont {Nakata}\ \emph {et~al.}(2021)\citenamefont {Nakata},
  \citenamefont {Zhao}, \citenamefont {Okuda}, \citenamefont {Bannai},
  \citenamefont {Suzuki}, \citenamefont {Tamiya}, \citenamefont {Heya},
  \citenamefont {Yan}, \citenamefont {Zuo}, \citenamefont {Tamate},
  \citenamefont {Tabuchi},\ and\ \citenamefont {Nakamura}}]{Nakata_2021}%
  \BibitemOpen
  \bibfield  {author} {\bibinfo {author} {\bibfnamefont {Y.}~\bibnamefont
  {Nakata}}, \bibinfo {author} {\bibfnamefont {D.}~\bibnamefont {Zhao}},
  \bibinfo {author} {\bibfnamefont {T.}~\bibnamefont {Okuda}}, \bibinfo
  {author} {\bibfnamefont {E.}~\bibnamefont {Bannai}}, \bibinfo {author}
  {\bibfnamefont {Y.}~\bibnamefont {Suzuki}}, \bibinfo {author} {\bibfnamefont
  {S.}~\bibnamefont {Tamiya}}, \bibinfo {author} {\bibfnamefont
  {K.}~\bibnamefont {Heya}}, \bibinfo {author} {\bibfnamefont {Z.}~\bibnamefont
  {Yan}}, \bibinfo {author} {\bibfnamefont {K.}~\bibnamefont {Zuo}}, \bibinfo
  {author} {\bibfnamefont {S.}~\bibnamefont {Tamate}}, \bibinfo {author}
  {\bibfnamefont {Y.}~\bibnamefont {Tabuchi}},\ and\ \bibinfo {author}
  {\bibfnamefont {Y.}~\bibnamefont {Nakamura}},\ }\bibfield  {title} {\bibinfo
  {title} {Quantum circuits for exact unitary $t$-designs and applications to
  higher-order randomized benchmarking},\ }\href
  {https://doi.org/10.1103/PRXQuantum.2.030339} {\bibfield  {journal} {\bibinfo
   {journal} {PRX Quantum}\ }\textbf {\bibinfo {volume} {2}},\ \bibinfo {pages}
  {030339} (\bibinfo {year} {2021})}\BibitemShut {NoStop}%
\bibitem [{\citenamefont {Haferkamp}(2022)}]{Haferkamp_2022}%
  \BibitemOpen
  \bibfield  {author} {\bibinfo {author} {\bibfnamefont {J.}~\bibnamefont
  {Haferkamp}},\ }\bibfield  {title} {\bibinfo {title} {Random quantum circuits
  are approximate unitary {$t$}-designs in depth
  {$O\left(nt^{5+o(1)}\right)$}},\ }\href
  {https://doi.org/10.22331/q-2022-09-08-795} {\bibfield  {journal} {\bibinfo
  {journal} {{Quantum}}\ }\textbf {\bibinfo {volume} {6}},\ \bibinfo {pages}
  {795} (\bibinfo {year} {2022})}\BibitemShut {NoStop}%
\bibitem [{\citenamefont {Tomamichel}(2012)}]{Tomamichel_2012}%
  \BibitemOpen
  \bibfield  {author} {\bibinfo {author} {\bibfnamefont {M.}~\bibnamefont
  {Tomamichel}},\ }\emph {\bibinfo {title} {A framework for non-asymptotic
  quantum information theory}},\ \href {https://doi.org/10.3929/ethz-a-7356080}
  {\bibinfo {type} {Doctoral thesis}},\ \bibinfo  {school} {ETH Zurich},
  \bibinfo {address} {Zürich} (\bibinfo {year} {2012}),\ \bibinfo {note}
  {diss., Eidgenössische Technische Hochschule ETH Zürich, Nr.
  20213.}\BibitemShut {Stop}%
\bibitem [{\citenamefont {Chiribella}\ \emph {et~al.}(2008)\citenamefont
  {Chiribella}, \citenamefont {D'Ariano},\ and\ \citenamefont
  {Perinotti}}]{Chiribella_2008}%
  \BibitemOpen
  \bibfield  {author} {\bibinfo {author} {\bibfnamefont {G.}~\bibnamefont
  {Chiribella}}, \bibinfo {author} {\bibfnamefont {G.~M.}\ \bibnamefont
  {D'Ariano}},\ and\ \bibinfo {author} {\bibfnamefont {P.}~\bibnamefont
  {Perinotti}},\ }\bibfield  {title} {\bibinfo {title} {Optimal cloning of
  unitary transformation},\ }\href
  {https://doi.org/10.1103/PhysRevLett.101.180504} {\bibfield  {journal}
  {\bibinfo  {journal} {Phys. Rev. Lett.}\ }\textbf {\bibinfo {volume} {101}},\
  \bibinfo {pages} {180504} (\bibinfo {year} {2008})}\BibitemShut {NoStop}%
\bibitem [{\citenamefont {Arecchi}\ \emph {et~al.}(1972)\citenamefont
  {Arecchi}, \citenamefont {Courtens}, \citenamefont {Gilmore},\ and\
  \citenamefont {Thomas}}]{Arecchi_1972}%
  \BibitemOpen
  \bibfield  {author} {\bibinfo {author} {\bibfnamefont {F.~T.}\ \bibnamefont
  {Arecchi}}, \bibinfo {author} {\bibfnamefont {E.}~\bibnamefont {Courtens}},
  \bibinfo {author} {\bibfnamefont {R.}~\bibnamefont {Gilmore}},\ and\ \bibinfo
  {author} {\bibfnamefont {H.}~\bibnamefont {Thomas}},\ }\bibfield  {title}
  {\bibinfo {title} {Atomic coherent states in quantum optics},\ }\href
  {https://doi.org/10.1103/PhysRevA.6.2211} {\bibfield  {journal} {\bibinfo
  {journal} {Phys. Rev. A}\ }\textbf {\bibinfo {volume} {6}},\ \bibinfo {pages}
  {2211} (\bibinfo {year} {1972})}\BibitemShut {NoStop}%
\bibitem [{\citenamefont {Yang}\ \emph {et~al.}(2014)\citenamefont {Yang},
  \citenamefont {Chiribella},\ and\ \citenamefont {Adesso}}]{Yang_2014}%
  \BibitemOpen
  \bibfield  {author} {\bibinfo {author} {\bibfnamefont {Y.}~\bibnamefont
  {Yang}}, \bibinfo {author} {\bibfnamefont {G.}~\bibnamefont {Chiribella}},\
  and\ \bibinfo {author} {\bibfnamefont {G.}~\bibnamefont {Adesso}},\
  }\bibfield  {title} {\bibinfo {title} {Certifying quantumness: Benchmarks for
  the optimal processing of generalized coherent and squeezed states},\ }\href
  {https://doi.org/10.1103/PhysRevA.90.042319} {\bibfield  {journal} {\bibinfo
  {journal} {Phys. Rev. A}\ }\textbf {\bibinfo {volume} {90}},\ \bibinfo
  {pages} {042319} (\bibinfo {year} {2014})}\BibitemShut {NoStop}%
\bibitem [{\citenamefont {Gottesman}(1998)}]{Gottesman_1998}%
  \BibitemOpen
  \bibfield  {author} {\bibinfo {author} {\bibfnamefont {D.}~\bibnamefont
  {Gottesman}},\ }\href@noop {} {\bibinfo {title} {The heisenberg
  representation of quantum computers}} (\bibinfo {year} {1998}),\ \Eprint
  {https://arxiv.org/abs/quant-ph/9807006} {arXiv:quant-ph/9807006 [quant-ph]}
  \BibitemShut {NoStop}%
\bibitem [{\citenamefont {Kueng}\ and\ \citenamefont
  {Gross}(2015)}]{Kueng_2015}%
  \BibitemOpen
  \bibfield  {author} {\bibinfo {author} {\bibfnamefont {R.}~\bibnamefont
  {Kueng}}\ and\ \bibinfo {author} {\bibfnamefont {D.}~\bibnamefont {Gross}},\
  }\href@noop {} {\bibinfo {title} {Qubit stabilizer states are complex
  projective 3-designs}} (\bibinfo {year} {2015}),\ \Eprint
  {https://arxiv.org/abs/1510.02767} {arXiv:1510.02767 [quant-ph]} \BibitemShut
  {NoStop}%
\bibitem [{\citenamefont {Zhu}(2017)}]{Zhu_2017}%
  \BibitemOpen
  \bibfield  {author} {\bibinfo {author} {\bibfnamefont {H.}~\bibnamefont
  {Zhu}},\ }\bibfield  {title} {\bibinfo {title} {Multiqubit clifford groups
  are unitary 3-designs},\ }\href {https://doi.org/10.1103/PhysRevA.96.062336}
  {\bibfield  {journal} {\bibinfo  {journal} {Phys. Rev. A}\ }\textbf {\bibinfo
  {volume} {96}},\ \bibinfo {pages} {062336} (\bibinfo {year}
  {2017})}\BibitemShut {NoStop}%
\bibitem [{\citenamefont {Lawrence}\ \emph {et~al.}(2002)\citenamefont
  {Lawrence}, \citenamefont {Brukner},\ and\ \citenamefont
  {Zeilinger}}]{Lawrence_2002}%
  \BibitemOpen
  \bibfield  {author} {\bibinfo {author} {\bibfnamefont {J.}~\bibnamefont
  {Lawrence}}, \bibinfo {author} {\bibfnamefont {i.~c.~v.}\ \bibnamefont
  {Brukner}},\ and\ \bibinfo {author} {\bibfnamefont {A.}~\bibnamefont
  {Zeilinger}},\ }\bibfield  {title} {\bibinfo {title} {Mutually unbiased
  binary observable sets on n qubits},\ }\href
  {https://doi.org/10.1103/PhysRevA.65.032320} {\bibfield  {journal} {\bibinfo
  {journal} {Phys. Rev. A}\ }\textbf {\bibinfo {volume} {65}},\ \bibinfo
  {pages} {032320} (\bibinfo {year} {2002})}\BibitemShut {NoStop}%
\bibitem [{\citenamefont {Wootters}\ and\ \citenamefont
  {Fields}(1989)}]{Wootters_1989}%
  \BibitemOpen
  \bibfield  {author} {\bibinfo {author} {\bibfnamefont {W.~K.}\ \bibnamefont
  {Wootters}}\ and\ \bibinfo {author} {\bibfnamefont {B.~D.}\ \bibnamefont
  {Fields}},\ }\bibfield  {title} {\bibinfo {title} {Optimal
  state-determination by mutually unbiased measurements},\ }\href
  {https://doi.org/10.1016/0003-4916(89)90322-9} {\bibfield  {journal}
  {\bibinfo  {journal} {Annals of Physics}\ }\textbf {\bibinfo {volume}
  {191}},\ \bibinfo {pages} {363} (\bibinfo {year} {1989})}\BibitemShut
  {NoStop}%
\bibitem [{\citenamefont {Gross}(2006)}]{Gross_2006}%
  \BibitemOpen
  \bibfield  {author} {\bibinfo {author} {\bibfnamefont {D.}~\bibnamefont
  {Gross}},\ }\bibfield  {title} {\bibinfo {title} {Hudson’s theorem for
  finite-dimensional quantum systems},\ }\href
  {https://doi.org/10.1063/1.2393152} {\bibfield  {journal} {\bibinfo
  {journal} {Journal of Mathematical Physics}\ }\textbf {\bibinfo {volume}
  {47}},\ \bibinfo {pages} {122107} (\bibinfo {year} {2006})}\BibitemShut
  {NoStop}%
\bibitem [{\citenamefont {Klappenecker}\ and\ \citenamefont
  {Rotteler}(2005)}]{Klappenecker_2005}%
  \BibitemOpen
  \bibfield  {author} {\bibinfo {author} {\bibfnamefont {A.}~\bibnamefont
  {Klappenecker}}\ and\ \bibinfo {author} {\bibfnamefont {M.}~\bibnamefont
  {Rotteler}},\ }\bibfield  {title} {\bibinfo {title} {Mutually unbiased bases
  are complex projective 2-designs},\ }in\ \href
  {https://doi.org/10.1109/ISIT.2005.1523643} {\emph {\bibinfo {booktitle}
  {Proceedings. International Symposium on Information Theory, 2005. ISIT
  2005.}}}\ (\bibinfo {year} {2005})\ pp.\ \bibinfo {pages}
  {1740--1744}\BibitemShut {NoStop}%
\bibitem [{\citenamefont {Zhu}\ \emph {et~al.}(2016)\citenamefont {Zhu},
  \citenamefont {Kueng}, \citenamefont {Grassl},\ and\ \citenamefont
  {Gross}}]{Zhu_2016}%
  \BibitemOpen
  \bibfield  {author} {\bibinfo {author} {\bibfnamefont {H.}~\bibnamefont
  {Zhu}}, \bibinfo {author} {\bibfnamefont {R.}~\bibnamefont {Kueng}}, \bibinfo
  {author} {\bibfnamefont {M.}~\bibnamefont {Grassl}},\ and\ \bibinfo {author}
  {\bibfnamefont {D.}~\bibnamefont {Gross}},\ }\href@noop {} {\bibinfo {title}
  {The clifford group fails gracefully to be a unitary 4-design}} (\bibinfo
  {year} {2016}),\ \Eprint {https://arxiv.org/abs/1609.08172} {arXiv:1609.08172
  [quant-ph]} \BibitemShut {NoStop}%
\bibitem [{\citenamefont {Chau}(2005)}]{Chau_2005}%
  \BibitemOpen
  \bibfield  {author} {\bibinfo {author} {\bibfnamefont {H.~F.}\ \bibnamefont
  {Chau}},\ }\bibfield  {title} {\bibinfo {title} {Unconditionally secure key
  distribution in higher dimensions by depolarization},\ }\href
  {https://doi.org/10.1109/TIT.2005.844076} {\bibfield  {journal} {\bibinfo
  {journal} {IEEE Transactions on Information Theory}\ }\textbf {\bibinfo
  {volume} {51}},\ \bibinfo {pages} {1451} (\bibinfo {year}
  {2005})}\BibitemShut {NoStop}%
\bibitem [{\citenamefont {Cleve}\ \emph {et~al.}(2016)\citenamefont {Cleve},
  \citenamefont {Leung}, \citenamefont {Liu},\ and\ \citenamefont
  {Wang}}]{Cleve_2016}%
  \BibitemOpen
  \bibfield  {author} {\bibinfo {author} {\bibfnamefont {R.}~\bibnamefont
  {Cleve}}, \bibinfo {author} {\bibfnamefont {D.}~\bibnamefont {Leung}},
  \bibinfo {author} {\bibfnamefont {L.}~\bibnamefont {Liu}},\ and\ \bibinfo
  {author} {\bibfnamefont {C.}~\bibnamefont {Wang}},\ }\bibfield  {title}
  {\bibinfo {title} {Near-linear constructions of exact unitary 2-designs},\
  }\href {https://doi.org/10.26421/QIC16.9-10-1} {\bibfield  {journal}
  {\bibinfo  {journal} {Quantum Info. Comput.}\ }\textbf {\bibinfo {volume}
  {16}},\ \bibinfo {pages} {721–756} (\bibinfo {year} {2016})}\BibitemShut
  {NoStop}%
\bibitem [{\citenamefont {Bravyi}\ \emph {et~al.}(2022)\citenamefont {Bravyi},
  \citenamefont {Latone},\ and\ \citenamefont {Maslov}}]{Bravyi_2022}%
  \BibitemOpen
  \bibfield  {author} {\bibinfo {author} {\bibfnamefont {S.}~\bibnamefont
  {Bravyi}}, \bibinfo {author} {\bibfnamefont {J.~A.}\ \bibnamefont {Latone}},\
  and\ \bibinfo {author} {\bibfnamefont {D.}~\bibnamefont {Maslov}},\
  }\bibfield  {title} {\bibinfo {title} {6-qubit optimal clifford circuits},\
  }\href {https://doi.org/10.1038/s41534-022-00583-7} {\bibfield  {journal}
  {\bibinfo  {journal} {npj Quantum Information}\ }\textbf {\bibinfo {volume}
  {8}},\ \bibinfo {pages} {79} (\bibinfo {year} {2022})}\BibitemShut {NoStop}%
\bibitem [{\citenamefont {Roy}\ and\ \citenamefont {Scott}(2009)}]{Roy_2009}%
  \BibitemOpen
  \bibfield  {author} {\bibinfo {author} {\bibfnamefont {A.}~\bibnamefont
  {Roy}}\ and\ \bibinfo {author} {\bibfnamefont {A.~J.}\ \bibnamefont
  {Scott}},\ }\bibfield  {title} {\bibinfo {title} {Unitary designs and
  codes},\ }\href@noop {} {\bibfield  {journal} {\bibinfo  {journal} {Designs,
  codes and cryptography}\ }\textbf {\bibinfo {volume} {53}},\ \bibinfo {pages}
  {13} (\bibinfo {year} {2009})}\BibitemShut {NoStop}%
\bibitem [{\citenamefont {Hayashi}(2006)}]{Hayashi_2006}%
  \BibitemOpen
  \bibfield  {author} {\bibinfo {author} {\bibfnamefont {M.}~\bibnamefont
  {Hayashi}},\ }\href {https://books.google.com.sg/books?id=UjDLl5sP7vEC}
  {\emph {\bibinfo {title} {Quantum Information: An Introduction}}}\ (\bibinfo
  {publisher} {Springer Berlin Heidelberg},\ \bibinfo {year}
  {2006})\BibitemShut {NoStop}%
\bibitem [{\citenamefont {Davis}\ \emph {et~al.}(2014)\citenamefont {Davis},
  \citenamefont {Rabinowitz},\ and\ \citenamefont {Rheinbolt}}]{Davis_2014}%
  \BibitemOpen
  \bibfield  {author} {\bibinfo {author} {\bibfnamefont {P.}~\bibnamefont
  {Davis}}, \bibinfo {author} {\bibfnamefont {P.}~\bibnamefont {Rabinowitz}},\
  and\ \bibinfo {author} {\bibfnamefont {W.}~\bibnamefont {Rheinbolt}},\ }\href
  {https://books.google.com.sg/books?id=mbLiBQAAQBAJ} {\emph {\bibinfo {title}
  {Methods of Numerical Integration}}},\ Computer Science and Applied
  Mathematics\ (\bibinfo  {publisher} {Elsevier Science},\ \bibinfo {year}
  {2014})\BibitemShut {NoStop}%
\bibitem [{\citenamefont {Abramowitz}\ and\ \citenamefont
  {Stegun}(1965)}]{Abramowitz_1965}%
  \BibitemOpen
  \bibfield  {author} {\bibinfo {author} {\bibfnamefont {M.}~\bibnamefont
  {Abramowitz}}\ and\ \bibinfo {author} {\bibfnamefont {I.}~\bibnamefont
  {Stegun}},\ }\href {https://books.google.com.sg/books?id=MtU8uP7XMvoC} {\emph
  {\bibinfo {title} {Handbook of Mathematical Functions: With Formulas, Graphs,
  and Mathematical Tables}}},\ Applied mathematics series\ (\bibinfo
  {publisher} {Dover Publications},\ \bibinfo {year} {1965})\BibitemShut
  {NoStop}%
\bibitem [{\citenamefont {Ouyang}\ \emph {et~al.}(2023)\citenamefont {Ouyang},
  \citenamefont {Goswami}, \citenamefont {Romero}, \citenamefont {Sanders},
  \citenamefont {Hsieh},\ and\ \citenamefont {Tomamichel}}]{Ouyang_2023}%
  \BibitemOpen
  \bibfield  {author} {\bibinfo {author} {\bibfnamefont {Y.}~\bibnamefont
  {Ouyang}}, \bibinfo {author} {\bibfnamefont {K.}~\bibnamefont {Goswami}},
  \bibinfo {author} {\bibfnamefont {J.}~\bibnamefont {Romero}}, \bibinfo
  {author} {\bibfnamefont {B.~C.}\ \bibnamefont {Sanders}}, \bibinfo {author}
  {\bibfnamefont {M.-H.}\ \bibnamefont {Hsieh}},\ and\ \bibinfo {author}
  {\bibfnamefont {M.}~\bibnamefont {Tomamichel}},\ }\href@noop {} {\bibinfo
  {title} {Approximate reconstructability of quantum states and noisy quantum
  secret sharing schemes}} (\bibinfo {year} {2023}),\ \Eprint
  {https://arxiv.org/abs/2302.02509} {arXiv:2302.02509 [quant-ph]} \BibitemShut
  {NoStop}%
\bibitem [{\citenamefont {Wilde}(2013)}]{wilde_2013}%
  \BibitemOpen
  \bibfield  {author} {\bibinfo {author} {\bibfnamefont {M.~M.}\ \bibnamefont
  {Wilde}},\ }\href {https://doi.org/10.1017/CBO9781139525343} {\emph {\bibinfo
  {title} {Quantum Information Theory}}}\ (\bibinfo  {publisher} {Cambridge
  University Press},\ \bibinfo {year} {2013})\BibitemShut {NoStop}%
\bibitem [{\citenamefont {Popescu}\ and\ \citenamefont
  {Rohrlich}(1994)}]{Popescu_1994}%
  \BibitemOpen
  \bibfield  {author} {\bibinfo {author} {\bibfnamefont {S.}~\bibnamefont
  {Popescu}}\ and\ \bibinfo {author} {\bibfnamefont {D.}~\bibnamefont
  {Rohrlich}},\ }\bibfield  {title} {\bibinfo {title} {Quantum nonlocality as
  an axiom},\ }\href {https://doi.org/10.1007/BF02058098} {\bibfield  {journal}
  {\bibinfo  {journal} {Foundations of Physics}\ }\textbf {\bibinfo {volume}
  {24}},\ \bibinfo {pages} {379} (\bibinfo {year} {1994})}\BibitemShut
  {NoStop}%
\bibitem [{\citenamefont {Masanes}\ \emph {et~al.}(2006)\citenamefont
  {Masanes}, \citenamefont {Acin},\ and\ \citenamefont {Gisin}}]{Masanes_2006}%
  \BibitemOpen
  \bibfield  {author} {\bibinfo {author} {\bibfnamefont {L.}~\bibnamefont
  {Masanes}}, \bibinfo {author} {\bibfnamefont {A.}~\bibnamefont {Acin}},\ and\
  \bibinfo {author} {\bibfnamefont {N.}~\bibnamefont {Gisin}},\ }\bibfield
  {title} {\bibinfo {title} {General properties of nonsignaling theories},\
  }\href {https://doi.org/10.1103/PhysRevA.73.012112} {\bibfield  {journal}
  {\bibinfo  {journal} {Phys. Rev. A}\ }\textbf {\bibinfo {volume} {73}},\
  \bibinfo {pages} {012112} (\bibinfo {year} {2006})}\BibitemShut {NoStop}%
\bibitem [{\citenamefont {Cavalcanti}\ \emph {et~al.}(2022)\citenamefont
  {Cavalcanti}, \citenamefont {Selby}, \citenamefont {Sikora}, \citenamefont
  {Galley},\ and\ \citenamefont {Sainz}}]{Cavalcanti_2022}%
  \BibitemOpen
  \bibfield  {author} {\bibinfo {author} {\bibfnamefont {P.~J.}\ \bibnamefont
  {Cavalcanti}}, \bibinfo {author} {\bibfnamefont {J.~H.}\ \bibnamefont
  {Selby}}, \bibinfo {author} {\bibfnamefont {J.}~\bibnamefont {Sikora}},
  \bibinfo {author} {\bibfnamefont {T.~D.}\ \bibnamefont {Galley}},\ and\
  \bibinfo {author} {\bibfnamefont {A.~B.}\ \bibnamefont {Sainz}},\ }\bibfield
  {title} {\bibinfo {title} {Post-quantum steering is a stronger-than-quantum
  resource for information processing},\ }\href
  {https://doi.org/10.1038/s41534-022-00574-8} {\bibfield  {journal} {\bibinfo
  {journal} {npj Quantum Information}\ }\textbf {\bibinfo {volume} {8}},\
  \bibinfo {pages} {76} (\bibinfo {year} {2022})}\BibitemShut {NoStop}%
\bibitem [{\citenamefont {Cruzeiro}\ \emph {et~al.}(2021)\citenamefont
  {Cruzeiro}, \citenamefont {Gisin},\ and\ \citenamefont
  {Popescu}}]{Cruzeiro_2021}%
  \BibitemOpen
  \bibfield  {author} {\bibinfo {author} {\bibfnamefont {E.~Z.}\ \bibnamefont
  {Cruzeiro}}, \bibinfo {author} {\bibfnamefont {N.}~\bibnamefont {Gisin}},\
  and\ \bibinfo {author} {\bibfnamefont {S.}~\bibnamefont {Popescu}},\
  }\href@noop {} {\bibinfo {title} {Blind steering in no-signalling theories}}
  (\bibinfo {year} {2021}),\ \Eprint {https://arxiv.org/abs/2007.15502}
  {arXiv:2007.15502 [quant-ph]} \BibitemShut {NoStop}%
\bibitem [{\citenamefont {Sainz}\ \emph {et~al.}(2018)\citenamefont {Sainz},
  \citenamefont {Aolita}, \citenamefont {Piani}, \citenamefont {Hoban},\ and\
  \citenamefont {Skrzypczyk}}]{Sainz_2018}%
  \BibitemOpen
  \bibfield  {author} {\bibinfo {author} {\bibfnamefont {A.~B.}\ \bibnamefont
  {Sainz}}, \bibinfo {author} {\bibfnamefont {L.}~\bibnamefont {Aolita}},
  \bibinfo {author} {\bibfnamefont {M.}~\bibnamefont {Piani}}, \bibinfo
  {author} {\bibfnamefont {M.~J.}\ \bibnamefont {Hoban}},\ and\ \bibinfo
  {author} {\bibfnamefont {P.}~\bibnamefont {Skrzypczyk}},\ }\bibfield  {title}
  {\bibinfo {title} {A formalism for steering with local quantum
  measurements},\ }\href {https://doi.org/10.1088/1367-2630/aad8df} {\bibfield
  {journal} {\bibinfo  {journal} {New Journal of Physics}\ }\textbf {\bibinfo
  {volume} {20}},\ \bibinfo {pages} {083040} (\bibinfo {year}
  {2018})}\BibitemShut {NoStop}%
\bibitem [{\citenamefont {Perelomov}(1977)}]{Perelomov_1977}%
  \BibitemOpen
  \bibfield  {author} {\bibinfo {author} {\bibfnamefont {A.~M.}\ \bibnamefont
  {Perelomov}},\ }\bibfield  {title} {\bibinfo {title} {Generalized coherent
  states and some of their applications},\ }\href
  {https://doi.org/10.1070/PU1977v020n09ABEH005459} {\bibfield  {journal}
  {\bibinfo  {journal} {Soviet Physics Uspekhi}\ }\textbf {\bibinfo {volume}
  {20}},\ \bibinfo {pages} {703} (\bibinfo {year} {1977})}\BibitemShut
  {NoStop}%
\bibitem [{\citenamefont {Radcliffe}(1971)}]{Radcliffe_1971}%
  \BibitemOpen
  \bibfield  {author} {\bibinfo {author} {\bibfnamefont {J.~M.}\ \bibnamefont
  {Radcliffe}},\ }\bibfield  {title} {\bibinfo {title} {Some properties of
  coherent spin states},\ }\href {https://doi.org/10.1088/0305-4470/4/3/009}
  {\bibfield  {journal} {\bibinfo  {journal} {Journal of Physics A: General
  Physics}\ }\textbf {\bibinfo {volume} {4}},\ \bibinfo {pages} {313} (\bibinfo
  {year} {1971})}\BibitemShut {NoStop}%
\end{thebibliography}%

\appendix

\section{Multinomial distribution and error function}\label{apd:multinomial}
We estimate
\begin{align}
    \sum_{\nu} \frac{m!}{d^m\beta^\nu}, 
\end{align}
when $m\gg dn^2$. First we apply Stirling's approximation 
\begin{align}
    \ln \frac{m!}{\beta!} \approx  m\ln m - \sum_i \beta_i\ln\beta_i +\ln\sqrt{\frac{m}{(2\pi)^{d-1}\prod_i\beta_i }} . 
\end{align}
To maximize the above quantity under the constraint of $\sum_i\beta_i=m$, we use a Lagrange multiplier $\gamma$. The first order expansion is 
\begin{align}
    \delta \Big(\ln \frac{m!}{\beta!} \Big)\approx -\sum_i \Big(\ln \beta_i + \frac{1}{2\beta_i}- \gamma \Big) \delta \beta_i. 
\end{align}
The maximum is achieved when $\beta_{\max}=\frac{m}{d} \mathbbm{1}$, where $\mathbbm{1}=(1,...,1)$. The second order expansion is 
\begin{align}
    \delta^2 \Big(\ln \frac{m!}{\beta!} \Big)\approx -\sum_i \frac{2\beta_i-1}{4\beta_i^2} (\delta\beta_i)^2. 
\end{align}
For $\beta \in [(\frac{m}{d}-\sqrt{\frac{m}{d}})\mathbbm{1}, (\frac{m}{d}+\sqrt{\frac{m}{d}})\mathbbm{1}]$, 
\begin{align}\label{eqn:multinoial_approx}
    \frac{m!}{\beta!} \approx \frac{d^{\frac{d}{2}+m}}{\sqrt{(2\pi m)^{d-1}}} e^{ - \frac{d}{2m}\sum_i(\beta_i - \frac{m}{d})^2}. 
\end{align}
When $ m\gg dn^2$, $\beta^\nu\in [(\frac{m}{d}-\sqrt{\frac{m}{d}})\mathbbm{1}, (\frac{m}{d}+\sqrt{\frac{m}{d}})\mathbbm{1}]$, and thus Eq.~\eqref{eqn:multinoial_approx} approximates $\beta^{\nu}$ well. We obtain
\begin{align}
    \sum_{\nu} \frac{m!}{d^m \beta^\nu!} \approx \sum_\nu \sqrt{\frac{d^d}{(2\pi m)^{d-1}}} e^{ - \frac{d}{2m}\sum_i(\beta_i^{\nu} - \frac{m}{d})^2}. 
\end{align}
Now we define 
\begin{align}
    f_i(\beta^\nu) = \sqrt{\frac{d}{(2\pi m)^{\frac{d-1}{d}}}} e^{-\frac{d}{2m(d-1)} \sum_{j\neq i} (\beta_j^\nu - \frac{m}{d})^2}, 
\end{align}
for $i\in \mathbbm{Z}_d$. Therefore, 
\begin{align}
    \sum_{\nu} \frac{m!}{d^m \beta!} \approx \sum_\nu \prod_i f_i. 
\end{align}
We apply Holder's inequality to obtain
\begin{align}
     \sum_\nu \prod_i f_i\leq \prod_i \left(\sum_\nu f_i(\beta^\nu)^d\right)^{\frac{1}{d}}.  
\end{align}
For all $\nu$ and $i$, there exists $\beta^{\nu,i}$ satisfying $\beta_j^{\nu,i}\in [\frac{m}{d}-\frac{n}{2},\frac{m}{d}+\frac{n}{2}]$ for and $\beta_j^{\nu,i}=\beta_j^{\nu} \mod (n+1)$ for $j\neq i$ such that 
\begin{align}
    f_i(\beta^\nu) \leq f_i(\beta^{\nu,i}). 
\end{align}
Therefore, 
\begin{align}
    \sum_\nu f_i(\beta^\nu)^d \leq \sum_\nu f_i(\beta^{\nu,i})^d \leq \sum_{\beta^{(i)}} f_i(\beta^{(i)})^d. 
\end{align}
where $\beta^{(i)}$ satisfies $\beta_j^{(i)}\in [\frac{m}{d}-\frac{n}{2},\frac{m}{d}+\frac{n}{2}]$ for $j\neq i$. In the last step, we replace the summation with integration to estimate
\begin{align}
    \sum_{\beta^{(i)}} \approx \prod_{j\neq i} \left(\int_{\frac{m}{d}-\frac{n}{2}}^{\frac{m}{d}+\frac{n}{2}} \d \beta_j^{(i)}\right). 
\end{align}
The integration is then a Gaussian integral, i.e.,
\begin{align}
    \sum_\nu \frac{m!}{d^m\beta^\nu!}\leq  d^{\frac{1}{2}}\left(1-\frac{1}{d}\right)^{\frac{d-1}{2}} \erf\left(\frac{dn}{2\sqrt{2(d-1)m}}\right)^{d-1}. 
\end{align}

\section{Integrals over the Haar measure}\label{apd:haar_measure}
We now show that 
\begin{align}
    \int \d U \proj{J_U}^{\otimes l} = \rho_0^l, 
\end{align}
where 
\begin{align}
    \rho_0^l & = \sum_{\lambda} \frac{|\mathbbm{Q}_\lambda^l|}{d^l |\mathbbm{P}_\lambda^l|} \Pi_\lambda^l, \\
    \Pi_\lambda & = P_\lambda^l \otimes P_\lambda^l \otimes \proj{\Psi_\lambda^l}, \\
    \ket{\Psi_\lambda} & = \frac{1}{\sqrt{|\mathbbm{Q}_\lambda^l|}} \sum_{q_\lambda^l} \ket{q_\lambda^l,\lambda}\otimes \ket{q_\lambda^l,\lambda}. 
\end{align}
As is implied by~\cite{Bacon_2006}, the Schur transform $U_{\Sch}$ which transforms the computational basis $\ket{i}$ to the Schur basis $\ket{p_\lambda^l,q_\lambda^l,\lambda}$ is made real. Thus the maximally entangled state $\ket{\Psi^{d,l}}$ satisfies
\begin{align}
    & \ket{\Psi^{d,l}} = (U_{\Sch}\otimes U_{\Sch}^*)\ket{\Psi^{d,l}} \nonumber \\
    & = \sum_{\lambda,p_\lambda^l, q_\lambda^l} \ket{p_\lambda^l,q_\lambda^l,\lambda}\otimes \ket{p_\lambda^l,q_\lambda^l,\lambda}. 
\end{align}
Therefore, 
\begin{widetext}
\begin{align}
    \int \d U \proj{J_U}^{\otimes l} & = \frac{1}{d^l}\sum_{\lambda, p_\lambda^l, q_\lambda^l} \sum_{\mu, r_\mu^l, s_\mu^l}\int \d U U^{\otimes l} \ketbra{p_\lambda^l, q_\lambda^l,\lambda}{r_\mu^l, s_\mu^l,\mu} (U^{\dagger })^{\otimes l} \otimes \ketbra{p_\lambda^l, q_\lambda^l,\lambda}{r_\mu^l, s_\mu^l,\mu} \nonumber \\
    & = \frac{1}{d^l} \sum_{\lambda, p_\lambda^l, q_\lambda^l} \sum_{\mu, r_\mu^l, s_\mu^l}\int \d U U_\lambda  \ketbra{p_\lambda^l,\lambda}{r_\mu^l,\mu} U_\mu^\dagger \otimes \ketbra{q_\lambda^l,\lambda}{s_\mu^l,\mu} \otimes \ketbra{p_\lambda^l,\lambda}{r_\mu^l,\mu} \otimes \ketbra{q_\lambda^l,\lambda}{s_\mu^l,\mu} \nonumber \\
    & = \frac{1}{d^l} \sum_{\lambda, p_\lambda^l} \sum_{q_\lambda^l,s_\lambda^l} \frac{1}{|\mathbbm{P}_\lambda^l|} P_{\lambda}^l \otimes \ketbra{q_\lambda^l,\lambda}{s_\lambda^l,\lambda} \otimes \ketbra{p_\lambda^l,\lambda}{p_\lambda^l,\lambda} \otimes \ketbra{q_\lambda^l,\lambda}{s_\lambda^l,\lambda}\nonumber \\
    & = \sum_{\lambda} \frac{|\mathbbm{Q}_\lambda^l|}{d^l |\mathbbm{P}_\lambda^l|} P_{\lambda}^l \otimes P_{\lambda}^l \otimes \proj{\Psi_\lambda^l} = \sum_{\lambda} \frac{|\mathbbm{Q}_\lambda^l|}{d^l |\mathbbm{P}_\lambda^l|} \Pi_\lambda^l, 
\end{align}
\end{widetext}
where we have transformed to the isomorphic space, i.e.,
\begin{align}
    \mathbbm{C}^{d^l} \simeq \bigoplus_\lambda \mathbbm{P}_\lambda^l \otimes \mathbbm{Q}_\lambda^l, 
\end{align}
in the second step and applied Schur's lemma, i.e.,
\begin{align}
    \int \d U U_\lambda \ketbra{p_\lambda^l,\lambda}{r_\mu^l,\mu} U_\mu^\dagger = \frac{1}{|\mathbbm{P}_\lambda^l|} \delta_{\lambda\mu} \delta_{p_\lambda^l r_\mu^l} P_\lambda^l, 
\end{align}
in the third step. One should be aware that $P_\lambda^l$ and $\ket{\Psi_\lambda^l}$ alone only make sense in $\bigoplus_\lambda \mathbbm{P}_{\lambda}^l \otimes \mathbbm{Q}_{\lambda}^l$. However, $\Pi_\lambda^l$ makes sense in both $\bigoplus_\lambda \mathbbm{P}_{\lambda}^l \otimes \mathbbm{Q}_{\lambda}^l$ and $\mathbbm{C}^{d^l}$, as 
\begin{align}
    \Pi_\lambda^l & = \frac{1}{|\mathbbm{Q}_\lambda^l|}\sum_{p_\lambda^l, r_\lambda^l, q_\lambda^l, s_\lambda^l }  \ketbra{p_\lambda^l,\lambda}{p_\lambda^l,\lambda}\otimes \ketbra{q_\lambda^l,\lambda}{s_\lambda^l,\lambda} \nonumber \\
    & \quad \quad \quad \quad \quad \quad \quad \otimes \ketbra{r_\lambda^l,\lambda}{r_\lambda^l,\lambda} \otimes \ketbra{q_\lambda^l,\lambda}{s_\lambda^l,\lambda} \nonumber \\
    & = \frac{1}{|\mathbbm{Q}_\lambda^l|} \sum_{p_\lambda^l, r_\lambda^l, q_\lambda^l, s_\lambda^l }  \ketbra{p_\lambda^l,q_\lambda^l,\lambda}{p_\lambda^l,s_\lambda^l,\lambda} \nonumber \\
    & \quad \quad \quad \quad \quad \quad \quad  \otimes \ketbra{r_\lambda^l,q_\lambda^l,\lambda}{r_\lambda^l,s_\lambda^l,\lambda}. 
\end{align}

\section{Pinching inequality}\label{apd:pinching}
Let $\{P_i\}$ be a set of projectors such that 
\begin{align}
    \sum_{i=1}^n P_i = \mathbbm{I}.
\end{align}
The pinching map $\mathcal{P}$ is defined as 
\begin{align}
    \mathcal{P}(\rho) = \sum_i P_i \rho P_i. 
\end{align}
The pinching inequality~\cite{Hayashi_2006} is 
\begin{align}
    \rho \leq n \mathcal{P}(\rho). 
\end{align}
Now suppose that the system is decomposed into two subsystems, $\A$ and $\B$. Let $P_i =  \proj{i}_{\A} \otimes \mathbbm{I}_{\B}$. The pinching inequality turns into
\begin{align}
    \rho_{\A\B} \leq \dim(\A) \mathcal{P}(\rho_{\A\B}). 
\end{align}
Notice that 
\begin{align}
    \mathcal{P}(\rho_{\A\B}) & = \sum_i \proj{i}_{\A} \otimes \mathbbm{I}_{\B} \rho_{\A\B}\proj{i}_{\A} \otimes \mathbbm{I}_{\B} \nonumber\\
    & = \sum_i \proj{i}_{\A} \otimes  \Tr_{\A} (\rho_{\A\B} \proj{i}_{\A} \otimes \mathbbm{I}_{\B} ) \nonumber \\
    & \leq \sum_i \mathbbm{I}_{\A} \otimes  \Tr_{\A} (\rho_{\A\B} \proj{i}_{\A} \otimes \mathbbm{I}_{\B} ) \nonumber \\
    & = \mathbbm{I}_{\A} \otimes \rho_{\B}. 
\end{align}
Therefore, 
\begin{align}
    \rho_{\A\B} \leq \dim(\A) \mathbbm{I}\otimes \rho_{\B}. 
\end{align}

\section{Rank inequality}\label{apd:rank}
Now we prove 
\begin{align}\label{eqn:rank_inequality}
    F(\rho,\sigma ) \leq N \max_i F(\rho, \proj{\psi_i}), 
\end{align}
where $\sigma=\sum_{i=1}^{N} p_i\proj{\psi_i}$. Let $\ket{\phi}$ be the purification of $\rho$ and 
\begin{align}
    \ket{\psi} = \sum_i \sqrt{p_i} \ket{\psi_i} \otimes \ket{i} = \sum_i \sqrt{p_i} \ket{\psi_i,i}, 
\end{align}
be the purification of $\sigma$. By Ulhmann's theorem, 
\begin{align}
    F(\rho,\sigma) & = \max_{\ket{\phi}}|\!\braket{\phi}{\psi}\!|^2 \nonumber \\
    & = \sum_{ij} \sqrt{p_i p_j}\braket{\phi}{\psi_i,i}\!\braket{\psi_j,j}{\phi}\nonumber \\
    & \leq \sum_{ij} \sqrt{p_i p_j} |\!\braket{\phi}{\psi_i,i}\!\braket{\psi_j,j}{\phi}\!|. 
\end{align}
Note that 
\begin{align}
    |\!\braket{\phi}{\psi_i,i}\!\braket{\psi_j,j}{\phi}\!| \leq \max_{i} |\!\braket{\phi}{\psi_i,i}\!|^2. 
\end{align}
Therefore, 
\begin{align}
    F(\rho,\sigma) \leq (\sum_i\sqrt{p_i} )^2\max_i |\!\braket{\phi}{\psi_i,i}\!|^2. 
\end{align}
Because $\sqrt{x}$ is a concave function, we obtain by Jensen's inequality
\begin{align}
    \sum_{i} \sqrt{p_i} \leq \sqrt{N}.
\end{align}
Therefore, 
\begin{align}
    F(\rho,\sigma) & \leq  N \max_i |\!\braket{\phi}{\psi_i,i}\!|^2 \nonumber \\
    & = N \max_i F(\proj{\phi}, \proj{\psi_i,i}). 
\end{align}
By data processing inequality, we obtain 
\begin{align}
    F(\rho,\sigma) \leq N \max_i F(\rho, \proj{\psi_i}). 
\end{align}

\section{Spin coherent design}\label{apd:spin_design}

We want to find $\{w_i\}$ and $\{(\theta_i,\phi_i)\}$ such that 
\begin{align}
    \int \d\Omega \proj{s,\theta,\phi}^{\otimes l} = \sum_{i} w_i \proj{s,\theta_i,\phi_i}^{\otimes l}, 
\end{align}
or equivalently 
\begin{align}\label{eqn:matrix_equation}
    \int \d\Omega \proj{ls,\theta,\phi} = \sum_{i} w_i \proj{ls,\theta_i,\phi_i}. 
\end{align}
According to~\cite{Arecchi_1972}, 
\begin{align}\label{eqn:basis_transform}
    \ket{ls,\theta,\phi} =\sum_{p=-ls}^{ls} \sqrt{\binom{2ls}{ls+p}}   & \sin^{ls+p} \frac{\theta}{2} \cos^{ls-p} \frac{\theta}{2} \nonumber \\
    & e^{ - \iu (ls+p)\phi } \ket{ls,p}. 
\end{align}
We substitute Eq.~\eqref{eqn:basis_transform} into Eq.~\eqref{eqn:matrix_equation}. In order to ensure that the matrix equation Eq.~\eqref{eqn:matrix_equation} holds, we equivalently ensure that the $(p,p')$ entry equation
\begin{align}
    & \int \d \phi \d \theta \sin\theta   \sin^{2ls+p+p'}  \frac{\theta}{2} \cos^{2ls-p-p'}  \frac{\theta}{2} e^{ - \iu (p-p')\phi }\nonumber  \\
    & = \sum_i  w_i  \sin^{2ls+p+p'}  \frac{\theta_i}{2} \cos^{2ls-p-p'}  \frac{\theta_i}{2} e^{ - \iu (p-p')\phi_i }, 
\end{align}
holds for any $p$ and $p'$.

One way to achieve the goal is to assume $\{w_i\} \simeq \{w_{\phi,k}\} \times \{w_{\theta,j}\}$ and $\{\theta_i,\phi_i\}\simeq \{\phi_k\} \times \{\theta_j\} $. With the assumption, we only need to ensure that for any $p$ and $p'$
\begin{align}\label{eqn:int_phi}
    \int_{0}^{2\pi} \d\phi  e^{ - \iu (p-p')\phi}  = \sum_k  w_{\phi,k}  e^{ - \iu (p-p')\phi_k }, 
\end{align}
and 
\begin{align}\label{eqn:int_theta}
    & \int_{0}^{\pi} \d\theta \sin\theta \sin^{2ls+p+p'}  \frac{\theta}{2} \cos^{2ls-p-p'} \frac{\theta}{2} \nonumber \\ 
    & = \sum_j  w_{\theta, j}  \sin^{2ls+p+p'}  \frac{\theta_j}{2} \cos^{2ls-p-p'}  \frac{\theta_j}{2}.
\end{align}

For $\phi$, it's quite obvious. First note that $p-p'$ is an integer due to the property of the angular momentum. Eq.\eqref{eqn:int_phi} holds for any $p$ and $p'$ when $\{w_{\phi,k} = \frac{1}{2ls+1}, k = 0,...,2ls \}$ and $\{\phi_{k} = \frac{2\pi k}{2ls+1}, k =0,..., 2ls\}$. Moreover, Eq.~\eqref{eqn:int_phi} is non-zero if and only if $p=p'$. 

For $\theta$, it's more complicated: First, notice that $p=p'$, otherwise the integration over $\phi$ will result in zero and there's no need to consider the integration over $\theta$. Second notice that $ls-p$ and $ls+p$ are both integers due to the property of the angular momentum. We replace $x = \cos\theta$ and denote the corresponding sets of weights and points by $\{w_{x,j}\}$ and $\{x_j\}$. For $p\geq 0$, 
\begin{align}
    & \sin^{2ls+2p}  \frac{\theta}{2} \cos^{2ls-2p} \frac{\theta}{2}\nonumber \\
    & =\frac{1}{2^{2ls}} \sin^{2ls-2p} \theta (1 - \cos \theta)^{2p} \nonumber \\ 
    & = \frac{1}{2^{2ls}} (1-x^2)^{ls-p} (1-x)^{2p} \nonumber \\
    & = \frac{1}{2^{2ls}} (1+x)^{ls-p} (1-x)^{ls+p},
\end{align}
and for $p<0$, 
\begin{align}
    & \sin^{2ls-2|p|}  \frac{\theta}{2} \cos^{2ls+2|p|} \frac{\theta}{2}\nonumber \\
    & =\frac{1}{2^{2ls}} \sin^{2ls-2|p|} \theta (1 + \cos \theta)^{2|p|} \nonumber \\ 
    & = \frac{1}{2^{2ls}} (1-x^2)^{ls-|p|} (1+x)^{2|p|} \nonumber \\
    & = \frac{1}{2^{2ls}} (1-x)^{ls+p} (1+x)^{ls-p}. 
\end{align}
Both functions are polynomials in the form of 
\begin{align}
    f_p(x) = \frac{1}{2^{2ls}} (1+x)^{ls-p} (1- x)^{ls+p}, 
\end{align}
with degree at most $2ls$. Eq.~\eqref{eqn:int_theta} thus becomes that for any $p$, it holds that 
\begin{align}
    \int_{-1}^{1} \d x \sqrt{1-x^2} f_p(x) = \sum_j w_{x,j} f_p(x_j). 
\end{align}
We choose $\{w_{x,j}\}$ and $\{x_j\}$ following the Gauss-Chebyshev quadrature rule with $|\{x_j\}|=\frac{2ls+1}{2}$. Given the weight $w(x)$ and the interval $[a,b]$, there exists a set of orthonormal polynomials $\{p_i(x),i=0,1,...,l\}$ ($i$ for degree $i$) such that
\begin{align}
    (p_i,p_j) = \int_a^b \d x w(x) p_i^*(x) p_j(x) = \delta_{ij}. 
\end{align}
Let $\{x_i,i=1,...,l\}$ denote the zeros of $p_l(x)$. There exists a set of weights $\{w_i,i=1,...,l\}$ such that 
\begin{align}
    \int_a^b \d x w(x) f(x) = \sum_{i=1}^{l} w_i f(x_i),  
\end{align}
for any polynomial $f(x)$ with degree at most $2l-1$~\cite{Davis_2014}. Here we use the Gauss-Chebyshev quadrature rule for $w(x) = \sqrt{1-x^2}$ and $[a,b]=[-1,1]$ and Chebyshev polynomials~\cite{Abramowitz_1965}. 

Combining both $\{w_{\phi,k}\}$ and $\{\phi_k\}$ for $\phi$ and $\{w_{\theta,j}\}$ and $\{\theta_j\}$ for $\theta$, we construct $\{w_{i}\}$ and $\{(\theta_i,\phi_i)\}$ for both $\theta$ and $\phi$ with $|\{(\theta_i,\phi_i)\}|= \frac{(2ls+1)^2}{2}$. 

\section{Trace distance and fidelity}\label{apd:trace_fidelity}
When at least one of $\rho$ and $\sigma$ is pure, it holds that  
\begin{align}
    \|\rho-\sigma\|_{\Tr} \geq 1 - F(\rho,\sigma).
\end{align}
The proof~\cite{Ouyang_2023} that the authors are aware of is as follows. Without loss of generality, we assume $\rho = \proj{\psi}$ is pure. The fidelity satisfies
\begin{align}
    F(\proj{\psi},\sigma) = \sandwich{\psi}{\sigma}{\psi}. 
\end{align}
As is shown in~\cite[Lemma 9.1.1]{wilde_2013}, the trace distance satisfies
\begin{align}
    \|\proj{\psi}-\sigma\|_{\Tr} = \max_{0\leq \Lambda\leq \mathbbm{I}} \Tr\left(\Lambda (\proj{\psi}-\sigma)\right). 
\end{align}
Let $\Lambda = \proj{\psi}$, we obtain
\begin{align}
    \|\proj{\psi}-\sigma\|_{\Tr} \geq 1 -  \sandwich{\psi}{\sigma}{\psi} = 1 - F(\proj{\psi},\sigma). 
\end{align}

\nocite{*}

\end{document}